\documentclass{article}%
\usepackage{amssymb}
\usepackage{amsmath}
\usepackage{makeidx}
\usepackage{amsfonts}
\usepackage{graphicx}
\usepackage{epsfig}%
\include{psfig.tex}
\usepackage{cite}%
\begin{document}

\author{Hartmut Wachter\thanks{E-Mail: Hartmut.Wachter@gmx.de}\\An der Schafscheuer 56\\D-91781 Wei\ss enburg, Federal Republic of Germany}
\title{Scattering of a particle on the $q$-de\-formed Euclidean space}
\maketitle
\date{}

\begin{abstract}
We develop a formalism for the scattering of a particle on the $q$%
-de\-\-formed Euclidean space. We write down $q$-ver\-sions of the
Lippmann-Schwinger equation. Their iterative solutions for a weak scattering
potential lead us to $q$-ver\-sions of the Born series. With the expressions
for the wave functions of the scattered particle, we can write down S-matrix
elements. We show that these S-ma\-trix elements satisfy unitarity conditions.
Considerations about the interaction picture for a quantum system in the
$q$-de\-\-formed Euclidean space and a discussion of a $q$-ver\-sion of
time-de\-\-pen\-\-dent perturbation theory conclude our studies.

\end{abstract}
\tableofcontents

\section{Introduction}

The Euclidean space and the Minkowski space show special symmetries, also
known as rotational symmetry, translational symmetry, or Lorentz symmetry.
These space-time symmetries are mathematically described by Lie groups.
Thus, the Euclidean space or the Minkowski space is a representation of
a Lie group.

In the 1980s, it became possible to deform Lie groups and matrix groups
together with their representation spaces within the category of Hopf algebras
\cite{Drinfeld:1985rx,Drinfeld:1986,1985LMaPh..10...63J,Woronowicz:1987b,Takeuchi:1990}%
. This insight led to the construction of $q$-de\-formed quantum groups and
quantum spaces. The latter include the $q$-Eu\-clidean space and the
$q$-Minkowski space \cite{Faddeev:1987ih,CarowWatamura:1990nk}. These
$q$-de\-formed quantum spaces are noncommutative algebras, i.~e., their
coordinate generators no longer commute. Two non-commuting coordinates cannot
be measured simultaneously with infinite precision because the measurement of
one coordinate interferes with the measurement results for the other
coordinates. Therefore, the accuracy of a position measurement should be
limited by a "smallest length"
\cite{Heisenberg:1930,Heisenberg:1938,Snyder:1947b} if space-time is described
by a $q$-de\-formed quantum space.

It has been argued that the problem with ultraviolet divergences in quantum
field theories does not arise in a discrete space-time since its smallest
length should constrain the values of momentum \cite{Hagar:2014}. With this in
mind, it is appealing to formulate a quantum theory or quantum field theory on
a $q$-de\-formed space. In Ref.~\cite{Wachter:2020B} and
Ref.~\cite{Wachter:2021B}, we have discussed $q$-ver\-sions of the
Schr\"{o}dinger equation for a free nonrelativistic particle and
$q$-ver\-sions of the free Klein-Gordon equation on $q$-Eu\-clidean space. We
have solved these $q$-de\-formed wave equations by using $q$-de\-formed
momentum eigenfunctions \cite{Wachter:2019A}. In this paper, we will study the
scattering of a particle on $q$-de\-formed Euclidean space. Our considerations are analogous to the undeformed case
\cite{Bjorken1Engl} since the time evolution operator in $q$-de\-formed
Euclidean space has the same structure as in the undeformed case
\cite{Wachter:2020A}.

Using the so-called star product formalism
\cite{Bayen:1977ha,1997q.alg.....9040K,Madore:2000en,Moyal:1949sk}, we can
develop a multivariable calculus on $q$-de\-formed Euclidean space
\cite{Wachter:2002A,Bauer:2003,Wachter:2004A,Wachter:2007A}. In the appendix,
we have summarized the basics of this $q$-de\-formed multivariable calculus as
far as it is necessary for our considerations. Moreover, we need to know the
solutions and the propagators for the wave equations describing a free
particle in $q$-de\-formed Euclidean space. We have summarized these results
from Ref.~\cite{Wachter:2020B} and Ref.~\cite{Wachter:2021B} in
Chap.~\ref{KapProSchrFel} of the present paper.

Using the propagators for the $q$-de\-formed wave equations of a free
particle, we can write down $q$-ver\-sions of the Lipp\-mann-Schwin\-ger
equation. Solving these $q$-de\-formed Lipp\-mann-Schwin\-ger equations by
iteration, we obtain perturbation expansions for the wave functions of a
scattered particle, i.~e., $q$-ver\-sions of the famous Born series (see
Chap.~\ref{ProWecKap}). With the help of\ these expansions, we write down
S-ma\-trix elements for the scattering of a $q$-de\-formed particle (see
Chap.~\ref{KapSMatDef}, Chap.~\ref{KapMomRepSma}, and Chap.~\ref{KapStrKleGorFelN}). These S-ma\-trix
elements again satisfy unitarity conditions. We show this in
Chap.~\ref{KapUniCon}. Finally, we discuss how to formulate our considerations
with the help of the interaction picture (see Chap.~\ref{IntPic}) or in the
framework of time-de\-pen\-dent perturbation theory (see
Chap.~\ref{KapZeiStoeThe}).

\section{Propagators for a free particle\label{KapProSchrFel}}

In this chapter, we summarize some results
from Ref.~\cite{Wachter:2020B} or Ref.~\cite{Wachter:2021B}.
As Hamiltonian operator for a free
nonrelativistic particle with mass $m$, we choose the following expression:
\begin{equation}
H_{0}=-(2\hspace{0.01in}m)^{-1}g_{AB}\hspace{0.01in}\partial^{A}\partial
^{B}=-(2\hspace{0.01in}m)^{-1}\partial^{A}\partial_{A}. \label{Ham2}%
\end{equation}
Since we have different actions of $q$-de\-formed partial derivatives [cf.
Eq.~(\ref{RegConAbl}) in App.~\ref{KapParDer}], we can write down the
following $q$-ver\-sions of the Schr\"{o}dinger equation for a free
nonrelativistic particle \cite{Wachter:2020A,Wachter:2020B}:%
\begin{align}
\text{i}\hspace{0.01in}\partial_{t}\triangleright\phi_{R}(\mathbf{x},t)  &
=H_{0}\triangleright\phi_{R}(\mathbf{x},t), & \phi_{L}(\mathbf{x}%
,t)\,\bar{\triangleleft}\,\partial_{t}\hspace{0.01in}\text{i}  &  =\phi
_{L}(\mathbf{x},t)\,\bar{\triangleleft}\,H_{0},\nonumber\\
\text{i}\hspace{0.01in}\partial_{t}\,\bar{\triangleright}\,\phi_{R}^{\ast
}(\mathbf{x},t)  &  =H_{0}\,\bar{\triangleright}\,\phi_{R}^{\ast}%
(\mathbf{x},t), & \phi_{L}^{\ast}(\mathbf{x},t)\triangleleft\partial
_{t}\hspace{0.01in}\text{i}  &  =\phi_{L}^{\ast}(\mathbf{x},t)\triangleleft
H_{0}. \label{SchGleQDef1N}%
\end{align}

We require that the solutions to the above Schr\"{o}\-dinger equations behave
as follows under quantum space conjugation:%
\begin{equation}
\overline{\phi_{L}(\mathbf{x},t)}=\phi_{R}(\mathbf{x},t),\text{\qquad
}\overline{\phi_{L}^{\ast}(\mathbf{x},t)}=\phi_{R}^{\ast}(\mathbf{x},t).
\label{KonEigSchWelWdh}%
\end{equation}
This condition ensures that quantum space conjugation transforms the
Schr\"{o}\-dinger equations on the left side of Eq.~(\ref{SchGleQDef1N}) into
the Schr\"{o}\-dinger equations on the right side of Eq.~(\ref{SchGleQDef1N})
and vice versa.

For plane wave solutions, the Schr\"{o}dinger equations above become%
\begin{align}
\text{i}\hspace{0.01in}\partial_{t}\triangleright u_{\hspace{0.01in}%
\mathbf{p}}(\mathbf{x},t)  &  =H_{0}\triangleright u_{\hspace{0.01in}%
\mathbf{p}}(\mathbf{x},t)=-(2\hspace{0.01in}m)^{-1}\hspace{0.01in}\partial
^{A}\partial_{A}\triangleright u_{\hspace{0.01in}\mathbf{p}}(\mathbf{x}%
,t)\nonumber\\
&  =u_{\hspace{0.01in}\mathbf{p}}(\mathbf{x},t)\circledast\mathbf{p}%
^{2}\hspace{0.01in}(2\hspace{0.01in}m)^{-1},\label{FreSchGlImp0}\\[0.08in]
(u^{\ast})_{\mathbf{p}}(\mathbf{x},t)\triangleleft\partial_{t}\hspace
{0.01in}\text{i}  &  =(u^{\ast})_{\mathbf{p}}(\mathbf{x},t)\triangleleft
H_{0}=-(u^{\ast})_{\mathbf{p}}\triangleleft\partial^{A}\partial_{A}%
\hspace{0.01in}(2\hspace{0.01in}m)^{-1}\nonumber\\
&  =(2\hspace{0.01in}m)^{-1}\hspace{0.01in}\mathbf{p}^{2}\circledast(u^{\ast
})_{\mathbf{p}}, \label{FreSchGlImp1}%
\end{align}
and%
\begin{align}
u^{\mathbf{p}}(\mathbf{x},t)\,\bar{\triangleleft}\,\partial_{t}\hspace
{0.01in}\text{i}  &  =u^{\mathbf{p}}(\mathbf{x},t)\,\bar{\triangleleft}%
\,H_{0}=-\hspace{0.01in}u^{\mathbf{p}}(\mathbf{x},t)\,\bar{\triangleleft
}\,\hat{\partial}^{A}\hat{\partial}_{A}\hspace{0.01in}(2\hspace{0.01in}%
m)^{-1}\nonumber\\
&  =(2\hspace{0.01in}m)^{-1}\mathbf{p}^{2}\circledast u^{\mathbf{p}%
}(\mathbf{x},t),\\[0.08in]
\text{i}\hspace{0.01in}\partial_{t}\,\bar{\triangleright}\,(u^{\ast
})^{\mathbf{p}}(\mathbf{x},t)  &  =H_{0}\,\bar{\triangleright}\,(u^{\ast
})^{\mathbf{p}}(\mathbf{x},t)=-(2\hspace{0.01in}m)^{-1}\hspace{0.01in}%
\hat{\partial}^{A}\hat{\partial}_{A}\,\bar{\triangleright}\,u_{\hspace
{0.01in}\mathbf{p}}(\mathbf{x},t)\nonumber\\
&  =(u^{\ast})^{\mathbf{p}}(\mathbf{x},t)\circledast\mathbf{p}^{2}%
\hspace{0.01in}(2\hspace{0.01in}m)^{-1}.
\end{align}
These plane wave solutions depend on $q$-de\-formed exponentials
\cite{Wachter:2020B}, i.~e.%
\begin{align}
u_{\hspace{0.01in}\mathbf{p}}(\mathbf{x},t)  &  =\operatorname*{vol}%
\nolimits^{-1/2}\exp_{q}(\mathbf{x}|\text{i}\mathbf{p})\circledast
\operatorname*{e}\nolimits^{-\text{i}t\varepsilon_{\mathbf{p}}},\nonumber\\
u^{\mathbf{p}}(\mathbf{x},t)  &  =\operatorname*{e}\nolimits^{\text{i}%
t\varepsilon_{\mathbf{p}}}\circledast\exp_{q}(\text{i}^{-1}\mathbf{p}%
|\hspace{0.01in}\mathbf{x})\operatorname*{vol}\nolimits^{-1/2},
\label{ConPlaWav0}%
\end{align}
and%
\begin{align}
(u^{\ast})_{\mathbf{p}}(\mathbf{x},t)  &  =\operatorname*{vol}\nolimits^{-1/2}%
\operatorname*{e}\nolimits^{\text{i}t\varepsilon_{\mathbf{p}}}\circledast
\exp_{q}^{\ast}(\text{i}\mathbf{p|\hspace{0.01in}x}),\nonumber\\
(u^{\ast})^{\mathbf{p}}(\mathbf{x},t)  &  =\operatorname*{vol}\nolimits^{-1/2}%
\exp_{q}^{\ast}(\mathbf{x}|\text{i}^{-1}\mathbf{p})\circledast
\operatorname*{e}\nolimits^{-\text{i}t\varepsilon_{\mathbf{p}}}.
\label{ConPlaWav1}%
\end{align}
The time-de\-pen\-dent phase factor and the volume element are given by%
\begin{equation}
\operatorname*{e}\nolimits^{\pm\text{i}t\varepsilon_{\mathbf{p}}}=\exp
(\pm\hspace{0.01in}\text{i}\hspace{0.01in}t\hspace{0.01in}\mathbf{p}%
^{2}(2\hspace{0.01in}m)^{-1})=\sum_{k\hspace{0.01in}=\hspace{0.01in}0}%
^{\infty}\frac{1}{k!}[\pm\hspace{0.01in}\text{i}\hspace{0.01in}t(2\hspace
{0.01in}m)^{-1}]^{k}\hspace{0.01in}\mathbf{p}^{2k} \label{TimDepPhaFac}%
\end{equation}
and%
\begin{equation}
\operatorname*{vol}=\int\text{d}_{q}^{3}p\int\text{d}_{q}^{3}x\hspace
{0.01in}\exp_{q}(\text{i}^{-1}\mathbf{p}|\mathbf{x}). \label{VolEleDef}%
\end{equation}
Note that powers of $\mathbf{p}^{2}\hspace{0.01in}(=g^{AB}\hspace{0.02in}%
p_{A}\circledast\hspace{0.01in}p_{B})$ are calculated by using the
star-prod\-uct (cf. App.~\ref{KapQuaZeiEle}):%
\begin{equation}
\mathbf{p}^{2k}=\hspace{0.01in}\overset{k-\text{times}}{\overbrace
{\mathbf{p}^{2}\circledast\ldots\circledast\mathbf{p}^{2}}}\hspace
{0.01in}=\sum_{l\hspace{0.01in}=\hspace{0.01in}0}^{k}\hspace{0.01in}%
q^{-2l}(-\lambda_{+})^{k-l}%
\genfrac{[}{]}{0pt}{}{k}{l}%
_{q^{4}}\,(\hspace{0.01in}p_{-})^{k-l}(\hspace{0.01in}p_{3})^{2l}%
(\hspace{0.01in}p_{+})^{k-l}. \label{EntPotP}%
\end{equation}
The $q$-de\-formed binomial coefficients are defined in complete analogy to
the undeformed case [also see Eq.~(\ref{qNumDef}) in App.~\ref{KapQuaZeiEle}]:%
\begin{equation}%
\genfrac{[}{]}{0pt}{}{n}{k}%
_{q}=\frac{[[n]]_{q}!}{[[n-k]]_{q}!\hspace{0.01in}[[k]]_{q}!}.
\label{qBinKoeBas}%
\end{equation}

The $q$-de\-formed plane wave solutions transform under quantum space
conjugation as follows:%
\begin{equation}
u^{\mathbf{p}}(\mathbf{x},t)=\overline{u_{\hspace{0.01in}\mathbf{p}%
}(\mathbf{x},t)},\qquad(u^{\ast})^{\mathbf{p}}(\mathbf{x},t)=\overline
{(u^{\ast})_{\mathbf{p}}(\mathbf{x},t)}. \label{KonEbeWel}%
\end{equation}
Moreover, they satisfy the orthonormality relations%
\begin{align}
\int\text{d}_{q}^{3}x\,(u^{\ast})_{\mathbf{p}}(\mathbf{x},t)\circledast
u_{\hspace{0.01in}\mathbf{p}^{\prime}}(\mathbf{x},t)  &  =\operatorname*{vol}%
\nolimits^{-1}\hspace{-0.01in}\delta_{q}^{\hspace{0.01in}3}((\ominus
\hspace{0.01in}\kappa^{-1}\mathbf{p})\oplus\mathbf{p}^{\prime}),\nonumber\\
\int\text{d}_{q}^{3}x\,u^{\mathbf{p}}(\mathbf{x},t)\circledast(u^{\ast
})^{\mathbf{p}^{\prime}}(\mathbf{x},t)  &  =\operatorname*{vol}\nolimits^{-1}%
\hspace{-0.01in}\delta_{q}^{\hspace{0.01in}3}(\hspace{0.01in}\mathbf{p}%
\oplus(\ominus\hspace{0.01in}\kappa^{-1}\mathbf{p}^{\prime})),
\label{OrtRelEbeWel1Schr}%
\end{align}
and the completeness relations%
\begin{align}
\int\text{d}_{q}^{3}p\,u_{\hspace{0.01in}\mathbf{p}}(\mathbf{x},t)\circledast
(u^{\ast})_{\mathbf{p}}(\hspace{0.01in}\mathbf{y},t)  &  =\operatorname*{vol}%
\nolimits^{-1}\hspace{-0.01in}\delta_{q}^{3}(\mathbf{x}\oplus(\ominus
\hspace{0.01in}\kappa^{-1}\mathbf{y})),\nonumber\\
\int\text{d}_{q}^{3}p\,(u^{\ast})^{\mathbf{p}}(\hspace{0.01in}\mathbf{y}%
,t)\circledast u^{\mathbf{p}}(\mathbf{x},t)  &  =\operatorname*{vol}%
\nolimits^{-1}\hspace{-0.01in}\delta_{q}^{3}((\ominus\hspace{0.01in}%
\kappa^{-1}\mathbf{y})\oplus\mathbf{x}). \label{VolRelZeiWelDreDim1}%
\end{align}
Note that the $q$-de\-formed delta function is defined as follows:%
\begin{equation}
\delta_{q}^{\hspace{0.01in}3}(\mathbf{x})=\int\text{d}_{q}^{3}p\hspace
{0.01in}\exp_{q}(\text{i}^{-1}\mathbf{p}|\mathbf{x}).
\end{equation}
The general solutions to the $q$-de\-formed Schr\"{o}dinger equations for a
free nonrelativistic particle can be written as expansions in terms of
$q$-de\-formed plane waves:%
\begin{align}
\phi_{R}(\mathbf{x},t)  &  =\int\text{d}_{q}^{3}p\,u_{\hspace{0.01in}%
\mathbf{p}}(\mathbf{x},t)\circledast c_{\hspace{0.01in}\mathbf{p}}, & \phi
_{L}^{\ast}(\mathbf{x},t)  &  =\int\text{d}_{q}^{3}p\,(c^{\ast})_{\mathbf{p}%
}\circledast(u^{\ast})_{\mathbf{p}}(\mathbf{x},t),\nonumber\\
\phi_{L}(\mathbf{x},t)  &  =\int\text{d}_{q}^{3}p\,c^{\hspace{0.01in}%
\mathbf{p}}\hspace{-0.01in}\circledast u^{\mathbf{p}}(\mathbf{x},t), &
\phi_{R}^{\ast}(\mathbf{x},t)  &  =\int\text{d}_{q}^{3}p\,(u^{\ast
})^{\mathbf{p}}(\mathbf{x},t)\circledast(c^{\ast})^{\mathbf{p}}.
\label{FreSolExp}%
\end{align}

With the $q$-de\-formed plane wave solutions, we can write down $q$-ver\-sions
of the propagator for a free nonrelativistic particle \cite{Wachter:2020B}. To
this end, we need the expressions%
\begin{align}
K_{R}(\mathbf{x}^{\hspace{0.01in}\prime}\hspace{-0.01in},t^{\hspace
{0.01in}\prime}\hspace{-0.01in};\mathbf{x},t)  &  =\int\text{d}_{q}%
^{3}p\,u_{\hspace{0.01in}\mathbf{p}}(\mathbf{x}^{\hspace{0.01in}\prime
},t^{\hspace{0.01in}\prime})\circledast(u^{\ast})_{\mathbf{p}}(\mathbf{x}%
,t),\nonumber\\
K_{L}(\mathbf{x},t;\mathbf{x}^{\hspace{0.01in}\prime}\hspace{-0.01in}%
,t^{\hspace{0.01in}\prime})  &  =\int\text{d}_{q}^{3}p\,(u^{\ast}%
)^{\mathbf{p}}(\mathbf{x},t)\circledast u^{\mathbf{p}}(\mathbf{x}%
^{\hspace{0.01in}\prime},t^{\hspace{0.01in}\prime}), \label{IntKer1}%
\end{align}
and%
\begin{align}
K_{R}^{\ast}(\mathbf{x}^{\hspace{0.01in}\prime}\hspace{-0.01in},t^{\hspace
{0.01in}\prime}\hspace{-0.01in};\mathbf{x},t)  &  =\int\text{d}_{q}%
^{3}p\,(u^{\ast})^{\mathbf{p}}(\mathbf{x}^{\hspace{0.01in}\prime}%
,t^{\hspace{0.01in}\prime})\circledast u^{\mathbf{p}}(\mathbf{x}%
,t),\nonumber\\
K_{L}^{\ast}(\mathbf{x},t;\mathbf{x}^{\hspace{0.01in}\prime}\hspace
{-0.01in},t^{\hspace{0.01in}\prime})  &  =\int\text{d}_{q}^{3}p\,u_{\hspace
{0.01in}\mathbf{p}}(\mathbf{x},t)\circledast(u^{\ast})_{\mathbf{p}}%
(\mathbf{x}^{\hspace{0.01in}\prime},t^{\hspace{0.01in}\prime}).
\label{IntKer2}%
\end{align}
The \textit{retarded propagators} take on the following form:%
\begin{align}
(K_{R})^{+}(\mathbf{x}^{\hspace{0.01in}\prime}\hspace{-0.01in},t^{\hspace
{0.01in}\prime}\hspace{-0.01in};\mathbf{x},t)  &  =\theta(t^{\hspace
{0.01in}\prime}\hspace{-0.02in}-t)\,K_{R}(\mathbf{x}^{\hspace{0.01in}\prime
}\hspace{-0.01in},t^{\hspace{0.01in}\prime}\hspace{-0.01in};\mathbf{x}%
,t),\nonumber\\
(K_{L})^{+}(\mathbf{x},t;\mathbf{x}^{\hspace{0.01in}\prime}\hspace
{-0.01in},t^{\hspace{0.01in}\prime})  &  =\theta(t^{\hspace{0.01in}\prime
}\hspace{-0.02in}-t)\,K_{L}(\mathbf{x},t;\mathbf{x}^{\hspace{0.01in}\prime
}\hspace{-0.01in},t^{\hspace{0.01in}\prime}),\label{DefRetPro1}\\[0.1in]
(K_{R}^{\ast})^{+}(\mathbf{x}^{\hspace{0.01in}\prime}\hspace{-0.01in}%
,t^{\hspace{0.01in}\prime}\hspace{-0.01in};\mathbf{x},t)  &  =\theta
(t^{\hspace{0.01in}\prime}\hspace{-0.02in}-t)\,K_{R}^{\ast}(\mathbf{x}%
^{\hspace{0.01in}\prime}\hspace{-0.01in},t^{\hspace{0.01in}\prime}%
\hspace{-0.01in};\mathbf{x},t),\nonumber\\
(K_{L}^{\ast})^{+}(\mathbf{x},t;\mathbf{x}^{\hspace{0.01in}\prime}%
\hspace{-0.01in},t^{\hspace{0.01in}\prime})  &  =\theta(t^{\hspace
{0.01in}\prime}\hspace{-0.02in}-t)\,K_{L}^{\ast}(\mathbf{x},t;\mathbf{x}%
^{\hspace{0.01in}\prime}\hspace{-0.01in},t^{\hspace{0.01in}\prime}).
\label{DefRetPro2}%
\end{align}
Note that $\theta(t)$ stands for the Heaviside function:%
\begin{equation}
\theta(t)=%
\begin{cases}
1 & \text{if }t\geq0,\\
0 & \text{otherwise}.
\end{cases}
\end{equation}
We can also write down \textit{advanced propagators} describing the
propagation of a wave function into the past:%
\begin{align}
(K_{R})^{-}(\mathbf{x}^{\hspace{0.01in}\prime},t^{\hspace{0.01in}\prime
}\hspace{-0.01in};\mathbf{x},t)  &  =\theta(t-t^{\hspace{0.01in}\prime
})\,K_{R}(\mathbf{x}^{\hspace{0.01in}\prime},t^{\hspace{0.01in}\prime}%
\hspace{-0.01in};\mathbf{x},t),\nonumber\\
(K_{L})^{-}(\mathbf{x},t;\mathbf{x}^{\hspace{0.01in}\prime},t^{\hspace
{0.01in}\prime})  &  =\theta(t-t^{\hspace{0.01in}\prime})\,K_{L}%
(\mathbf{x},t;\mathbf{x}^{\hspace{0.01in}\prime},t^{\hspace{0.01in}\prime
}),\label{DefAdvPro1}\\[0.1in]
(K_{R}^{\ast})^{-}(\mathbf{x}^{\hspace{0.01in}\prime},t^{\hspace{0.01in}%
\prime}\hspace{-0.01in};\mathbf{x},t)  &  =\theta(t-t^{\hspace{0.01in}\prime
})\,K_{R}^{\ast}(\mathbf{x}^{\hspace{0.01in}\prime},t^{\hspace{0.01in}\prime
}\hspace{-0.01in};\mathbf{x},t),\nonumber\\
(K_{L}^{\ast})^{-}(\mathbf{x},t;\mathbf{x}^{\hspace{0.01in}\prime}%
\hspace{-0.01in},t^{\hspace{0.01in}\prime})  &  =\theta(t-t^{\hspace
{0.01in}\prime})\,K_{L}^{\ast}(\mathbf{x},t;\mathbf{x}^{\hspace{0.01in}\prime
}\hspace{-0.01in},t^{\hspace{0.01in}\prime}). \label{DefAdvPro2}%
\end{align}
The conjugation properties of the $q$-de\-formed plane waves [cf.
Eq.~(\ref{KonEbeWel})] and the expressions in Eqs.~(\ref{IntKer1}) and
(\ref{IntKer2}) give%
\begin{align}
\overline{(K_{R})^{\pm}(\mathbf{x}^{\hspace{0.01in}\prime}\hspace
{-0.01in},t^{\hspace{0.01in}\prime}\hspace{-0.01in};\mathbf{x},t)}  &
=(K_{L})^{\pm}(\mathbf{x},t;\mathbf{x}^{\hspace{0.01in}\prime}\hspace
{-0.01in},t^{\hspace{0.01in}\prime}),\nonumber\\
(\overline{K_{L}^{\ast})^{\pm}(\mathbf{x},t;\mathbf{x}^{\hspace{0.01in}\prime
}\hspace{-0.01in},t^{\hspace{0.01in}\prime})}  &  =(K_{R}^{\ast})^{\pm
}(\mathbf{x}^{\hspace{0.01in}\prime}\hspace{-0.01in},t^{\hspace{0.01in}\prime
}\hspace{-0.01in};\mathbf{x},t). \label{KonFreProNR}%
\end{align}

The $q$-de\-formed propagators for a free nonrelativistic particle satisfy the
equations \cite{Wachter:2020B}%
\begin{align}
(\text{i\hspace{0.01in}}\partial_{t^{\prime}}-H_{0}^{\prime})\triangleright
(K_{R})^{\pm}(\mathbf{x}^{\hspace{0.01in}\prime},t^{\hspace{0.01in}\prime
};\mathbf{x},t)  &  =\pm\,\text{i}\operatorname*{vol}\nolimits^{-1}%
\hspace{-0.01in}\delta(t^{\hspace{0.01in}\prime}\hspace{-0.02in}%
-t)\,\delta_{q}^{3}(\mathbf{x}^{\hspace{0.01in}\prime}\oplus(\ominus
\hspace{0.01in}\kappa^{-1}\mathbf{x})),\nonumber\\
(K_{L})^{\pm}(\mathbf{x},t\text{\hspace{0.01in}};\mathbf{x}^{\hspace
{0.01in}\prime}\hspace{-0.01in},t^{\hspace{0.01in}\prime})\,\bar
{\triangleleft}\,(\partial_{t^{\prime}}\text{i}-H_{0}^{\prime})  &
=\mp\,\text{i}\operatorname*{vol}\nolimits^{-1}\hspace{-0.01in}\delta
(t-t^{\hspace{0.01in}\prime})\,\delta_{q}^{3}((\ominus\hspace{0.01in}%
\kappa^{-1}\mathbf{x})\oplus\mathbf{x}^{\hspace{0.01in}\prime}),
\label{SchrGlGreHabAnf}%
\end{align}
or%
\begin{align}
(\text{i\hspace{0.01in}}\partial_{t^{\prime}}-H_{0}^{\prime})\,\bar
{\triangleright}\,(K_{R}^{\ast})^{\pm}(\mathbf{x}^{\hspace{0.01in}\prime
}\hspace{-0.01in},t^{\hspace{0.01in}\prime}\hspace{-0.01in};\mathbf{x},t)  &
=\pm\,\text{i}\operatorname*{vol}\nolimits^{-1}\hspace{-0.01in}\delta
(t^{\hspace{0.01in}\prime}\hspace{-0.02in}-t)\,\delta_{q}^{3}((\ominus
\hspace{0.01in}\kappa^{-1}\mathbf{x}^{\hspace{0.01in}\prime})\oplus
\mathbf{x}),\nonumber\\
(K_{L}^{\ast})^{\pm}(\mathbf{x},t\text{\hspace{0.01in}};\mathbf{x}%
^{\hspace{0.01in}\prime}\hspace{-0.01in},t^{\hspace{0.01in}\prime
})\triangleleft(\partial_{t^{\prime}}\text{i}-H_{0}^{\prime})  &
=\mp\,\text{i}\operatorname*{vol}\nolimits^{-1}\hspace{-0.01in}\delta
(t-t^{\hspace{0.01in}\prime})\,\delta_{q}^{3}(\mathbf{x}\oplus(\ominus
\hspace{0.01in}\kappa^{-1}\mathbf{x}^{\hspace{0.01in}\prime})).
\label{SchrGlGreHabEnd}%
\end{align}
These equations
enable us to obtain solutions to the following inhomogeneous wave equations:%
\begin{align}
(\text{i\hspace{0.01in}}\partial_{t}-H_{0})\triangleright(\psi_{R})^{\pm
}(\mathbf{x},t)  &  =\varrho(\mathbf{x},t),\nonumber\\
(\psi_{L})^{\pm}(\mathbf{x},t)\,\bar{\triangleleft}\,(\partial_{t}%
\text{\hspace{0.01in}i\hspace{0.01in}}-H_{0})  &  =\varrho(\mathbf{x}%
,t),\label{InhSchrGleAll}\\[0.1in]
(\text{i\hspace{0.01in}}\partial_{t}-H_{0})\,\bar{\triangleright}\,(\psi
_{R}^{\ast})^{\pm}(\mathbf{x},t)  &  =\varrho(\mathbf{x},t),\nonumber\\
(\psi_{L}^{\ast})^{\pm}(\mathbf{x},t)\triangleleft(\partial_{t}\text{\hspace
{0.01in}i\hspace{0.01in}}-H_{0})  &  =\varrho(\mathbf{x},t).
\end{align}
The solutions to the equations above can be written in the following form \cite{Wachter:2020B}:%
\begin{align}
(\psi_{R})^{\pm}(\mathbf{x},t)  &  =\mp\,\text{i}\int\text{d}t^{\hspace
{0.01in}\prime}\hspace{-0.01in}\int\text{d}_{q}^{3}x^{\prime}\,(K_{R})^{\pm
}(\mathbf{x},t;\mathbf{x}^{\hspace{0.01in}\prime}\hspace{-0.01in}%
,t^{\hspace{0.01in}\prime})\circledast\varrho(\mathbf{x}^{\hspace
{0.01in}\prime}\hspace{-0.01in},t^{\hspace{0.01in}\prime}),\nonumber\\
(\psi_{L})^{\pm}(\mathbf{x},t)  &  =\pm\,\text{i}\int\text{d}t^{\hspace
{0.01in}\prime}\hspace{-0.01in}\int\text{d}_{q}^{3}x^{\prime}\,\varrho
(\mathbf{x}^{\hspace{0.01in}\prime},t^{\hspace{0.01in}\prime})\circledast
(K_{L})^{\pm}(\mathbf{x}^{\hspace{0.01in}\prime}\hspace{-0.01in}%
,t^{\hspace{0.01in}\prime}\hspace{-0.01in};\mathbf{x}%
,t),\label{SolInhSchr1Hab}\\[0.1in]
(\psi_{R}^{\ast})^{\pm}(\mathbf{x},t)  &  =\mp\,\text{i}\int\text{d}%
t^{\hspace{0.01in}\prime}\hspace{-0.01in}\int\text{d}_{q}^{3}x^{\prime
}\,(K_{R}^{\ast})^{\pm}(\mathbf{x},t;\mathbf{x}^{\hspace{0.01in}\prime}%
\hspace{-0.01in},t^{\hspace{0.01in}\prime})\circledast\varrho(\mathbf{x}%
^{\hspace{0.01in}\prime}\hspace{-0.01in},t^{\hspace{0.01in}\prime
}),\nonumber\\
(\psi_{L}^{\ast})^{\pm}(\mathbf{x},t)  &  =\pm\,\text{i}\int\text{d}%
t^{\hspace{0.01in}\prime}\hspace{-0.01in}\int\text{d}_{q}^{3}x^{\prime
}\,\varrho(\mathbf{x}^{\hspace{0.01in}\prime}\hspace{-0.01in},t^{\hspace
{0.01in}\prime})\circledast(K_{L}^{\ast})^{\pm}(\mathbf{x}^{\hspace
{0.01in}\prime}\hspace{-0.01in},t^{\hspace{0.01in}\prime}\hspace
{-0.01in};\mathbf{x},t). \label{SolInhSchr2Hab}%
\end{align}

Alternatively, we can consider a $q$-ver\-sion of the Klein-Gor\-don equation
in dealing with a free particle on the $q$-de\-formed Euclidean space
\cite{Wachter:2021B}:%
\begin{equation}
c^{-2}\partial_{t}^{\hspace{0.01in}2}\triangleright\varphi_{R}-\hspace
{-0.01in}\nabla_{q}^{\hspace{0.01in}2}\triangleright\varphi_{R}+(m\hspace
{0.01in}c)^{2}\hspace{0.01in}\varphi_{R}=0. \label{KleGorGleLin}%
\end{equation}
Note that the $q$-de\-formed Laplace operator $\nabla_{q}^{\hspace{0.01in}2}$
depends on the metric of the three-di\-men\-sion\-al $q$-de\-formed Euclidean
space [see Eq.~(\ref{DreDimMet}) in\ App.~\ref{KapQuaZeiEle}]:%
\begin{equation}
\nabla_{q}^{\hspace{0.01in}2}=\partial^{A}\partial_{A}=g^{AB}\hspace
{0.01in}\partial_{B}\hspace{0.01in}\partial_{A}.
\end{equation}
By conjugating Eq.~(\ref{KleGorGleLin}), we obtain another $q$-ver\-sion of
the Klein-Gor\-don equation:%
\begin{equation}
\varphi_{L}\,\bar{\triangleleft}\,\hspace{0.01in}\partial_{t}^{\hspace
{0.01in}2}c^{-2}\hspace{-0.01in}-\varphi_{L}\,\bar{\triangleleft}%
\,\hspace{0.01in}\nabla_{q}^{\hspace{0.01in}2}+\varphi_{L}\hspace
{0.01in}(m\hspace{0.01in}c)^{2}\hspace{-0.01in}=0. \label{KleGorGleRec}%
\end{equation}
Accordingly, the wave function $\varphi_{R}$ transforms into $\varphi_{L}$ by
conjugation:%
\begin{equation}
\overline{\varphi_{R}}=\varphi_{L}. \label{KonGKWel1}%
\end{equation}

There are two types of left-ac\-tions and two types of right-ac\-tions for
$q$-de\-formed partial derivatives [see Eq.~(\ref{UnkOpeDarAbl}) and
Eq.~(\ref{KonOpeDarAbl}) in\ App.~\ref{KapParDer}]. Thus, we get further
$q$-ver\-sions of the Klein-Gor\-don equation by applying the following
substitutions in Eq.~(\ref{KleGorGleLin}) or Eq.~(\ref{KleGorGleRec}):%
\begin{equation}
\triangleright\,\leftrightarrow\,\bar{\triangleright},\qquad\bar
{\triangleleft}\,\leftrightarrow\,\triangleleft,\qquad\varphi_{R}%
\leftrightarrow\varphi_{R}^{\ast},\qquad\varphi_{L}\leftrightarrow\varphi
_{L}^{\ast}. \label{ErsKleGor}%
\end{equation}
This way, we have%
\begin{align}
c^{-2}\partial_{t}^{\hspace{0.01in}2}\,\bar{\triangleright}\,\varphi_{R}%
^{\ast}-\nabla_{q}^{\hspace{0.01in}2}\,\bar{\triangleright}\,\varphi_{R}%
^{\ast}+(m\hspace{0.01in}c)^{2}\hspace{0.01in}\varphi_{R}^{\ast}  &
=0,\nonumber\\[0.03in]
\varphi_{L}^{\ast}\hspace{-0.01in}\triangleleft\partial_{t}^{\hspace{0.01in}%
2}c^{-2}\hspace{-0.01in}-\varphi_{L}^{\ast}\hspace{-0.01in}\triangleleft
\nabla_{q}^{\hspace{0.01in}2}+\varphi_{L}^{\ast}\hspace{0.01in}(m\hspace
{0.01in}c)^{2}  &  =0, \label{KleGorGleLin2}%
\end{align}
where $\varphi_{R}^{\ast}$ transforms into $\varphi_{L}^{\ast}$ by
conjugation:%
\begin{equation}
\overline{\varphi_{R}^{\ast}}=\varphi_{L}^{\ast}. \label{KonGKWel2}%
\end{equation}

There are plane wave solutions to our different $q$-ver\-sions of the
Klein-Gor\-don equation. For example, the plane waves%
\begin{align}
\varphi_{\mathbf{p}}(\mathbf{x},t)  &  =\frac{c}{\sqrt{2}}\,u_{\hspace
{0.01in}\mathbf{p}}(\mathbf{x})\circledast\exp(-\text{i\hspace{0.01in}%
}tE_{\mathbf{p}})\circledast E_{\mathbf{p}}^{\hspace{0.01in}-1/2},\nonumber\\
\varphi^{\hspace{0.01in}\mathbf{p}}(\mathbf{x},t)  &  =\frac{c}{\sqrt{2}%
}\,E_{\mathbf{p}}^{\hspace{0.01in}-1/2}\hspace{-0.01in}\circledast
\exp(\text{i\hspace{0.01in}}tE_{\mathbf{p}})\circledast u^{\mathbf{p}%
}(\mathbf{x}) \label{EbeWelKGF}%
\end{align}
are subject to%
\begin{align}
c^{-2}\partial_{t}^{\hspace{0.01in}2}\triangleright\varphi_{\mathbf{p}}  &
=\nabla_{q}^{\hspace{0.01in}2}\triangleright\varphi_{\mathbf{p}}%
-(m\hspace{0.01in}c)^{2}\varphi_{\mathbf{p}},\nonumber\\
\varphi^{\hspace{0.01in}\mathbf{p}}\,\bar{\triangleleft}\,\hspace
{0.01in}\partial_{t}^{\hspace{0.01in}2}c^{-2}  &  =\varphi^{\hspace
{0.01in}\mathbf{p}}\,\bar{\triangleleft}\,\hspace{0.01in}\nabla_{q}%
^{\hspace{0.01in}2}\hspace{0.01in}-\varphi^{\hspace{0.01in}\mathbf{p}%
}(m\hspace{0.01in}c)^{2}. \label{qKleGorGleImp1}%
\end{align}
The time-de\-pen\-dent phase factors are given by%
\begin{equation}
\exp(\pm\hspace{0.01in}\text{i\hspace{0.01in}}tE_{\mathbf{p}})=\sum
_{n\hspace{0.01in}=\hspace{0.01in}0}^{\infty}\frac{(\pm\hspace{0.01in}%
\text{i\hspace{0.01in}}tE_{\mathbf{p}})^{n}}{n!}.
\end{equation}
Note that we have to calculate the powers of the energy of a $q$-de\-formed
scalar particle by the following formula [also see Eq.~(\ref{EntPotP})]:%
\begin{equation}
E_{\mathbf{p}}^{\hspace{0.01in}\alpha}=c^{\alpha}\,(\hspace{0.01in}%
\mathbf{p}^{2}+(m\hspace{0.01in}c)^{2})^{\alpha/2}=c^{\alpha}\sum
_{k\hspace{0.01in}=\hspace{0.01in}0}^{\infty}\binom{\alpha/2}{k}%
\,\mathbf{p}^{2k}(m\hspace{0.01in}c)^{\alpha-2k}.
\end{equation}
Moreover, we have dual plane waves:%
\begin{align}
(\varphi^{\ast})_{\mathbf{p}}(\mathbf{x},t)  &  =\frac{c}{\sqrt{2}%
}\,E_{\mathbf{p}}^{\hspace{0.01in}-1/2}\hspace{-0.01in}\circledast
\exp(\text{i}\hspace{0.01in}tE_{\mathbf{p}})\circledast(u^{\ast})_{\mathbf{p}%
}(\mathbf{x}),\nonumber\\
(\varphi^{\ast})^{\mathbf{p}}(\mathbf{x},t)  &  =\frac{c}{\sqrt{2}}%
\hspace{0.01in}(u^{\ast})^{\mathbf{p}}(\mathbf{x})\circledast\exp
(-\text{i\hspace{0.01in}}tE_{\mathbf{p}})\circledast E_{\mathbf{p}}%
^{\hspace{0.01in}-1/2}. \label{EbeWelKGFSte}%
\end{align}
They are solutions to the $q$-de\-formed Klein-Gor\-don equations given in
Eq.~(\ref{KleGorGleLin2}):%
\begin{align}
(\varphi^{\ast})_{\mathbf{p}}\hspace{-0.01in}\triangleleft\partial
_{t}^{\hspace{0.01in}2}c^{-2}-(\varphi^{\ast})_{\mathbf{p}}\triangleleft
\nabla_{q}^{\hspace{0.01in}2}+(\varphi^{\ast})_{\mathbf{p}}\hspace
{0.01in}(m\hspace{0.01in}c)^{2}\hspace{-0.01in}  &  =0,\nonumber\\[0.03in]
c^{-2}\partial_{t}^{\hspace{0.01in}2}\,\hspace{0.01in}\bar{\triangleright
}\,(\varphi^{\ast})^{\mathbf{p}}-\nabla_{q}^{\hspace{0.01in}2}\,\hspace
{0.01in}\bar{\triangleright}\,(\varphi^{\ast})^{\mathbf{p}}+(m\hspace
{0.01in}c)^{2}\hspace{0.01in}(\varphi^{\ast})^{\mathbf{p}}\hspace{-0.01in}  &
=0. \label{qKleGorGleImp3}%
\end{align}

We can write the general solutions to the $q$-de\-formed Klein-Gor\-don
equations as expansions in terms of plane wave solutions, i.~e.%
\begin{align}
\varphi_{R}(\mathbf{x},t)  &  =(\varphi_{R})^{[+]}+(\varphi_{R})^{[-]}%
=\sum_{\varepsilon\hspace{0.01in}=\hspace{0.01in}\pm}\int\text{d}_{q}%
^{3}\hspace{0.01in}p\,\varphi_{\mathbf{p}}(\varepsilon\hspace{0.01in}%
\mathbf{x},\varepsilon\hspace{0.01in}t)\circledast\hspace{-0.01in}%
f_{\mathbf{p}}^{[\varepsilon]},\nonumber\\[0.03in]
\varphi_{L}(\mathbf{x},t)  &  =(\varphi_{L})^{[+]}+(\varphi_{L})^{[-]}%
=\sum_{\varepsilon\hspace{0.01in}=\hspace{0.01in}\pm}\int\text{d}_{q}%
^{3}\hspace{0.01in}p\,f_{[\varepsilon]}^{\mathbf{p}}\circledast\varphi
^{\hspace{0.01in}\mathbf{p}}(\varepsilon\hspace{0.01in}\mathbf{x}%
,\varepsilon\hspace{0.01in}t), \label{EntWicKGFR}%
\end{align}
and%
\begin{align}
\varphi_{L}^{\ast}(\mathbf{x},t)  &  =(\varphi_{L}^{\ast})^{[+]}+(\varphi
_{L}^{\ast})^{[-]}=\sum_{\varepsilon\hspace{0.01in}=\hspace{0.01in}\pm}%
\int\text{d}_{q}^{3}\hspace{0.01in}p\,h_{\hspace{0.01in}\mathbf{p}%
}^{[\varepsilon]}\circledast(\varphi^{\ast})_{\mathbf{p}}(\varepsilon
\hspace{0.01in}\mathbf{x},\varepsilon\hspace{0.01in}t),\nonumber\\[0.03in]
\varphi_{R}^{\ast}(\mathbf{x},t)  &  =(\varphi_{R}^{\ast})^{[+]}+(\varphi
_{R}^{\ast})^{[-]}=\sum_{\varepsilon\hspace{0.01in}=\hspace{0.01in}\pm}%
\int\text{d}_{q}^{3}\hspace{0.01in}p\,(\varphi^{\ast})^{\mathbf{p}%
}(\varepsilon\hspace{0.01in}\mathbf{x},\varepsilon\hspace{0.01in}t)\circledast
h_{[\varepsilon]}^{\mathbf{p}}, \label{EntKGFLSte}%
\end{align}
The plane waves solutions to our $q$-de\-formed Klein-Gordon equations satisfy
the \textit{orthogonality relations }\cite{Wachter:2021B}%
\begin{align}
&  \text{i\hspace{0.01in}}c^{-2}\hspace{-0.02in}\int\text{d}_{q}%
^{3}x\,(\varphi^{\ast})_{\mathbf{p}}(\mathbf{x},\varepsilon\hspace
{0.01in}t)\triangleleft\partial_{t}\circledast\varphi_{\mathbf{p}^{\prime}%
}(\mathbf{x},\varepsilon^{\prime}\hspace{0.01in}t)\nonumber\\
&  +\text{i\hspace{0.01in}}c^{-2}\hspace{-0.02in}\int\text{d}_{q}%
^{3}x\,(\varphi^{\ast})_{\mathbf{p}}(\mathbf{x},\varepsilon\hspace
{0.01in}t)\circledast\partial_{t}\triangleright\varphi_{\mathbf{p}^{\prime}%
}(\mathbf{x},\varepsilon^{\prime}\hspace{0.01in}t)\nonumber\\
&  \qquad\qquad=\varepsilon\hspace{0.01in}\delta_{\varepsilon,\varepsilon
^{\prime}}\operatorname*{vol}\nolimits^{-1}\hspace{-0.01in}\delta_{q}%
^{3}((\ominus\hspace{0.01in}\kappa^{-1}\mathbf{p})\oplus\mathbf{p}^{\prime})
\label{OrtRelKleGorPhiSte}%
\end{align}
and%
\begin{align}
&  \text{i\hspace{0.01in}}c^{-2}\hspace{-0.02in}\int\text{d}_{q}^{3}%
x\,\varphi^{\hspace{0.01in}\mathbf{p}}(\mathbf{x},\varepsilon\hspace
{0.01in}t)\,\bar{\triangleleft}\,\partial_{t}\circledast(\varphi^{\ast
})^{\mathbf{p}^{\prime}}(\mathbf{x},\varepsilon^{\prime}\hspace{0.01in}%
t)\nonumber\\
&  +\text{i\hspace{0.01in}}c^{-2}\hspace{-0.02in}\int\text{d}_{q}%
^{3}x\,\varphi^{\hspace{0.01in}\mathbf{p}}(\mathbf{x},\varepsilon
\hspace{0.01in}t)\circledast\partial_{t}\,\bar{\triangleright}\,(\varphi
^{\ast})^{\mathbf{p}^{\prime}}(\mathbf{x},\varepsilon^{\prime}\hspace
{0.01in}t)\nonumber\\
&  \qquad\qquad=\varepsilon\hspace{0.01in}\delta_{\varepsilon,\varepsilon
^{\prime}}\operatorname*{vol}\nolimits^{-1}\hspace{-0.01in}\delta_{q}%
^{3}(\hspace{0.01in}\mathbf{p}\oplus(\ominus\hspace{0.01in}\kappa
^{-1}\mathbf{p}^{\prime})). \label{OrtRelKleGorPhiSte2}%
\end{align}

We need to formulate the propagators for the $q$-de\-formed Klein-Gor\-don
equations so that the positive energy solution runs forward in time while the
negative energy solution runs backward in time. Propagators with these
properties are given by%
\begin{align}
\Delta_{R}(\mathbf{x}^{\hspace{0.01in}\prime}\hspace{-0.01in},t^{\hspace
{0.01in}\prime}\hspace{-0.01in};\mathbf{x},t)  &  =\theta(t^{\hspace
{0.01in}\prime}\hspace{-0.02in}-t)\int\text{d}_{q}^{3}\hspace{0.01in}%
p\,\varphi_{\mathbf{p}}(\mathbf{x}^{\hspace{0.01in}\prime}\hspace
{-0.01in},t^{\hspace{0.01in}\prime})\circledast(\varphi^{\ast})_{\mathbf{p}%
}(\mathbf{x},t)\nonumber\\
&  \hspace{0.15in}+\theta(t-t^{\hspace{0.01in}\prime})\int\text{d}_{q}%
^{3}\hspace{0.01in}p\,\varphi_{\mathbf{p}}(\mathbf{x}^{\hspace{0.01in}\prime
}\hspace{-0.01in},-\hspace{0.01in}t^{\hspace{0.01in}\prime})\circledast
(\varphi^{\ast})_{\mathbf{p}}(\mathbf{x},-\hspace{0.01in}t)
\label{EntFreProKleGorFelDelL}%
\end{align}
and%
\begin{align}
\Delta_{L}(\mathbf{x},t;\mathbf{x}^{\hspace{0.01in}\prime}\hspace
{-0.01in},t^{\hspace{0.01in}\prime})  &  =\theta(t^{\hspace{0.01in}\prime
}\hspace{-0.02in}-t)\int\text{d}_{q}^{3}\hspace{0.01in}p\,(\varphi^{\ast
})^{\mathbf{p}}(\mathbf{x},t)\circledast\varphi^{\hspace{0.01in}\mathbf{p}%
}(\mathbf{x}^{\hspace{0.01in}\prime}\hspace{-0.01in},t^{\hspace{0.01in}\prime
})\nonumber\\
&  \hspace{0.15in}+\theta(t-t^{\hspace{0.01in}\prime})\int\text{d}_{q}%
^{3}\hspace{0.01in}p\,(\varphi^{\ast})^{\mathbf{p}}(\mathbf{x},-\hspace
{0.01in}t)\circledast\varphi^{\hspace{0.01in}\mathbf{p}}(\mathbf{x}%
^{\hspace{0.01in}\prime}\hspace{-0.01in},-\hspace{0.01in}t^{\hspace
{0.01in}\prime}). \label{EntFreProKleGorFelDelR}%
\end{align}
We can also introduce so-called dual propagators, i.~e.%
\begin{align}
\Delta_{R}^{\ast}(\mathbf{x}^{\hspace{0.01in}\prime}\hspace{-0.01in}%
,t^{\hspace{0.01in}\prime}\hspace{-0.01in};\mathbf{x},t)=  &  \hspace
{0.03in}\theta(t^{\hspace{0.01in}\prime}\hspace{-0.02in}-t)\int\text{d}%
_{q}^{3}\hspace{0.01in}p\,(\varphi^{\ast})^{\mathbf{p}}(\mathbf{x}%
^{\hspace{0.01in}\prime}\hspace{-0.01in},t^{\hspace{0.01in}\prime}%
)\circledast\varphi^{\hspace{0.01in}\mathbf{p}}(\mathbf{x},t)\nonumber\\
&  +\theta(t-t^{\hspace{0.01in}\prime})\int\text{d}_{q}^{3}\hspace
{0.01in}p\,(\varphi^{\ast})^{\mathbf{p}}(\mathbf{x}^{\hspace{0.01in}\prime
}\hspace{-0.01in},-\hspace{0.01in}t^{\hspace{0.01in}\prime})\circledast
\varphi^{\hspace{0.01in}\mathbf{p}}(\mathbf{x},-\hspace{0.01in}t)
\end{align}
and%
\begin{align}
\Delta_{L}^{\ast}(\mathbf{x},t;\mathbf{x}^{\hspace{0.01in}\prime}%
\hspace{-0.01in},t^{\hspace{0.01in}\prime})=  &  \hspace{0.03in}%
\theta(t^{\hspace{0.01in}\prime}\hspace{-0.02in}-t)\int\text{d}_{q}^{3}%
\hspace{0.01in}p\,\varphi_{\mathbf{p}}(\mathbf{x},t)\circledast(\varphi^{\ast
})_{\mathbf{p}}(\mathbf{x}^{\hspace{0.01in}\prime}\hspace{-0.01in}%
,t^{\hspace{0.01in}\prime})\nonumber\\
&  +\theta(t-t^{\hspace{0.01in}\prime})\int\text{d}_{q}^{3}\hspace
{0.01in}p\,\varphi_{\mathbf{p}}(\mathbf{x},-\hspace{0.01in}t)\circledast
(\varphi^{\ast})_{\mathbf{p}}(\mathbf{x}^{\hspace{0.01in}\prime}%
\hspace{-0.01in},-\hspace{0.01in}t^{\hspace{0.01in}\prime})
\label{EntFreProKleGorFelDelRSte}%
\end{align}
The propagators for the $q$-de\-formed Klein-Gor\-don equations transform into
each other by conjugation:%
\begin{align}
\overline{\Delta_{R}(\mathbf{x}^{\hspace{0.01in}\prime}\hspace{-0.01in}%
,t^{\hspace{0.01in}\prime}\hspace{-0.01in};\mathbf{x},t)}  &  =\Delta
_{L}(\mathbf{x},t;\mathbf{x}^{\hspace{0.01in}\prime}\hspace{-0.01in}%
,t^{\hspace{0.01in}\prime}),\nonumber\\
\overline{\Delta_{L}^{\ast}(\mathbf{x}^{\hspace{0.01in}\prime}\hspace
{-0.01in},t^{\hspace{0.01in}\prime}\hspace{-0.01in};\mathbf{x},t)}  &
=\Delta_{R}^{\ast}(\mathbf{x},t;\mathbf{x}^{\hspace{0.01in}\prime}%
\hspace{-0.01in},t^{\hspace{0.01in}\prime}).
\end{align}

The propagators for the $q$-de\-formed Klein-Gor\-don equations
satisfy the equations%
\begin{align}
&  ((m\hspace{0.01in}c)^{2}+c^{-2}\partial_{t}^{\hspace{0.01in}2}-\nabla
_{q}^{\hspace{0.01in}2})\triangleright\Delta_{R}(\mathbf{x},t;\mathbf{x}%
^{\hspace{0.01in}\prime}\hspace{-0.01in},t^{\hspace{0.01in}\prime
})=\nonumber\\
&  \qquad\qquad=-\text{\hspace{0.01in}i\hspace{0.01in}}\operatorname*{vol}%
\nolimits^{-1}\hspace{-0.01in}\delta(t-t^{\hspace{0.01in}\prime})\,\delta
_{q}^{3}(\mathbf{x}\oplus(\ominus\hspace{0.01in}\kappa^{-1}\mathbf{x}%
^{\hspace{0.01in}\prime})),\label{DifGleKGFLR0}\\[0.07in]
&  \Delta_{L}(\mathbf{x}^{\hspace{0.01in}\prime}\hspace{-0.01in}%
,t^{\hspace{0.01in}\prime}\hspace{-0.01in};\mathbf{x},t)\,\bar{\triangleleft
}\,(\partial_{t}^{\hspace{0.01in}2}c^{-2}\hspace{-0.01in}-\nabla_{q}%
^{\hspace{0.01in}2}+(m\hspace{0.01in}c)^{2})=\nonumber\\
&  \qquad\qquad=\text{i\hspace{0.01in}}\operatorname*{vol}\nolimits^{-1}%
\hspace{-0.01in}\delta_{q}^{3}((\ominus\hspace{0.01in}\kappa^{-1}%
\mathbf{x}^{\hspace{0.01in}\prime})\oplus\mathbf{x})\,\delta(t^{\hspace
{0.01in}\prime}-t), \label{DifGleKGFLR}%
\end{align}
or%
\begin{align}
&  \Delta_{L}^{\ast}(\mathbf{x}^{\hspace{0.01in}\prime}\hspace{-0.01in}%
,t^{\hspace{0.01in}\prime}\hspace{-0.01in};\mathbf{x},t)\,\bar{\triangleleft
}\,(\partial_{t}^{\hspace{0.01in}2}c^{-2}\hspace{-0.01in}-\nabla_{q}%
^{\hspace{0.01in}2}+(m\hspace{0.01in}c)^{2})=\nonumber\\
&  \qquad\qquad=\text{i\hspace{0.01in}}\operatorname*{vol}\nolimits^{-1}%
\hspace{-0.01in}\delta_{q}^{3}((\ominus\hspace{0.01in}\kappa^{-1}%
\mathbf{x}^{\hspace{0.01in}\prime})\oplus\mathbf{x})\,\delta(t^{\hspace
{0.01in}\prime}-t),\\[0.07in]
&  ((m\hspace{0.01in}c)^{2}+c^{-2}\partial_{t}^{\hspace{0.01in}2}-\nabla
_{q}^{\hspace{0.01in}2})\triangleright\Delta_{R}^{\ast}(\mathbf{x}%
,t;\mathbf{x}^{\hspace{0.01in}\prime}\hspace{-0.01in},t^{\hspace{0.01in}%
\prime})=\nonumber\\
&  \qquad\qquad=-\text{\hspace{0.01in}i\hspace{0.01in}}\operatorname*{vol}%
\nolimits^{-1}\hspace{-0.01in}\delta(t-t^{\hspace{0.01in}\prime})\,\delta
_{q}^{3}(\mathbf{x}\oplus(\ominus\hspace{0.01in}\kappa^{-1}\mathbf{x}%
^{\hspace{0.01in}\prime})).
\end{align}
Using the above identities, we can show that the functions%
\begin{align}
\phi_{R}(\mathbf{x},t)  &  =\varphi_{R}(\mathbf{x},t)+\text{i}\int\text{d}%
^{3}x^{\prime}\text{d}t^{\hspace{0.01in}\prime}\,\Delta_{R}(\mathbf{x}%
,t;\mathbf{x}^{\hspace{0.01in}\prime}\hspace{-0.01in},t^{\hspace{0.01in}%
\prime})\circledast\mathcal{\varrho}(\mathbf{x}^{\hspace{0.01in}\prime}%
\hspace{-0.01in},t^{\hspace{0.01in}\prime}),\nonumber\\[0.12in]
\phi_{L}(\mathbf{x},t)  &  =\varphi_{L}(\mathbf{x},t)-\text{i}\int\text{d}%
^{3}x^{\prime}\text{d}t^{\hspace{0.01in}\prime}\,\mathcal{\varrho}%
(\mathbf{x}^{\hspace{0.01in}\prime}\hspace{-0.01in},t^{\hspace{0.01in}\prime
})\circledast\Delta_{L}(\mathbf{x}^{\hspace{0.01in}\prime}\hspace
{-0.01in},t^{\hspace{0.01in}\prime}\hspace{-0.01in};\mathbf{x},t)
\label{LoeInhKleGorGle}%
\end{align}
are solutions to the following $q$-de\-formed inhomogenous Klein-Gor\-don
equations:%
\begin{align}
c^{-2}\partial_{t}^{\hspace{0.01in}2}\triangleright\phi_{R}-\nabla
_{q}^{\hspace{0.01in}2}\triangleright\phi_{R}+(m\hspace{0.01in}c)^{2}%
\hspace{0.01in}\phi_{R}  &  =\varrho,\nonumber\\[0.05in]
\phi_{L}\,\bar{\triangleleft}\,\hspace{0.01in}\partial_{t}^{\hspace{0.01in}%
2}c^{-2}\hspace{-0.01in}-\phi_{L}\,\bar{\triangleleft}\,\hspace{0.01in}%
\nabla_{q}^{\hspace{0.01in}2}+\phi_{L}\hspace{0.01in}(m\hspace{0.01in}c)^{2}
&  =\mathcal{\varrho}.
\end{align}
Note that $\varphi_{R}$ and $\varphi_{L}$ in Eq.~(\ref{LoeInhKleGorGle}) are
solutions to the free $q$-de\-formed Klein-Gor\-don equations given in
Eqs.~(\ref{KleGorGleLin}) and (\ref{KleGorGleRec}). Similarly, the functions%
\begin{align}
\phi_{R}^{\ast}(\mathbf{x},t)  &  =\varphi_{R}^{\ast}(\mathbf{x}%
,t)+\text{i}\int\text{d}^{3}x^{\prime}\text{d}t^{\hspace{0.01in}\prime
}\,\Delta_{R}^{\ast}(\mathbf{x},t;\mathbf{x}^{\hspace{0.01in}\prime}%
\hspace{-0.01in},t^{\hspace{0.01in}\prime})\circledast\mathcal{\varrho
}(\mathbf{x}^{\hspace{0.01in}\prime}\hspace{-0.01in},t^{\hspace{0.01in}\prime
}),\nonumber\\[0.12in]
\phi_{L}^{\ast}(\mathbf{x},t)  &  =\varphi_{L}^{\ast}(\mathbf{x}%
,t)-\text{i}\int\text{d}^{3}x^{\prime}\text{d}t^{\hspace{0.01in}\prime
}\,\mathcal{\varrho}(\mathbf{x}^{\hspace{0.01in}\prime}\hspace{-0.01in}%
,t^{\hspace{0.01in}\prime})\circledast\Delta_{L}^{\ast}(\mathbf{x}%
^{\hspace{0.01in}\prime}\hspace{-0.01in},t^{\hspace{0.01in}\prime}%
\hspace{-0.01in};\mathbf{x},t) \label{LoeInhKleGorGleLRSte}%
\end{align}
satisfy the following $q$-de\-formed versions of the inhomogenous
Klein-Gor\-don equation:%
\begin{align}
c^{-2}\partial_{t}^{\hspace{0.01in}2}\,\hspace{0.01in}\bar{\triangleright
}\,\phi_{R}^{\ast}-\nabla_{q}^{\hspace{0.01in}2}\,\bar{\triangleright}%
\,\phi_{R}^{\ast}+(m\hspace{0.01in}c)^{2}\hspace{0.01in}\phi_{R}^{\ast}  &
=\mathcal{\varrho},\nonumber\\[0.05in]
\phi_{L}^{\ast}\hspace{-0.01in}\triangleleft\partial_{t}^{\hspace{0.01in}%
2}c^{-2}\hspace{-0.01in}-\phi_{L}^{\ast}\hspace{-0.01in}\triangleleft
\nabla_{q}^{\hspace{0.01in}2}+\phi_{L}^{\ast}\hspace{0.01in}(m\hspace
{0.01in}c)^{2}  &  =\mathcal{\varrho}.
\end{align}
Once again, $\varphi_{R}^{\ast}$ and $\varphi_{L}^{\ast}$ are solutions to the
free $q$-de\-formed Klein-Gor\-don equations given in Eq.~(\ref{KleGorGleLin2}).

\section{Lipp\-mann-Schwin\-ger equations\label{ProWecKap}}

The Hamiltonian operator of a scattering experiment is the sum of the
Hamiltonian operator $H_{0}$ of a free particle and a potential $V$:%
\begin{equation}
H=H_{0}+V(\mathbf{x}). \label{ZerHH0Wec}%
\end{equation}
The potential $V$ transforms as a scalar under rotations in the $q$-de\-formed
Euclidean space. It also shows trivial braiding properties. Moreover, $V$ has
to be real so that no particles are absorbed:%
\begin{equation}
\overline{V(\mathbf{x})}=V(\mathbf{x}).
\end{equation}
We assume $V$ to vanish at infinity. With the Hamiltonian operator in
Eq.~(\ref{ZerHH0Wec}), we get the $q$-de\-formed Schr\"{o}dinger equations%
\begin{align}
\text{i\hspace{0.01in}}\partial_{t}\triangleright\psi_{R}(\mathbf{x}%
,t)-H_{0}\triangleright\psi_{R}(\mathbf{x},t)  &  =V(\mathbf{x})\circledast
\psi_{R}(\mathbf{x},t),\nonumber\\
\text{i\hspace{0.01in}}\partial_{t}\,\bar{\triangleright}\,\psi_{R}^{\ast
}(\mathbf{x},t)-H_{0}\,\bar{\triangleright}\,\psi_{R}^{\ast}(\mathbf{x},t)  &
=V(\mathbf{x})\circledast\psi_{R}^{\ast}(\mathbf{x},t), \label{SchGleStr1}%
\end{align}
and%
\begin{align}
\psi_{L}(\mathbf{x},t)\,\bar{\triangleleft}\,\partial_{t}\text{\hspace
{0.01in}i}-\psi_{L}(\mathbf{x},t)\,\bar{\triangleleft}\,H_{0}  &  =\psi
_{L}(\mathbf{x},t)\circledast V(\mathbf{x}),\nonumber\\
\psi_{L}^{\ast}(\mathbf{x},t)\triangleleft\partial_{t}\text{\hspace{0.01in}%
i}-\psi_{L}^{\ast}(\mathbf{x},t)\triangleleft H_{0}  &  =\psi_{L}^{\ast
}(\mathbf{x},t)\circledast V(\mathbf{x}). \label{SchGleStr2}%
\end{align}

In a scattering experiment, a particle starts at $t=-\infty$ as a free
particle. After the interaction, the particle at $t=\infty$ will emerge as a
free particle again. For this reason, the solutions to the $q$-de\-formed
Schr\"{o}dinger equations above have to satisfy the following \textit{boundary
conditions}:\footnote{We have also taken into account that particles could go
backwards in time.}%
\begin{align}
\lim_{t\hspace{0.01in}\rightarrow\hspace{0.01in}\mp\infty}\hspace
{0.01in}[(\psi_{L/R})^{\pm}(\mathbf{x},t)-\phi_{L/R}(\mathbf{x},t)]  &
=0,\nonumber\\
\lim_{t\hspace{0.01in}\rightarrow\hspace{0.01in}\mp\infty}\hspace
{0.01in}[(\psi_{L/R}^{\ast})^{\pm}(\mathbf{x},t)-\phi_{L/R}^{\ast}%
(\mathbf{x},t)]  &  =0. \label{ConBoun2H}%
\end{align}
Recall that $\phi_{R}$, $\phi_{L}$, $\phi_{R}^{\ast}$, and $\phi_{L}^{\ast}$
denote solutions to the $q$-de\-formed Schr\"{o}dinger equations for a free
particle [cf. Eq.~(\ref{SchGleQDef1N})\ of Chap.~\ref{KapProSchrFel}].

From Chap.~\ref{KapProSchrFel}, we know how to solve $q$-ver\-sions of
inhomogeneous Schr\"{o}dinger equations by using Green's functions. This
method enables us to derive integral equations for the solutions of the
Schr\"{o}dinger equations in (\ref{SchGleStr1}) and (\ref{SchGleStr2}). By
doing so, we get $q$-ana\-logs of the \textit{Lipp\-mann-Schwin\-ger
equations}. Concretely, we have to replace the inhomogeneity $\varrho$ in
Eq.~(\ref{SolInhSchr1Hab}) or Eq.~(\ref{SolInhSchr2Hab}) of
Chap.~\ref{KapProSchrFel} with the expressions on the right-hand side of the
Schr\"{o}dinger equations in (\ref{SchGleStr1}) and (\ref{SchGleStr2}). Since
we can add a solution of the free Schr\"{o}dinger equation to a solution of an
inhomogeneous Schr\"{o}dinger equation, we finally obtain%
\begin{align}
&  (\psi_{R})^{\pm}(\mathbf{x},t)=\phi_{R}(\mathbf{x},t)\nonumber\\
&  \qquad\mp\text{i}\hspace{-0.01in}\int\text{d}t^{\hspace{0.01in}\prime
}\hspace{-0.01in}\int\text{d}_{q}^{3}x^{\hspace{0.01in}\prime}\hspace
{0.01in}(K_{R})^{\pm}(\mathbf{x},t;\mathbf{x}^{\hspace{0.01in}\prime}%
\hspace{-0.01in},t^{\hspace{0.01in}\prime})\circledast V(\mathbf{x}%
^{\hspace{0.01in}\prime})\circledast(\psi_{R})^{\pm}(\mathbf{x}^{\hspace
{0.01in}\prime}\hspace{-0.01in},t^{\hspace{0.01in}\prime}),\nonumber\\[0.1in]
&  (\psi_{R}^{\ast})^{\pm}(\mathbf{x},t)=\phi_{R}^{\ast}(\mathbf{x}%
,t)\nonumber\\
&  \qquad\mp\text{i}\hspace{-0.01in}\int\text{d}t^{\hspace{0.01in}\prime
}\hspace{-0.01in}\int\text{d}_{q}^{3}x^{\hspace{0.01in}\prime}\hspace
{0.01in}(K_{R}^{\hspace{0.01in}\ast})^{\pm}(\mathbf{x},t;\mathbf{x}%
^{\hspace{0.01in}\prime}\hspace{-0.01in},t^{\hspace{0.01in}\prime})\circledast
V(\mathbf{x}^{\hspace{0.01in}\prime})\circledast(\psi_{R}^{\ast})^{\pm
}(\mathbf{x}^{\hspace{0.01in}\prime}\hspace{-0.01in},t^{\hspace{0.01in}\prime
}), \label{LipSchGle1}%
\end{align}
and%
\begin{align}
&  (\psi_{L})^{\pm}(\mathbf{x},t)=\phi_{L}(\mathbf{x},t)\nonumber\\
&  \qquad\pm\text{i}\hspace{-0.01in}\int\text{d}t^{\hspace{0.01in}\prime
}\hspace{-0.02in}\int\text{d}_{q}^{3}x^{\hspace{0.01in}\prime}\hspace
{0.01in}(\psi_{L})^{\pm}(\mathbf{x}^{\hspace{0.01in}\prime}\hspace
{-0.01in},t^{\hspace{0.01in}\prime})\circledast V(\mathbf{x}^{\hspace
{0.01in}\prime})\circledast(K_{L})^{\pm}(\mathbf{x}^{\hspace{0.01in}\prime
}\hspace{-0.01in},t^{\hspace{0.01in}\prime};\mathbf{x},t),\nonumber\\[0.1in]
&  (\psi_{L}^{\ast})^{\pm}(\mathbf{x},t)=\phi_{L}^{\ast}(\mathbf{x}%
,t)\nonumber\\
&  \qquad\pm\text{i}\hspace{-0.01in}\int\text{d}t^{\hspace{0.01in}\prime
}\hspace{-0.02in}\int\text{d}_{q}^{3}x^{\hspace{0.01in}\prime}\hspace
{0.01in}(\psi_{L}^{\ast})^{\pm}(\mathbf{x}^{\hspace{0.01in}\prime}%
\hspace{-0.01in},t^{\hspace{0.01in}\prime})\circledast V(\mathbf{x}%
^{\hspace{0.01in}\prime})\circledast(K_{L}^{\hspace{0.01in}\ast})^{\pm
}(\mathbf{x}^{\hspace{0.01in}\prime}\hspace{-0.01in},t^{\hspace{0.01in}\prime
};\mathbf{x},t). \label{LipSchGle4}%
\end{align}
The above $q$-de\-formed Lipp\-mann-Schwin\-ger equations are compatible with
the boundary conditions in Eq.~(\ref{ConBoun2H}), as can be shown by using the
identities\footnote{In the following, we use the notation $x=(\mathbf{x},t)$
or $x^{\hspace{0.01in}\prime}=(\mathbf{x}^{\hspace{0.01in}\prime}%
\hspace{-0.01in},t^{\hspace{0.01in}\prime})$.}%
\begin{align}
\lim_{t\hspace{0.01in}\rightarrow\hspace{0.01in}\mp\infty}(K_{R})^{\pm
}(x;x^{\hspace{0.01in}\prime})  &  =\lim_{t\hspace{0.01in}\rightarrow
\hspace{0.01in}\mp\infty}\theta(\pm\hspace{0.01in}t\mp t^{\hspace
{0.01in}\prime})\,K_{R}(x;x^{\hspace{0.01in}\prime})=0,\nonumber\\
\lim_{t\hspace{0.01in}\rightarrow\hspace{0.01in}\mp\infty}(K_{R}%
^{\hspace{0.01in}\ast})^{\pm}(x;x^{\hspace{0.01in}\prime})  &  =\lim
_{t\hspace{0.01in}\rightarrow\hspace{0.01in}\mp\infty}\theta(\pm
\hspace{0.01in}t\mp t^{\hspace{0.01in}\prime})\,K_{R}^{\hspace{0.01in}\ast
}(x;x^{\hspace{0.01in}\prime})=0,
\end{align}
and%
\begin{align}
\lim_{t\hspace{0.01in}\rightarrow\hspace{0.01in}\mp\infty}(K_{L})^{\pm
}(x^{\hspace{0.01in}\prime}\hspace{-0.01in};x)  &  =\lim_{t\hspace
{0.01in}\rightarrow\hspace{0.01in}\mp\infty}\theta(\pm\hspace{0.01in}t\mp
t^{\hspace{0.01in}\prime})\,K_{L}(x^{\hspace{0.01in}\prime}\hspace
{-0.01in};x)=0,\nonumber\\
\lim_{t\hspace{0.01in}\rightarrow\hspace{0.01in}\mp\infty}(K_{L}%
^{\hspace{0.01in}\ast})^{\pm}(x^{\hspace{0.01in}\prime}\hspace{-0.01in};x)  &
=\lim_{t\hspace{0.01in}\rightarrow\hspace{0.01in}\mp\infty}\theta(\pm
\hspace{0.01in}t\mp t^{\hspace{0.01in}\prime})\,K_{L}^{\hspace{0.01in}\ast
}(x^{\hspace{0.01in}\prime}\hspace{-0.01in};x)=0.
\end{align}

We aim to solve the $q$-de\-formed Lipp\-mann-Schwin\-ger equations by
iteration. To this end, we introduce new Green's functions defined by the
following equations \cite{Bjorken1Engl}:%
\begin{align}
(\psi_{R})^{\pm}(x)  &  =\lim_{t^{\prime}\rightarrow\hspace{0.01in}\mp\infty
}\int\text{d}_{q}^{3}x^{\hspace{0.01in}\prime}\hspace{0.01in}(G_{R})^{\pm
}(x;x^{\hspace{0.01in}\prime})\circledast\phi_{R}(x^{\hspace{0.01in}\prime
}),\nonumber\\
(\psi_{R}^{\ast})^{\pm}(x)  &  =\lim_{t^{\prime}\rightarrow\hspace{0.01in}%
\mp\infty}\int\text{d}_{q}^{3}x^{\hspace{0.01in}\prime}\hspace{0.01in}%
(G_{R}^{\hspace{0.01in}\ast})^{\pm}(x;x^{\hspace{0.01in}\prime})\circledast
\phi_{R}^{\ast}(x^{\hspace{0.01in}\prime}),\label{DefGreGesInh1}\\[0.1in]
(\psi_{L})^{\pm}(x)  &  =\lim_{t^{\prime}\rightarrow\hspace{0.01in}\mp\infty
}\int\text{d}_{q}^{3}x^{\hspace{0.01in}\prime}\hspace{0.01in}\phi
_{L}(x^{\hspace{0.01in}\prime})\circledast(G_{L})^{\pm}(x^{\hspace
{0.01in}\prime}\hspace{-0.01in};x),\nonumber\\
(\psi_{L}^{\ast})^{\pm}(x)  &  =\lim_{t^{\prime}\rightarrow\hspace{0.01in}%
\mp\infty}\int\text{d}_{q}^{3}x^{\hspace{0.01in}\prime}\hspace{0.01in}\phi
_{L}^{\ast}(x^{\hspace{0.01in}\prime})\circledast(G_{L}^{\hspace{0.01in}\ast
})^{\pm}(x^{\hspace{0.01in}\prime}\hspace{-0.01in};x). \label{DefGreGesInh2}%
\end{align}
Using these identities, we rewrite the right-hand side of each $q$-de\-formed
Lipp\-mann-Schwin\-ger equation. We get, for example:%
\begin{align}
(\psi_{R})^{\pm}(x)=  &  \lim_{t^{\prime}\rightarrow\hspace{0.01in}\mp\infty
}\int\text{d}_{q}^{3}x^{\hspace{0.01in}\prime}\hspace{0.01in}(K_{R})^{\pm
}(x;x^{\hspace{0.01in}\prime})\circledast\phi_{R}(x^{\hspace{0.01in}\prime
})\nonumber\\
&  \mp\text{i}\lim_{t^{\prime}\rightarrow\hspace{0.01in}\mp\infty}\int
\text{d}_{q}^{3}x^{\hspace{0.01in}\prime}\hspace{-0.02in}\int\text{d}%
t^{\hspace{0.01in}\prime\prime}\hspace{-0.02in}\int\text{d}_{q}^{3}%
x^{\hspace{0.01in}\prime\prime}\,(K_{R})^{\pm}(x;x^{\hspace{0.01in}%
\prime\prime})\circledast V(\mathbf{x}^{\hspace{0.01in}\prime\prime
})\nonumber\\
&  \qquad\qquad\qquad\qquad\circledast(G_{R})^{\pm}(x^{\hspace{0.01in}%
\prime\prime}\hspace{-0.01in};x^{\hspace{0.01in}\prime})\circledast\phi
_{R}(x^{\hspace{0.01in}\prime}). \label{HerGesGre1}%
\end{align}
Note that the first expression on the right-hand side of the above equation is
a consequence of the following identity \cite{Wachter:2020B}:%
\begin{align}
\phi_{R}(\mathbf{x},t)  
&  =\lim_{t^{\prime}\rightarrow\hspace{0.01in}\mp\infty}\int\text{d}_{q}%
^{3}x^{\hspace{0.01in}\prime}\hspace{0.01in}(K_{R})^{\pm}(\mathbf{x}%
,t;\mathbf{x}^{\hspace{0.01in}\prime}\hspace{-0.01in},t^{\hspace{0.01in}%
\prime})\circledast\phi_{R}(\mathbf{x}^{\hspace{0.01in}\prime}\hspace
{-0.01in},t^{\hspace{0.01in}\prime}). \label{PhiTraKL}%
\end{align}
Comparing Eq.~(\ref{HerGesGre1}) with the first identity in
Eq.~(\ref{DefGreGesInh1}), we can read off a Lipp\-mann-Schwin\-ger equation
for the Green's function $(G_{R})^{+}$ or $(G_{R})^{-}$:%
\begin{align}
&  (G_{R})^{\pm}(x;x^{\hspace{0.01in}\prime})=(K_{R})^{\pm}(x;x^{\hspace
{0.01in}\prime})\nonumber\\
&  \qquad\mp\text{i\hspace{-0.01in}}\int\text{d}t^{\hspace{0.01in}\prime
\prime}\hspace{-0.02in}\int\text{d}_{q}^{3}x^{\hspace{0.01in}\prime\prime
}\hspace{0.01in}(K_{R})^{\pm}(x;x^{\hspace{0.01in}\prime\prime})\circledast
V(\mathbf{x}^{\hspace{0.01in}\prime\prime})\circledast(G_{R})^{\pm}%
(x^{\hspace{0.01in}\prime\prime}\hspace{-0.01in};x^{\hspace{0.01in}\prime}).
\label{LipSchwGleGreGLGesDrei}%
\end{align}
In the same way, we get:%
\begin{align}
&  (G_{L}^{\hspace{0.01in}\ast})^{\pm}(x^{\hspace{0.01in}\prime}%
\hspace{-0.01in};x)=(K_{L}^{\hspace{0.01in}\ast})^{\pm}(x^{\hspace
{0.01in}\prime}\hspace{-0.01in};x)\nonumber\\
&  \qquad\pm\text{i\hspace{-0.01in}}\int\text{d}t^{\hspace{0.01in}\prime
\prime}\hspace{-0.02in}\int\text{d}_{q}^{3}x^{\hspace{0.01in}\prime\prime
}\hspace{0.01in}(G_{L}^{\hspace{0.01in}\ast})^{\pm}(x;x^{\hspace{0.01in}%
\prime\prime})\circledast V(\mathbf{x}^{\hspace{0.01in}\prime\prime
})\circledast(K_{L}^{\hspace{0.01in}\ast})^{\pm}(x^{\hspace{0.01in}%
\prime\prime}\hspace{-0.01in};x^{\hspace{0.01in}\prime}).
\label{LipSchwGleGreGRGesDrei}%
\end{align}
We obtain the Lipp\-mann-Schwin\-ger equations for the other Green's functions
in Eqs.~(\ref{DefGreGesInh1}) and (\ref{DefGreGesInh2}) by applying the
following substitutions:%
\begin{align}
G_{R}  &  \rightarrow G_{R}^{\hspace{0.01in}\ast},\qquad K_{R} \rightarrow K_{R}^{\hspace
{0.01in}\ast},\nonumber\\
G_{L}^{\hspace{0.01in}\ast}  &  \rightarrow G_{L},\qquad K_{L}^{\hspace
{0.01in}\ast}\rightarrow K_{L}. \label{ErsRegGreFktGK}%
\end{align}

The new Green's functions satisfy Schr\"{o}dinger equations for the
Hamiltonian operator $H$ with a $q$-de\-formed delta function as
inhomogeneity, i.~e.%
\begin{align}
(\text{i\hspace{0.01in}}\partial_{t}-H)\triangleright(G_{R})^{\pm
}(x;x^{\hspace{0.01in}\prime})  &  =\pm\,\text{i}\operatorname*{vol}%
\nolimits^{-1}\hspace{-0.01in}\delta(\pm\hspace{0.01in}t\mp t^{\hspace
{0.01in}\prime})\,\delta_{q}^{3}(\mathbf{x}\oplus(\ominus\hspace{0.01in}%
\kappa^{-1}\mathbf{x}^{\hspace{0.01in}\prime})),\nonumber\\
(\text{i\hspace{0.01in}}\partial_{t}-H)\,\bar{\triangleright}\,(G_{R}%
^{\hspace{0.01in}\ast})^{\pm}(x;x^{\hspace{0.01in}\prime})  &  =\pm
\,\text{i}\operatorname*{vol}\nolimits^{-1}\hspace{-0.01in}\delta(\pm
\hspace{0.01in}t\mp t^{\hspace{0.01in}\prime})\,\delta_{q}^{3}((\ominus
\hspace{0.01in}\kappa^{-1}\mathbf{x})\oplus\mathbf{x}^{\hspace{0.01in}\prime
}), \label{ChaIdeSchrGesGreFkt1}%
\end{align}
and%
\begin{align}
(G_{L})^{\pm}(x^{\hspace{0.01in}\prime}\hspace{-0.01in};x)\,\bar
{\triangleleft}\,(\partial_{t}\text{\hspace{0.01in}i}-H)  &  =\mp
\,\text{i}\operatorname*{vol}\nolimits^{-1}\hspace{-0.01in}\delta(\pm
\hspace{0.01in}t\mp t^{\hspace{0.01in}\prime})\,\delta_{q}^{3}((\ominus
\hspace{0.01in}\kappa^{-1}\mathbf{x}^{\hspace{0.01in}\prime})\oplus
\mathbf{x}),\nonumber\\
(G_{L}^{\hspace{0.01in}\ast})^{\pm}(x^{\hspace{0.01in}\prime}\hspace
{-0.01in};x)\triangleleft(\partial_{t}\text{\hspace{0.01in}i}-H)  &
=\mp\,\text{i}\operatorname*{vol}\nolimits^{-1}\hspace{-0.01in}\delta
(\pm\hspace{0.01in}t\mp t^{\hspace{0.01in}\prime})\,\delta_{q}^{3}%
(\mathbf{x}^{\hspace{0.01in}\prime}\oplus(\ominus\hspace{0.01in}\kappa
^{-1}\mathbf{x})). \label{ChaIdeSchrGesGreFkt2}%
\end{align}
By way of example, we show how to derive these identities. First, we apply the
operator i\hspace{0.01in}$\partial_{t}-H$ to the Lipp\-mann-Schwin\-ger
equation for $(G_{R})^{\pm}$ [see Eq.~(\ref{LipSchwGleGreGLGesDrei})]:%
\begin{align}
&  (\text{i\hspace{0.01in}}\partial_{t}-H)\triangleright(G_{R})^{\pm
}(x;x^{\hspace{0.01in}\prime})=\nonumber\\
&  \qquad=(\text{i\hspace{0.01in}}\partial_{t}-H_{0})\triangleright
(K_{R})^{\pm}(x;x^{\hspace{0.01in}\prime})-V(\mathbf{x})\circledast
(K_{R})^{\pm}(x;x^{\hspace{0.01in}\prime})\nonumber\\
&  \qquad\hspace*{0.15in}\mp\text{i}\int\text{d}t^{\hspace{0.01in}\prime
\prime}\hspace{0.01in}\text{d}_{q}^{3}x^{\hspace{0.01in}\prime\prime}%
\hspace{0.01in}(\text{i\hspace{0.01in}}\partial_{t}-H_{0})\triangleright
(K_{R})^{\pm}(x;x^{\hspace{0.01in}\prime\prime})\circledast V(\mathbf{x}%
^{\hspace{0.01in}\prime\prime})\circledast(G_{R})^{\pm}(x^{\hspace
{0.01in}\prime\prime}\hspace{-0.01in};x^{\hspace{0.01in}\prime})\nonumber\\
&  \qquad\hspace*{0.15in}\pm\text{i}\int\text{d}t^{\hspace{0.01in}\prime
\prime}\hspace{0.01in}\text{d}_{q}^{3}x^{\hspace{0.01in}\prime\prime
}\,V(\mathbf{x})\circledast(K_{R})^{\pm}(x;x^{\hspace{0.01in}\prime\prime
})\circledast V(\mathbf{x}^{\hspace{0.01in}\prime\prime})\circledast
(G_{R})^{\pm}(x^{\hspace{0.01in}\prime\prime}\hspace{-0.01in};x^{\hspace
{0.01in}\prime}).
\end{align}
Next, we use the identities for the free Schr\"{o}dinger propagator
$(K_{R})^{\pm}$ given in Eq.~(\ref{SchrGlGreHabAnf}) of
Chap.~\ref{KapProSchrFel}:%
\begin{align}
&  (\text{\hspace{0.01in}i\hspace{0.01in}}\partial_{t}-H)\triangleright
(G_{R})^{\pm}(x;x^{\hspace{0.01in}\prime})=\nonumber\\
&  \hspace*{0.15in}=\pm\,\text{i}\operatorname*{vol}\nolimits^{-1}%
\hspace{-0.01in}\delta(t-t^{\hspace{0.01in}\prime})\,\delta_{q}^{3}%
(\mathbf{x}\oplus(\ominus\hspace{0.01in}\kappa^{-1}\mathbf{x}^{\hspace
{0.01in}\prime}))-V(\mathbf{x})\circledast(K_{R})^{\pm}(x;x^{\hspace
{0.01in}\prime})\nonumber\\
&  \hspace*{0.15in}\hspace*{0.15in}+\operatorname*{vol}\nolimits^{-1}%
\hspace{-0.02in}\int\text{d}t^{\hspace{0.01in}\prime\prime}\hspace
{0.01in}\text{d}_{q}^{3}x^{\hspace{0.01in}\prime\prime}\,\delta(t-t^{\hspace
{0.01in}\prime\prime})\,\delta_{q}^{3}(\mathbf{x}\oplus(\ominus\hspace
{0.01in}\kappa^{-1}\mathbf{x}^{\hspace{0.01in}\prime\prime}))\circledast
V(\mathbf{x}^{\hspace{0.01in}\prime\prime})\circledast(G_{R})^{\pm}%
(x^{\hspace{0.01in}\prime\prime}\hspace{-0.01in};x^{\hspace{0.01in}\prime
})\nonumber\\
&  \hspace*{0.15in}\hspace*{0.15in}\pm\text{i\hspace{0.01in}}V(\mathbf{x}%
)\circledast\int\text{d}t^{\hspace{0.01in}\prime\prime}\hspace{0.01in}%
\text{d}_{q}^{3}x^{\hspace{0.01in}\prime\prime}\hspace{0.01in}(K_{R})^{\pm
}(x;x^{\hspace{0.01in}\prime\prime})\circledast V(\mathbf{x}^{\hspace
{0.01in}\prime\prime})\circledast(G_{R})^{\pm}(x^{\hspace{0.01in}\prime\prime
}\hspace{-0.01in};x^{\hspace{0.01in}\prime})\nonumber\\
&  \hspace*{0.15in}=\pm\,\text{i}\operatorname*{vol}\nolimits^{-1}%
\hspace{-0.01in}\delta(t-t^{\hspace{0.01in}\prime})\,\delta_{q}^{3}%
(\mathbf{x}\oplus(\ominus\hspace{0.01in}\kappa^{-1}\mathbf{x}^{\hspace
{0.01in}\prime})).
\end{align}
For a better understanding, we note the following: After calculating the
integrals of the expressions with the delta functions, all terms depending on
the interaction potential $V$ cancel out due to
Eq.~(\ref{LipSchwGleGreGLGesDrei}).

The Green's functions for the $q$-de\-formed Schr\"{o}dinger
equations with interaction satisfy the identities%
\begin{align}
\theta(\pm\hspace*{0.01in}t\mp t^{\hspace{0.01in}\prime})(\psi_{R})^{\pm}(x)
&  =\hspace{-0.01in}\int\text{d}_{q}^{3}x^{\hspace{0.01in}\prime}%
\hspace{0.01in}(G_{R})^{\pm}(x;x^{\hspace{0.01in}\prime})\circledast(\psi
_{R})^{\pm}(x^{\hspace{0.01in}\prime}),\nonumber\\
\theta(\pm\hspace*{0.01in}t\mp t^{\hspace{0.01in}\prime})(\psi_{R}^{\ast
})^{\pm}(x)  &  =\hspace{-0.01in}\int\text{d}_{q}^{3}x^{\hspace{0.01in}\prime
}\hspace{0.01in}(G_{R}^{\hspace{0.01in}\ast})^{\pm}(x;x^{\hspace{0.01in}%
\prime})\circledast(\psi_{L}^{\ast})^{\pm}(x^{\hspace{0.01in}\prime}),
\label{NeuDefGreGesTheNicRel1}%
\end{align}
and%
\begin{align}
\theta(\pm\hspace*{0.01in}t\mp t^{\hspace{0.01in}\prime})(\psi_{L})^{\pm}(x)
&  =\hspace{-0.01in}\int\text{d}_{q}^{3}x^{\hspace{0.01in}\prime}%
\hspace{0.01in}(\psi_{L})^{\pm}(x^{\hspace{0.01in}\prime})\circledast
(G_{L})^{\pm}(x^{\hspace{0.01in}\prime}\hspace{-0.01in};x),\nonumber\\
\theta(\pm\hspace*{0.01in}t\mp t^{\hspace{0.01in}\prime})(\psi_{L}^{\ast
})^{\pm}(x)  &  =\hspace{-0.01in}\int\text{d}_{q}^{3}x^{\hspace{0.01in}\prime
}\hspace{0.01in}(\psi_{L}^{\ast})^{\pm}(x^{\hspace{0.01in}\prime}%
)\circledast(G_{L}^{\hspace{0.01in}\ast})^{\pm}(x^{\hspace{0.01in}\prime
}\hspace{-0.01in};x). \label{NeuDefGreGesTheNicRel2}%
\end{align}
Indeed, if we take the limit $t^{\prime}\rightarrow\mp\infty$ on both sides of
the above identities and consider the boundary conditions in
Eq.~(\ref{ConBoun2H}), we regain the identities in
Eq.~(\ref{DefGreGesInh1}) or Eq.~(\ref{DefGreGesInh2}).

We can also use Eqs.~(\ref{ChaIdeSchrGesGreFkt1}) and
(\ref{ChaIdeSchrGesGreFkt2}) to derive the formulas in
Eq.~(\ref{NeuDefGreGesTheNicRel1}) and Eq.~(\ref{NeuDefGreGesTheNicRel2}). We
show this by an example, applying the operator i$\hspace{0.01in}\partial
_{t}-H$ to both sides of the first identity of
Eq.~(\ref{NeuDefGreGesTheNicRel1}):%
\begin{align}
&  \int\text{d}_{q}^{3}x^{\hspace{0.01in}\prime}\hspace{0.01in}(\text{i\hspace
{0.01in}}\partial_{t}-H)\triangleright(G_{R})^{\pm}(x;x^{\hspace{0.01in}%
\prime})\circledast(\psi_{R})^{\pm}(x^{\hspace{0.01in}\prime})=\nonumber\\
&  \qquad=(\text{i\hspace{0.01in}}\partial_{t}-H)\triangleright\theta
(\pm\hspace*{0.01in}t\mp t^{\hspace{0.01in}\prime})\hspace{0.01in}(\psi
_{R})^{\pm}(x)=\pm\hspace*{0.01in}\text{i\hspace{0.01in}}\delta(\pm
\hspace*{0.01in}t\mp t^{\hspace{0.01in}\prime})\hspace{0.01in}(\psi_{R})^{\pm
}(x)\nonumber\\
&  \qquad=\pm\hspace*{0.01in}\text{i\hspace{0.01in}}\delta(\pm\hspace
*{0pt}t\mp t^{\hspace{0.01in}\prime})\operatorname*{vol}\nolimits^{-1}%
\hspace{-0.02in}\int\text{d}_{q}^{3}x^{\hspace{0.01in}\prime}\hspace
{0.01in}\delta_{q}^{3}(\mathbf{x}\oplus(\ominus\hspace{0.01in}\kappa
^{-1}\mathbf{x}^{\hspace{0.01in}\prime}))\circledast(\psi_{R})^{\pm
}(x^{\hspace{0.01in}\prime}).
\end{align}
The result above implies the first identity in Eq.~(\ref{ChaIdeSchrGesGreFkt1}).

The propagators for the Schr\"{o}dinger equations with interaction can be
combined in the following way:%
\begin{align}
(G_{R})^{\pm}(\mathbf{x},t;\mathbf{x}^{\hspace{0.01in}\prime}\hspace
{-0.01in},t^{\hspace{0.01in}\prime})  &  =\int\text{d}^{3}x^{\hspace
{0.01in}\prime\prime}\hspace{0.01in}(G_{R})^{\pm}(\mathbf{x},t;\mathbf{x}%
^{\hspace{0.01in}\prime\prime}\hspace{-0.01in},t^{\hspace{0.01in}\prime\prime
})\circledast(G_{R})^{\pm}(\mathbf{x}^{\hspace{0.01in}\prime\prime}%
\hspace{-0.01in},t^{\hspace{0.01in}\prime\prime}\hspace{-0.01in}%
;\mathbf{x}^{\hspace{0.01in}\prime}\hspace{-0.01in},t^{\hspace{0.01in}\prime
}),\nonumber\\
(G_{L})^{\pm}(\mathbf{x}^{\hspace{0.01in}\prime}\hspace{-0.01in}%
,t^{\hspace{0.01in}\prime}\hspace{-0.01in};\mathbf{x},t)  &  =\int\text{d}%
_{q}^{3}x^{\hspace{0.01in}\prime\prime}\hspace{0.01in}(G_{L})^{\pm}%
(\mathbf{x}^{\hspace{0.01in}\prime},t^{\hspace{0.01in}\prime};\mathbf{x}%
^{\hspace{0.01in}\prime\prime}\hspace{-0.01in},t^{\hspace{0.01in}\prime\prime
})\circledast(G_{L})^{\pm}(\mathbf{x}^{\hspace{0.01in}\prime\prime}%
\hspace{-0.01in},t^{\hspace{0.01in}\prime\prime}\hspace{-0.01in}%
;\mathbf{x},t). \label{KomEigGreFktGes}%
\end{align}
For the intermediate point $t^{\hspace{0.01in}\prime\prime}$, it is
$t^{\hspace{0.01in}\prime}\hspace{-0.01in}\leq t^{\hspace{0.01in}\prime\prime
}\hspace{-0.01in}\leq t$ in the case of retarded propagators, and $t\leq
t^{\hspace{0.01in}\prime\prime}\hspace{-0.01in}\leq t^{\hspace{0.01in}\prime}$
in the case of advanced ones. We can prove the above rules by applying the
identities in Eqs.~(\ref{NeuDefGreGesTheNicRel1}) and
(\ref{NeuDefGreGesTheNicRel2}) twice. We obtain further identities if we
substitute $(G_{R})^{\pm}$ by $(G_{R}^{\hspace{0.01in}\ast})^{\pm}$ or
$(G_{L})^{\pm}$ by $(G_{L}^{\hspace{0.01in}\ast})^{\pm}$.

We consider the case that a retarded propagator transforms a wave function
from time $t^{\hspace{0.01in}\prime}$ to a later time $t$. An advanced
propagator pulls the new wave function from time $t$ back to the initial time
$t^{\hspace{0.01in}\prime}$. This way, we regain the original wave function:%
\begin{gather}
\psi_{R}(x^{\hspace{0.01in}\prime})=\int\text{d}_{q}^{3}x\,(G_{R}%
)^{-}(x^{\hspace{0.01in}\prime}\hspace{-0.01in};x)\circledast\psi
_{R}(x)\nonumber\\
=\int\text{d}_{q}^{3}x^{\hspace{0.01in}\prime\prime}\hspace{-0.02in}%
\int\text{d}_{q}^{3}x\,(G_{R})^{-}(\mathbf{x}^{\hspace{0.01in}\prime}%
\hspace{-0.01in},t^{\hspace{0.01in}\prime}\hspace{-0.01in};\mathbf{x}%
,t)\circledast(G_{R})^{+}(\mathbf{x},t;\mathbf{x}^{\hspace{0.01in}\prime
\prime}\hspace{-0.01in},t^{\hspace{0.01in}\prime})\circledast\psi
_{R}(\mathbf{x}^{\hspace{0.01in}\prime\prime}\hspace{-0.01in},t^{\hspace
{0.01in}\prime}).
\end{gather}
Comparing this result with the identity%
\begin{equation}
\psi_{R}(\mathbf{x}^{\hspace{0.01in}\prime}\hspace{-0.01in},t^{\hspace
{0.01in}\prime})=\operatorname*{vol}\nolimits^{-1}\hspace{-0.02in}\int
\text{d}_{q}^{3}x^{\hspace{0.01in}\prime\prime}\,\delta_{q}^{3}(\mathbf{x}%
^{\hspace{0.01in}\prime}\oplus(\ominus\hspace{0.01in}\kappa^{-1}%
\mathbf{x}^{\hspace{0.01in}\prime\prime}))\circledast\psi_{R}(\mathbf{x}%
^{\hspace{0.01in}\prime\prime}\hspace{-0.01in},t^{\hspace{0.01in}\prime}),
\end{equation}
we obtain for $t^{\hspace{0.01in}\prime}\hspace{-0.01in}\leq t$:%
\begin{equation}
\int\text{d}_{q}^{3}x\,(G_{R})^{-}(\mathbf{x}^{\hspace{0.01in}\prime}%
\hspace{-0.01in},t^{\hspace{0.01in}\prime}\hspace{-0.01in};\mathbf{x}%
,t)\circledast(G_{R})^{+}(\mathbf{x},t;\mathbf{x}^{\hspace{0.01in}\prime
\prime}\hspace{-0.01in},t^{\hspace{0.01in}\prime})=\operatorname*{vol}%
\nolimits^{-1}\hspace{-0.01in}\delta_{q}^{3}(\mathbf{x}^{\hspace{0.01in}%
\prime}\oplus(\ominus\hspace{0.01in}\kappa^{-1}\mathbf{x}^{\hspace
{0.01in}\prime\prime})). \label{KomGreDel1}%
\end{equation}
In the same manner, we obtain for $t^{\hspace{0.01in}\prime}\hspace
{-0.01in}\geq t$:%
\begin{equation}
\int\text{d}_{q}^{3}x\,(G_{R})^{+}(\mathbf{x}^{\hspace{0.01in}\prime}%
\hspace{-0.01in},t^{\hspace{0.01in}\prime}\hspace{-0.01in};\mathbf{x}%
,t)\circledast(G_{R})^{-}(\mathbf{x},t;\mathbf{x}^{\hspace{0.01in}\prime
\prime}\hspace{-0.01in},t^{\hspace{0.01in}\prime})=\operatorname*{vol}%
\nolimits^{-1}\hspace{-0.01in}\delta_{q}^{3}(\mathbf{x}^{\hspace{0.01in}%
\prime}\oplus(\ominus\hspace{0.01in}\kappa^{-1}\mathbf{x}^{\hspace
{0.01in}\prime\prime})).
\end{equation}
Similar arguments lead to%
\begin{equation}
\int\text{d}_{q}^{3}x\,(G_{L}^{\hspace{0.01in}\ast})^{\pm}(\mathbf{x}%
^{\hspace{0.01in}\prime\prime}\hspace{-0.01in},t^{\hspace{0.01in}\prime
}\hspace{-0.01in};\mathbf{x},t)\circledast(G_{L}^{\hspace{0.01in}\ast})^{\mp
}(\mathbf{x},t;\mathbf{x}^{\hspace{0.01in}\prime}\hspace{-0.01in}%
,t^{\hspace{0.01in}\prime})=\operatorname*{vol}\nolimits^{-1}\hspace
{-0.01in}\delta_{q}^{3}((\ominus\hspace{0.01in}\kappa^{-1}\mathbf{x}%
^{\hspace{0.01in}\prime\prime})\oplus\mathbf{x}^{\hspace{0.01in}\prime})
\label{KomGreDel3}%
\end{equation}
with $t^{\hspace{0.01in}\prime}\hspace{-0.01in}\leq t$ for the upper sign and
$t^{\hspace{0.01in}\prime}\hspace{-0.01in}\geq t$ for the lower sign. For the
other $q$-ver\-sions of Green's functions, we get corresponding relations if
we substitute $(G_{R})^{\pm}$ by $(G_{R}^{\hspace{0.01in}\ast})^{\pm}$ or
$(G_{L}^{\hspace{0.01in}\ast})^{\pm}$ by $(G_{L})^{\pm}$ in the above identities.

If the interaction potential $V$ is sufficiently small, we can solve the
Lipp\-mann-Schwin\-ger equations in (\ref{LipSchGle1}) and
(\ref{LipSchGle4}) iteratively. To this end, we consider the solution of a
free $q$-de\-formed Schr\"{o}dinger equation as an approximate solution and
plug it into the right-hand side of the corresponding $q$-de\-formed
Lipp\-mann-Schwin\-ger equation. Doing so, we get a second approximate
solution, which we again plug into the right-hand side of the $q$-de\-formed
Lipp\-mann-Schwin\-ger equation. Repeating this procedure, we finally obtain
$q$-ver\-sions of the \textit{Born series}, i.~e.%
\begin{align}
(\psi_{R})^{\pm}(x)=  &  \hspace{0.04in}\phi_{R}(x)+\text{i}^{\mp1}%
\hspace{-0.02in}\int\text{d}t^{\hspace{0.01in}\prime}\hspace{-0.02in}%
\int\text{d}_{q}^{3}x^{\hspace{0.01in}\prime}\hspace{0.01in}(K_{R})^{\pm
}(x;x^{\hspace{0.01in}\prime})\circledast V(\mathbf{x}^{\hspace{0.01in}\prime
})\circledast\phi_{R}(x^{\hspace{0.01in}\prime})\nonumber\\
&  \hspace{0.02in}+\text{i}^{\mp\hspace{0.01in}2}\int\text{d}t^{\hspace
{0.01in}\prime}\hspace{-0.02in}\int\text{d}_{q}^{3}x^{\hspace{0.01in}\prime
}\hspace{-0.02in}\int\text{d}t^{\hspace{0.01in}\prime\prime}\hspace
{-0.02in}\int\text{d}_{q}^{3}x^{\hspace{0.01in}\prime\prime}\hspace
{0.01in}(K_{R})^{\pm}(x;x^{\hspace{0.01in}\prime})\circledast V(\mathbf{x}%
^{\hspace{0.01in}\prime})\nonumber\\
&  \hspace{0.03in}\qquad\qquad\qquad\circledast(K_{R})^{\pm}(x^{\hspace
{0.01in}\prime}\hspace{-0.01in};x^{\hspace{0.01in}\prime\prime})\circledast
V(\mathbf{x}^{\hspace{0.01in}\prime\prime})\circledast\phi_{R}(x^{\hspace
{0.01in}\prime\prime})+\ldots\label{LipSchGlePsiL}%
\end{align}
and%
\begin{align}
(\psi_{L}^{\ast})^{\pm}(x)=  &  \hspace{0.04in}\phi_{L}^{\ast}(x)+\text{i}%
^{\pm1}\hspace{-0.02in}\int\text{d}t^{\hspace{0.01in}\prime}\hspace
{-0.02in}\int\text{d}_{q}^{3}x^{\hspace{0.01in}\prime}\,\phi_{L}^{\ast
}(x^{\hspace{0.01in}\prime})\circledast V(\mathbf{x}^{\hspace{0.01in}\prime
})\circledast(K_{L}^{\hspace{0.01in}\ast})^{\pm}(x^{\hspace{0.01in}\prime
}\hspace{-0.01in};x)\nonumber\\
&  \hspace{0.02in}+\text{i}^{\pm\hspace{0.01in}2}\hspace{-0.02in}\int
\text{d}t^{\hspace{0.01in}\prime}\hspace{-0.02in}\int\text{d}_{q}%
^{3}x^{\hspace{0.01in}\prime}\hspace{-0.02in}\int\text{d}t^{\hspace
{0.01in}\prime\prime}\hspace{-0.02in}\int\text{d}_{q}^{3}x^{\hspace
{0.01in}\prime\prime}\hspace{0.01in}\phi_{L}^{\ast}(x^{\hspace{0.01in}%
\prime\prime})\circledast V(\mathbf{x}^{\hspace{0.01in}\prime\prime
})\nonumber\\
&  \hspace{0.03in}\qquad\qquad\qquad\circledast(K_{L}^{\hspace{0.01in}\ast
})^{\pm}(x^{\hspace{0.01in}\prime\prime}\hspace{-0.01in};x^{\hspace
{0.01in}\prime})\circledast V(\mathbf{x}^{\hspace{0.01in}\prime}%
)\circledast(K_{L}^{\hspace{0.01in}\ast})^{\pm}(x^{\hspace{0.01in}\prime
}\hspace{-0.01in};x)+\ldots\label{LipSchGlePsiRSte}%
\end{align}
We get the expansions for the other $q$-ver\-sions of the
Lipp\-mann-Schwin\-ger equation by applying the following substitutions to the
above identities:%
\begin{align}
\psi_{R}  &  \rightarrow\psi_{R}^{\ast},\qquad\phi_{R}\rightarrow\phi
_{R}^{\ast},\qquad K_{R}\rightarrow K_{R}^{\ast},\nonumber\\
\psi_{L}^{\ast}  &  \rightarrow\psi_{L},\qquad\phi_{L}^{\ast}\rightarrow
\phi_{L},\qquad K_{L}^{\ast}\rightarrow K_{L}.
\end{align}

We can also solve the $q$-de\-formed Lipp\-mann-Schwin\-ger equations for
Green's functions by iteration. For example, if we plug in the free propagator
$(K_{R})^{\pm}$ on the right-hand side of
Eq.~(\ref{LipSchwGleGreGLGesDrei}) as a first approximation for the Green's function $(G_{R})^{\pm}$,
we get a second approximation for $(G_{R})^{\pm}$, which we plug in again on
the right-hand side of Eq.~(\ref{LipSchwGleGreGLGesDrei}), and so on. This
way, we have%
\begin{align}
&  (G_{R})^{\pm}(x;x^{\hspace{0.01in}\prime})=(K_{R})^{\pm}(x;x^{\hspace
{0.01in}\prime})\nonumber\\
&  \qquad+\text{i}^{\mp1}\hspace{-0.02in}\int\text{d}t^{\hspace{0.01in}%
\prime\prime}\hspace{-0.02in}\int\text{d}_{q}^{3}x^{\hspace{0.01in}%
\prime\prime}\hspace{0.01in}(K_{R})^{\pm}(x;x^{\hspace{0.01in}\prime\prime
})\circledast V(\mathbf{x}^{\hspace{0.01in}\prime\prime})\circledast
(K_{R})^{\pm}(x^{\hspace{0.01in}\prime\prime}\hspace{-0.01in};x^{\hspace
{0.01in}\prime})\nonumber\\
&  \qquad+\text{i}^{\mp\hspace{0.01in}2}\hspace{-0.02in}\int\text{d}%
t^{\hspace{0.01in}\prime\prime}\hspace{-0.02in}\int\text{d}_{q}^{3}%
x^{\hspace{0.01in}\prime\prime}\hspace{-0.02in}\int\text{d}t^{\hspace
{0.01in}\prime\prime\prime}\hspace{-0.02in}\int\text{d}_{q}^{3}x^{\hspace
{0.01in}\prime\prime\prime}\hspace{0.01in}(K_{R})^{\pm}(x;x^{\hspace
{0.01in}\prime\prime\prime})\circledast V(\mathbf{x}^{\hspace{0.01in}%
\prime\prime\prime})\nonumber\\
&  \qquad\quad\qquad\qquad\circledast(K_{R})^{\pm}(x^{\hspace{0.01in}%
\prime\prime\prime}\hspace{-0.01in};x^{\hspace{0.01in}\prime\prime
})\circledast V(\mathbf{x}^{\hspace{0.01in}\prime\prime})\circledast
(K_{R})^{\pm}(x^{\hspace{0.01in}\prime\prime}\hspace{-0.01in};x^{\hspace
{0.01in}\prime})+\ldots\label{BorReiGL}%
\end{align}
and%
\begin{align}
&  (G_{L}^{\ast})^{\pm}(x^{\hspace{0.01in}\prime}\hspace{-0.01in}%
;x)=(K_{L}^{\hspace{0.01in}\ast})^{\pm}(x^{\hspace{0.01in}\prime}%
\hspace{-0.01in};x)\nonumber\\
&  \qquad+\text{i}^{\pm1}\hspace{-0.02in}\int\text{d}t^{\hspace{0.01in}%
\prime\prime}\hspace{-0.02in}\int\text{d}_{q}^{3}x^{\hspace{0.01in}%
\prime\prime}\hspace{0.01in}(K_{L}^{\hspace{0.01in}\ast})^{\pm}(x^{\hspace
{0.01in}\prime}\hspace{-0.01in};x^{\hspace{0.01in}\prime\prime})\circledast
V(\mathbf{x}^{\hspace{0.01in}\prime\prime})\circledast(K_{L}^{\hspace
{0.01in}\ast})^{\pm}(x^{\hspace{0.01in}\prime\prime}\hspace{-0.01in}%
;x)\nonumber\\
&  \qquad+\text{i}^{\pm\hspace{0.01in}2}\hspace{-0.02in}\int\text{d}%
t^{\hspace{0.01in}\prime\prime}\hspace{-0.02in}\int\text{d}_{q}^{3}%
x^{\hspace{0.01in}\prime\prime}\hspace{-0.02in}\int\text{d}t^{\hspace
{0.01in}\prime\prime\prime}\hspace{-0.02in}\int\text{d}_{q}^{3}x^{\hspace
{0.01in}\prime\prime\prime}\hspace{0.01in}(K_{L}^{\hspace{0.01in}\ast})^{\pm
}(x^{\hspace{0.01in}\prime}\hspace{-0.01in};x^{\hspace{0.01in}\prime\prime
})\circledast V(\mathbf{x}^{\hspace{0.01in}\prime\prime})\nonumber\\
&  \qquad\quad\qquad\qquad\circledast(K_{L}^{\hspace{0.01in}\ast})^{\pm
}(x^{\hspace{0.01in}\prime\prime}\hspace{-0.01in};x^{\hspace{0.01in}%
\prime\prime\prime})\circledast V(\mathbf{x}^{\hspace{0.01in}\prime
\prime\prime})\circledast(K_{L}^{\hspace{0.01in}\ast})^{\pm}(x^{\hspace
{0.01in}\prime\prime\prime}\hspace{-0.01in};x)+\ldots
\end{align}
To obtain the corresponding expansions for the other $q$-ver\-sions of Green's
functions, we apply the substitutions of Eq.~(\ref{ErsRegGreFktGK}) to the
above identities again. If we compare the expansions for $(G_{R})^{\pm}$ and
$(G_{L}^{\hspace{0.01in}\ast})^{\pm}$ and those for $(G_{L})^{\pm}$ and
$(G_{R}^{\hspace{0.01in}\ast})^{\pm}$, we find the following identifications:%
\begin{align}
(G_{R})^{\pm}(x;x^{\hspace{0.01in}\prime})  &  =(G_{L}^{\hspace{0.01in}\ast
})^{\mp}(x;x^{\hspace{0.01in}\prime}),\nonumber\\
(G_{L})^{\pm}(x;x^{\hspace{0.01in}\prime})  &  =(G_{R}^{\hspace{0.01in}\ast
})^{\mp}(x;x^{\hspace{0.01in}\prime}). \label{ZusGreAvaRet}%
\end{align}

Next, we study how Green's functions for the Schr\"{o}dinger equations with
interaction behave under conjugation. If we conjugate
Eq.~(\ref{LipSchwGleGreGLGesDrei}) and take into account the conjugation
properties of the free propagators [cf. Eq.~(\ref{KonFreProNR}) in
Chap.~\ref{KapProSchrFel}], we obtain%
\begin{align}
&  \overline{(G_{R})^{\pm}(x;x^{\hspace{0.01in}\prime})}=(K_{L})^{\pm
}(x^{\hspace{0.01in}\prime}\hspace{-0.01in};x)\nonumber\\
&  \qquad\pm\text{i\hspace{-0.01in}}\int\text{d}t^{\hspace{0.01in}\prime
\prime}\hspace{-0.02in}\int\text{d}_{q}^{3}x^{\hspace{0.01in}\prime\prime
}\,\overline{(G_{R})^{\pm}(x^{\hspace{0.01in}\prime\prime}\hspace{-0.01in}%
;x)}\circledast V(\mathbf{x}^{\hspace{0.01in}\prime\prime})\circledast
(K_{L})^{\pm}(x^{\hspace{0.01in}\prime\prime}\hspace{-0.01in};x^{\hspace
{0.01in}\prime})
\end{align}
for an interaction potential with $\overline{V(\mathbf{x})}=V(\mathbf{x})$.
This result implies:%
\begin{equation}
\overline{(G_{R})^{\pm}(x;x^{\hspace{0.01in}\prime})}=(G_{L})^{\pm}%
(x^{\hspace{0.01in}\prime}\hspace{-0.01in};x). \label{KonGreGes1}%
\end{equation}
In the same way, we get:%
\begin{equation}
\overline{(G_{R}^{\hspace{0.01in}\ast})^{\pm}(x;x^{\hspace{0.01in}\prime}%
)}=(G_{L}^{\hspace{0.01in}\ast})^{\pm}(x^{\hspace{0.01in}\prime}%
\hspace{-0.01in};x). \label{KonGreGes2}%
\end{equation}

The $q$-de\-formed propagators for a free nonrelativistic particle can be
expanded in terms of plane waves [also see Eqs.~(\ref{IntKer1}) and
(\ref{IntKer2}) in Chap.~\ref{KapProSchrFel}]. Such expansions also exist for
the propagators of the $q$-de\-formed Schr\"{o}dinger equations with
interaction. To show this, we consider the expressions for the scattered waves
in Eqs.~(\ref{DefGreGesInh1}) and (\ref{DefGreGesInh2}). We replace the free
solutions in the expressions for the scattered waves with the plane waves
given in Eqs.~(\ref{ConPlaWav0}) and (\ref{ConPlaWav1}) of
Chap.~\ref{KapProSchrFel}. This way, we obtain%
\begin{align}
U_{\mathbf{p}}^{\pm}(\mathbf{x},t)  &  =\lim_{t^{\prime}\rightarrow
\hspace{0.01in}\mp\infty}\int\text{d}_{q}^{3}x^{\hspace{0.01in}\prime}%
\hspace{0.01in}(G_{R})^{\pm}(\mathbf{x},t;\mathbf{x}^{\hspace{0.01in}\prime
}\hspace{-0.01in},t^{\hspace{0.01in}\prime})\circledast u_{\hspace
{0.01in}\mathbf{p}}(\mathbf{x}^{\hspace{0.01in}\prime}\hspace{-0.01in}%
,t^{\hspace{0.01in}\prime}),\nonumber\\
(U^{\ast})_{\mathbf{p}}^{\pm}(\mathbf{x},t)  &  =\lim_{t^{\prime}%
\rightarrow\hspace{0.01in}\mp\infty}\int\text{d}_{q}^{3}x^{\hspace
{0.01in}\prime}\hspace{0.01in}(u^{\ast})_{\mathbf{p}}(\mathbf{x}%
^{\hspace{0.01in}\prime}\hspace{-0.01in},t^{\hspace{0.01in}\prime}%
)\circledast(G_{L}^{\hspace{0.01in}\ast})^{\pm}(\mathbf{x}^{\hspace
{0.01in}\prime}\hspace{-0.01in},t^{\hspace{0.01in}\prime}\hspace
{-0.01in};\mathbf{x},t), \label{DefGroUp}%
\end{align}
or%
\begin{align}
U_{\pm}^{\mathbf{p}}(\mathbf{x},t)  &  =\lim_{t^{\prime}\rightarrow
\hspace{0.01in}\mp\infty}\int\text{d}_{q}^{3}x^{\hspace{0.01in}\prime}%
\hspace{0.01in}u^{\mathbf{p}}(\mathbf{x}^{\hspace{0.01in}\prime}%
\hspace{-0.01in},t^{\hspace{0.01in}\prime})\circledast(G_{L})^{\pm}%
(\mathbf{x}^{\hspace{0.01in}\prime}\hspace{-0.01in},t^{\hspace{0.01in}\prime
}\hspace{-0.01in};\mathbf{x},t)\nonumber\\
(U^{\ast})_{\pm}^{\mathbf{p}}(\mathbf{x},t)  &  =\lim_{t^{\prime}%
\rightarrow\hspace{0.01in}\mp\infty}\int\text{d}_{q}^{3}x^{\hspace
{0.01in}\prime}\hspace{0.01in}(G_{R}^{\hspace{0.01in}\ast})^{\pm}%
(\mathbf{x},t;\mathbf{x}^{\hspace{0.01in}\prime}\hspace{-0.01in}%
,t^{\hspace{0.01in}\prime})\circledast(u^{\ast})^{\mathbf{p}}(\mathbf{x}%
^{\hspace{0.01in}\prime}\hspace{-0.01in},t^{\hspace{0.01in}\prime}).
\label{DefGroUSteP}%
\end{align}
These expressions for the scattered waves satisfy the orthonormality relations%
\begin{align}
\int\text{d}_{q}^{3}x\,(U^{\ast})_{\mathbf{p}^{\prime}}^{\pm}(\mathbf{x}%
,t)\circledast U_{\mathbf{p}}^{\pm}(\mathbf{x},t)  &  =\int\text{d}_{q}%
^{3}x\,(u^{\ast})_{\mathbf{p}^{\prime}}(\mathbf{x},t)\circledast
u_{\hspace{0.01in}\mathbf{p}}(\mathbf{x},t)\nonumber\\
&  =\operatorname*{vol}\nolimits^{-1}\hspace{-0.01in}\delta_{q}^{3}%
((\ominus\hspace{0.01in}\kappa^{-1}\mathbf{p}^{\prime})\oplus\mathbf{p}),
\label{OrtNorRelGroU1}%
\end{align}
or%
\begin{align}
\int\text{d}_{q}^{3}x\,U_{\pm}^{\mathbf{p}}(\mathbf{x},t)\circledast(U^{\ast
})_{\pm}^{\mathbf{p}^{\prime}}\hspace{-0.01in}(\mathbf{x},t)  &  =\int
\text{d}_{q}^{3}x\,u^{\mathbf{p}}(\mathbf{x},t)\circledast(u^{\ast
})^{\mathbf{p}^{\prime}}\hspace{-0.01in}(\mathbf{x},t)\nonumber\\
&  =\operatorname*{vol}\nolimits^{-1}\hspace{-0.01in}\delta_{q}^{3}%
(\hspace{0.01in}\mathbf{p}\oplus(\ominus\hspace{0.01in}\kappa^{1}%
\mathbf{p}^{\prime})). \label{OrtNorRelGroU2}%
\end{align}
Since the wave functions in Eqs.~(\ref{DefGroUp}) and (\ref{DefGroUSteP})
arise from the scattering of plane waves, we can prove the above
orthonormality relations with a calculation similar to that in
Eq.~(\ref{ErhNorGreBew}) of the next chapter.

We show that the wave functions in Eq.~(\ref{DefGroUp}) or
Eq.~(\ref{DefGroUSteP}) also form a complete system of functions. The general
solutions to the $q$-de\-formed Schr\"{o}dinger equations for a free
nonrelativistic particle can be written as expansions in terms of
$q$-de\-formed plane waves. For example, we have\footnote{See
Eq.~(\ref{FreSolExp}) in Chap.~\ref{KapProSchrFel}.}%
\begin{equation}
\phi_{R}(\mathbf{x},t)=\int\text{d}_{q}^{3}p\,u_{\hspace{0.01in}\mathbf{p}%
}(\mathbf{x},t)\circledast c_{\hspace{0.01in}\mathbf{p}}.
\label{FouEntFreSchGleWdhCp}%
\end{equation}
Plugging this expansion into the first identity of Eq.~(\ref{DefGreGesInh1})
and taking into account the definition of $U_{\mathbf{p}}^{(\pm)}$\ [see
Eq.~(\ref{DefGroUp})], we find:%
\begin{align}
(\psi_{R})^{\pm}(x)  &  =\lim_{t^{\prime}\rightarrow\hspace{0.01in}\mp\infty
}\int\text{d}_{q}^{3}x^{\prime}\hspace{0.01in}(G_{R})^{\pm}(x;x^{\hspace
{0.01in}\prime})\circledast\phi_{R}(x^{\hspace{0.01in}\prime})=\nonumber\\
&  =\lim_{t^{\prime}\rightarrow\hspace{0.01in}\mp\infty}\int\text{d}_{q}%
^{3}x^{\prime}\hspace{0.01in}(G_{R})^{\pm}(x;x^{\hspace{0.01in}\prime
})\circledast\hspace{-0.01in}\int\text{d}_{q}^{3}p\,u_{\hspace{0.01in}%
\mathbf{p}}(x^{\hspace{0.01in}\prime})\circledast c_{\hspace{0.01in}%
\mathbf{p}}\nonumber\\
&  =\int\text{d}_{q}^{3}p\lim_{t^{\prime}\rightarrow\hspace{0.01in}\mp\infty
}\int\text{d}_{q}^{3}x^{\hspace{0.01in}\prime}\hspace{0.01in}(G_{R})^{\pm
}(x;x^{\hspace{0.01in}\prime})\circledast u_{\hspace{0.01in}\mathbf{p}%
}(x^{\hspace{0.01in}\prime})\circledast c_{\hspace{0.01in}\mathbf{p}%
}\nonumber\\
&  =\int\text{d}_{q}^{3}p\,U_{\mathbf{p}}^{\pm}(x)\circledast c_{\hspace
{0.01in}\mathbf{p}}. \label{EntStrWelL1}%
\end{align}
In the same way, we get the expansions%
\begin{equation}
(\psi_{R}^{\ast})^{\pm}(x)=\int\text{d}_{q}^{3}p\,(U^{\ast})_{\pm}%
^{\mathbf{p}}(x)\circledast(c^{\ast})^{\mathbf{p}}, \label{EntStrWelLSte}%
\end{equation}
and%
\begin{align}
(\psi_{L})^{\pm}(\mathbf{x},t)  &  =\int\text{d}_{q}^{3}p\,c^{\hspace
{0.01in}\mathbf{p}}\hspace{-0.01in}\circledast U_{\pm}^{\mathbf{p}}%
(\mathbf{x},t),\nonumber\\
(\psi_{L}^{\ast})^{\pm}(\mathbf{x},t)  &  =\int\text{d}_{q}^{3}p\,(c^{\ast
})_{\mathbf{p}}\circledast(U^{\ast})_{\mathbf{p}}^{\pm}(\mathbf{x},t)
\label{EntStrWelRSte}%
\end{align}
if we have%
\begin{equation}
\phi_{R}^{\ast}(\mathbf{x},t)=\int\text{d}_{q}^{3}p\,(u^{\ast})^{\mathbf{p}%
}(\mathbf{x},t)\circledast(c^{\ast})^{\mathbf{p}},
\label{FouEntFreSchGleWdhLSte}%
\end{equation}
and%
\begin{align}
\phi_{L}(\mathbf{x},t)  &  =\int\text{d}_{q}^{3}p\,c^{\hspace{0.01in}%
\mathbf{p}}\hspace{-0.01in}\circledast u^{\mathbf{p}}(\mathbf{x}%
,t),\nonumber\\
\phi_{L}^{\ast}(\mathbf{x},t)  &  =\int\text{d}_{q}^{3}p\,(c^{\ast
})_{\mathbf{p}}\circledast(u^{\ast})_{\mathbf{p}}(\mathbf{x},t).
\label{FouEntFreSchGleWdh3}%
\end{align}
We calculate the coefficients in the above expansions using the orthonormality
relations in Eqs.~(\ref{OrtNorRelGroU1}) and (\ref{OrtNorRelGroU2}). For
example, we have:%
\begin{align}
\int\text{d}_{q}^{3}x\,(U^{\ast})_{\mathbf{p}}^{\pm}(\mathbf{x},t)\circledast
(\psi_{R})^{\pm}(\mathbf{x},t)  &  =\int\text{d}_{q}^{3}p^{\prime}%
\hspace{-0.02in}\int\text{d}_{q}^{3}x\,(U^{\ast})_{\mathbf{p}}^{\pm
}(\mathbf{x},t)\circledast U_{\mathbf{p}^{\prime}}^{\pm}(\mathbf{x}%
,t)\circledast c_{\hspace{0.01in}\mathbf{p}^{\prime}}\nonumber\\
&  =\int\text{d}_{q}^{3}p^{\prime}\hspace{0.01in}\operatorname*{vol}%
\nolimits^{-1}\hspace{-0.01in}\delta_{q}^{3}((\ominus\hspace{0.01in}%
\kappa^{-1}\mathbf{p})\oplus\mathbf{p}^{\prime})\circledast c_{\hspace
{0.01in}\mathbf{p}^{\prime}}\nonumber\\
&  =c\hspace{0.01in}_{\mathbf{p}}. \label{BerCpEntPro}%
\end{align}
By similar reasoning, we find%
\begin{equation}
\int\text{d}_{q}^{3}x\,(\psi_{L})^{\pm}(\mathbf{x},t)\circledast(U^{\ast
})_{\pm}^{\mathbf{p}}(\mathbf{x},t)=c^{\hspace{0.01in}\mathbf{p}},
\end{equation}
and%
\begin{align}
\int\text{d}_{q}^{3}x\,U_{\pm}^{\mathbf{p}}(\mathbf{x},t)\circledast(\psi
_{R}^{\ast})^{\pm}(\mathbf{x},t)  &  =(c^{\ast})^{\mathbf{p}},\nonumber\\
\int\text{d}_{q}^{3}x\,(\psi_{L}^{\ast})^{\pm}(\mathbf{x},t)\circledast
U_{\mathbf{p}}^{\pm}(\mathbf{x},t)  &  =(c^{\ast})_{\mathbf{p}}.
\end{align}

Using the scattered wave functions in Eq.~(\ref{DefGroUp}) or
Eq.~(\ref{DefGroUSteP}), we can write down expressions for the Green's
functions to Schr\"{o}dinger equations with an interaction. To this end, we
plug the integral expression for $c_{\hspace{0.01in}\mathbf{p}}$ given in
Eq.~(\ref{BerCpEntPro}) into the expansion in Eq.~(\ref{EntStrWelL1}):%
\begin{align}
(\psi_{R})^{\pm}(\mathbf{x},t)  &  =\int\text{d}_{q}^{3}p\,U_{\mathbf{p}}%
^{\pm}(\mathbf{x},t)\circledast c_{\hspace{0.01in}\mathbf{p}}\nonumber\\
&  =\int\text{d}_{q}^{3}x^{\prime}\hspace{-0.02in}\int\text{d}_{q}%
^{3}p\,U_{\mathbf{p}}^{\pm}(\mathbf{x},t)\circledast(U^{\ast})_{\mathbf{p}%
}^{\pm}(\mathbf{x}^{\hspace{0.01in}\prime}\hspace{-0.01in},t^{\hspace
{0.01in}\prime})\circledast(\psi_{R})^{\pm}(\mathbf{x}^{\hspace{0.01in}\prime
}\hspace{-0.01in},t^{\hspace{0.01in}\prime}).
\end{align}
By comparing this result with the first identity in
Eq.~(\ref{NeuDefGreGesTheNicRel1}), we find%
\begin{equation}
(G_{R})^{\pm}(\mathbf{x},t;\mathbf{x}^{\hspace{0.01in}\prime}\hspace
{-0.01in},t^{\hspace{0.01in}\prime})=\int\text{d}_{q}^{3}p\,U_{\mathbf{p}%
}^{\pm}(\mathbf{x},t)\circledast(U^{\ast})_{\mathbf{p}}^{\pm}(\mathbf{x}%
^{\hspace{0.01in}\prime}\hspace{-0.01in},t^{\hspace{0.01in}\prime}),
\end{equation}
where $t>t^{\hspace{0.01in}\prime}$ for $(G_{R})^{+}$ and $t<t^{\hspace
{0.01in}\prime}$ for $(G_{R})^{-}$. By similar reasoning, we find%
\begin{equation}
(G_{R}^{\ast})^{\pm}(\mathbf{x},t;\mathbf{x}^{\hspace{0.01in}\prime
},t^{\hspace{0.01in}\prime})=\int\text{d}_{q}^{3}p\,(U^{\ast})_{\pm
}^{\mathbf{p}}(\mathbf{x},t)\circledast U_{\pm}^{\mathbf{p}}(\mathbf{x}%
^{\hspace{0.01in}\prime}\hspace{-0.01in},t^{\hspace{0.01in}\prime}),
\end{equation}
and%
\begin{align}
(G_{L})^{\pm}(\mathbf{x}^{\hspace{0.01in}\prime}\hspace{-0.01in}%
,t^{\hspace{0.01in}\prime}\hspace{-0.01in};\mathbf{x},t)  &  =\int\text{d}%
_{q}^{3}p\,(U^{\ast})_{\pm}^{\mathbf{p}}(\mathbf{x}^{\hspace{0.01in}\prime
}\hspace{-0.01in},t^{\hspace{0.01in}\prime})\circledast U_{\pm}^{\mathbf{p}%
}(\mathbf{x},t),\nonumber\\
(G_{L}^{\ast})^{\pm}(\mathbf{x}^{\hspace{0.01in}\prime}\hspace{-0.01in}%
,t^{\hspace{0.01in}\prime}\hspace{-0.01in};\mathbf{x},t)  &  =\int\text{d}%
_{q}^{3}p\,U_{\mathbf{p}}^{\pm}(\mathbf{x}^{\hspace{0.01in}\prime}%
\hspace{-0.01in},t^{\hspace{0.01in}\prime})\circledast(U^{\ast})_{\mathbf{p}%
}^{\pm}(\mathbf{x},t),
\end{align}
where $t>t^{\hspace{0.01in}\prime}$ for the retarded Green's functions and
$t<t^{\hspace{0.01in}\prime}$ for the advanced Green's functions.

\section{Scattering matrices\label{KapStrMatN}}

\subsection{Definition\label{KapSMatDef}}

An initially free particle can change its state by scattering. The
\textit{S-ma\-trix} relates the initial state and the final state of the
particle undergoing a scattering process.

The wave function $(\psi_{R})^{\pm}$ represents the state which emerges from a
free state with wave function $\phi_{R}$ by scattering. If the wave function
$\phi_{L}^{\ast}$ describes the free particle state after scattering, we have
the following scattering amplitude \cite{Bjorken1Engl}:%
\begin{align}
(S_{R})^{\pm}(\phi,\psi)  &  =\lim_{t\hspace{0.01in}\rightarrow\hspace
{0.01in}\pm\infty}\int\text{d}_{q}^{3}x\,\phi_{L}^{\ast}(x)\circledast
(\psi_{R})^{\pm}(x)\nonumber\\[0.08in]
&  =\lim_{\substack{t\hspace{0.01in}\rightarrow\pm\hspace{0.01in}%
\infty\\t^{\prime}\rightarrow\hspace{0.01in}\mp\infty}}\int\text{d}_{q}%
^{3}x\hspace{-0.01in}\int\text{d}_{q}^{3}x^{\hspace{0.01in}\prime}%
\hspace{0.01in}\phi_{L}^{\ast}(x)\circledast(G_{R})^{\pm}(x;x^{\hspace
{0.01in}\prime})\circledast\phi_{R}(x^{\hspace{0.01in}\prime}).
\label{SMatDef1}%
\end{align}
Note that the second identity follows from the first formula in
Eq.~(\ref{DefGreGesInh1}) of the previous chapter. Eq.~(\ref{SMatDef1}) shows
that the elements of the S-ma\textbf{\-}trix $(S_{R})^{\pm}$ are identical to
the matrix elements of the Green's function $(G_{R})^{\pm}$. The same applies
to the other versions of Green's functions, i.~e.%
\begin{align}
(S_{L})^{\pm}(\psi,\phi)  &  =\lim_{t\hspace{0.01in}\rightarrow\hspace
{0.01in}\pm\infty}\int\text{d}_{q}^{3}x\,(\psi_{L})^{\pm}(x)\circledast
\phi_{R}^{\ast}(x)\nonumber\\
&  =\lim_{\substack{t\hspace{0.01in}\rightarrow\hspace{0.01in}\pm
\infty\\t^{\prime}\rightarrow\hspace{0.01in}\mp\infty}}\int\text{d}_{q}%
^{3}x\hspace{-0.01in}\int\text{d}_{q}^{3}x^{\hspace{0.01in}\prime}%
\hspace{0.01in}\phi_{L}(x^{\hspace{0.01in}\prime})\circledast(G_{L})^{\pm
}(x^{\hspace{0.01in}\prime}\hspace{-0.01in};x)\circledast\phi_{R}^{\ast
}(x),\\[0.1in]
(S_{R}^{\ast})^{\pm}(\phi,\psi)  &  =\lim_{t\hspace{0.01in}\rightarrow
\hspace{0.01in}\pm\infty}\int\text{d}_{q}^{3}x\,\phi_{L}(x)\circledast
(\psi_{R}^{\ast})^{\pm}(x)\nonumber\\
&  =\lim_{\substack{t\hspace{0.01in}\rightarrow\hspace{0.01in}\pm
\infty\\t^{\prime}\rightarrow\mp\infty}}\int\text{d}_{q}^{3}x\hspace
{-0.01in}\int\text{d}_{q}^{3}x^{\hspace{0.01in}\prime}\hspace{0.01in}\phi
_{L}(x)\circledast(G_{R}^{\ast})^{\pm}(x;x^{\hspace{0.01in}\prime}%
)\circledast\phi_{R}^{\ast}(x^{\hspace{0.01in}\prime}), \label{SMatDef3}%
\end{align}
and%
\begin{align}
(S_{L}^{\ast})^{\pm}(\psi,\phi)  &  =\lim_{t\hspace{0.01in}\rightarrow
\hspace{0.01in}\pm\infty}\int\text{d}_{q}^{3}x\,(\psi_{L}^{\ast})^{\pm
}(x)\circledast\phi_{R}(x)\nonumber\\
&  =\lim_{\substack{t\hspace{0.01in}\rightarrow\hspace{0.01in}\pm
\infty\\t^{\prime}\rightarrow\hspace{0.01in}\mp\infty}}\int\text{d}_{q}%
^{3}x\hspace{-0.01in}\int\text{d}_{q}^{3}x^{\hspace{0.01in}\prime}%
\hspace{0.01in}\phi_{L}^{\ast}(x^{\hspace{0.01in}\prime})\circledast
(G_{L}^{\ast})^{\pm}(x^{\hspace{0.01in}\prime}\hspace{-0.01in};x)\circledast
\phi_{R}(x). \label{SMatDef4}%
\end{align}
If we compare the different versions of S-ma\textbf{\-}trix elements and take
into account Eq.~(\ref{ZusGreAvaRet}) from the previous chapter, we can find
the following identities:%
\begin{equation}
(S_{R})^{\pm}(\phi,\psi)=(S_{L}^{\ast})^{\mp}(\psi,\phi),\qquad(S_{L})^{\pm
}(\psi,\phi)=(S_{R}^{\ast})^{\mp}(\phi,\psi).
\end{equation}

Next, we study the conjugation properties of the S-ma\textbf{\-}trix elements.
Conjugating the expressions in Eqs.~(\ref{SMatDef1})-(\ref{SMatDef4}) and
taking into account Eq.~(\ref{KonEigSchWelWdh}) of Chap.~\ref{KapProSchrFel}
as well as Eqs.~(\ref{KonGreGes1}) and (\ref{KonGreGes2}) of the previous
chapter, we find:%
\begin{equation}
\overline{(S_{R})^{\pm}(\phi,\psi)}=(S_{L})^{\pm}(\psi,\phi),\qquad
\overline{(S_{R}^{\ast})^{\pm}(\phi,\psi)}=(S_{L}^{\ast})^{\pm}(\psi,\phi).
\end{equation}

The conjugation properties of the free wave functions carry over to the
scattered wave functions. Using the series expansions in
Eqs.~(\ref{LipSchGlePsiL}) and (\ref{LipSchGlePsiRSte}) of the previous
chapter and the conjugation properties of the free propagators [see
Eq.~(\ref{KonFreProNR}) of Chap.~\ref{KapProSchrFel}], we can verify the
following identities:%
\begin{equation}
\overline{\psi_{L}(\mathbf{x},t)}=\psi_{R}(\mathbf{x},t),\qquad\overline
{\psi_{L}^{\ast}(\mathbf{x},t)}=\psi_{R}^{\ast}(\mathbf{x},t).
\end{equation}

\subsection{Momentum representation \label{KapMomRepSma}}

To get S-ma\-trix elements in a basis of momentum eigenfunctions, we replace
the solutions to the free Schr\"{o}dinger equations with their expansions in
terms of plane waves [cf. Eqs.~(\ref{FouEntFreSchGleWdhCp}),
(\ref{FouEntFreSchGleWdhLSte}), and (\ref{FouEntFreSchGleWdh3}) of the previous 
chapter]. This way, we have%
\begin{align}
(S_{R})^{\pm}(\phi,\psi)  &  =\lim_{\substack{t\hspace{0.01in}\rightarrow
\hspace{0.01in}\pm\infty\\t^{\prime}\rightarrow\hspace{0.01in}\mp\infty}%
}\int\text{d}_{q}^{3}x\hspace{-0.01in}\int\text{d}_{q}^{3}x^{\hspace
{0.01in}\prime}\hspace{0.01in}\phi_{L}^{\ast}(x)\circledast(G_{R})^{\pm
}(x;x^{\hspace{0.01in}\prime})\circledast\phi_{R}(x^{\hspace{0.01in}\prime
})\nonumber\\
&  =\int\text{d}_{q}^{3}p\,\text{d}_{q}^{3}p^{\prime}\hspace{0.01in}(c^{\ast
})_{\mathbf{p}}\circledast(S_{R})_{\mathbf{p\hspace{0.01in}p}^{\prime}}^{\pm
}\circledast c_{\hspace{0.01in}\mathbf{p}^{\prime}},
\end{align}
where%
\begin{equation}
(S_{R})_{\mathbf{p\hspace{0.01in}p}^{\prime}}^{\pm}=\lim_{\substack{t\hspace
{0.01in}\rightarrow\hspace{0.01in}\pm\infty\\t^{\prime}\rightarrow
\hspace{0.01in}\mp\infty}}\int\text{d}_{q}^{3}x\,\text{d}_{q}^{3}%
x^{\hspace{0.01in}\prime}\hspace{0.01in}(u^{\ast})_{\mathbf{p}}(x)\circledast
(G_{R})^{\pm}(x;x^{\hspace{0.01in}\prime})\circledast u_{\hspace
{0.01in}\mathbf{p}^{\prime}}(x^{\hspace{0.01in}\prime}). \label{SMatImpDar1}%
\end{equation}
If we apply these considerations to the other versions of the S-ma\textbf{\-}%
trix, we also get%
\begin{equation}
(S_{L})^{\pm}(\psi,\phi)=\int\text{d}_{q}^{3}p^{\prime}\hspace{0.01in}%
\text{d}_{q}^{3}p\,c^{\hspace{0.01in}\mathbf{p}^{\prime}}\hspace
{-0.02in}\circledast(S_{L})_{\mathbf{p}^{\prime}\mathbf{p}}^{\pm}%
\circledast(c^{\ast})^{\mathbf{p}}%
\end{equation}
with%
\begin{equation}
(S_{L})_{\mathbf{p}^{\prime}\mathbf{p}}^{\pm}=\lim_{\substack{t\mathbf{\hspace
{0.01in}}\rightarrow\mathbf{\hspace{0.01in}}\pm\infty\\t^{\prime}%
\rightarrow\mathbf{\hspace{0.01in}}\mp\infty}}\int\text{d}_{q}^{3}%
x^{\mathbf{\hspace{0.01in}}\prime}\hspace{0.01in}\text{d}_{q}^{3}%
x\,u^{\mathbf{p}^{\prime}}\hspace{-0.02in}(x^{\mathbf{\hspace{0.01in}}\prime
})\circledast(G_{L})^{\pm}(x^{\mathbf{\hspace{0.01in}}\prime}\hspace
{-0.01in};x)\circledast(u^{\ast})^{\mathbf{p}}(x),
\end{equation}
and%
\begin{align}
(S_{R}^{\ast})^{\pm}(\phi,\psi)  &  =\int\text{d}_{q}^{3}p\,\text{d}_{q}%
^{3}p^{\prime}\hspace{0.01in}c^{\hspace{0.01in}\mathbf{p}}\circledast
(S_{R}^{\ast})_{\mathbf{p\hspace{0.01in}p}^{\prime}}^{\pm}\hspace
{-0.01in}\circledast(c^{\ast})^{\mathbf{p}^{\prime}},\nonumber\\
(S_{L}^{\ast})^{\pm}(\psi,\phi)  &  =\int\text{d}_{q}^{3}p^{\prime}%
\hspace{0.01in}\text{d}_{q}^{3}p\,(c^{\ast})_{\mathbf{p}^{\prime}}%
\circledast(S_{L}^{\ast})_{\mathbf{p}^{\prime}\mathbf{p}}^{\pm}\circledast
c_{\hspace{0.01in}\mathbf{p}}%
\end{align}
with%
\begin{align}
(S_{R}^{\ast})_{\mathbf{p\hspace{0.01in}p}^{\prime}}^{\pm}  &  =\lim
_{\substack{t\hspace{0.01in}\rightarrow\hspace{0.01in}\pm\infty\\t^{\prime
}\rightarrow\hspace{0.01in}\mp\infty}}\int\text{d}_{q}^{3}x\,\text{d}_{q}%
^{3}x^{\hspace{0.01in}\prime}\hspace{0.01in}u^{\mathbf{p}}(x)\circledast
(G_{R}^{\hspace{0.01in}\ast})^{\pm}(x;x^{\hspace{0.01in}\prime})\circledast
(u^{\ast})^{\mathbf{p}^{\prime}}(x^{\hspace{0.01in}\prime}),\nonumber\\
(S_{L}^{\ast})_{\mathbf{p}^{\prime}\mathbf{p}}^{\pm}  &  =\lim
_{\substack{t\hspace{0.01in}\rightarrow\hspace{0.01in}\pm\infty\\t^{\prime
}\rightarrow\hspace{0.01in}\mp\infty}}\int\text{d}_{q}^{3}x^{\hspace
{0.01in}\prime}\hspace{0.01in}\text{d}_{q}^{3}x\,(u^{\ast})_{\mathbf{p}%
^{\prime}}(x^{\hspace{0.01in}\prime})\circledast(G_{L}^{\hspace{0.01in}\ast
})^{\pm}(x^{\hspace{0.01in}\prime}\hspace{-0.01in};x)\circledast
u_{\hspace{0.01in}\mathbf{p}}(x). \label{SMatImpDar4}%
\end{align}

Comparing the expressions for the different types of S-ma\textbf{\-}trix
elements and considerung Eq.~(\ref{ZusGreAvaRet}) of Chap.~\ref{ProWecKap}, we
find the following identities:%
\begin{equation}
(S_{R})_{\mathbf{p\hspace{0.01in}p}^{\prime}}^{\pm}=(S_{L}^{\ast
})_{\mathbf{p\hspace{0.01in}p}^{\prime}}^{\mp},\qquad(S_{L})_{\mathbf{p\hspace
{0.01in}p}^{\prime}}^{\pm}=(S_{R}^{\ast})_{\mathbf{p\hspace{0.01in}p}^{\prime
}}^{\mp}. \label{IdeSMatLR}%
\end{equation}
The conjugation properties of momentum eigenfunctions and Green's functions
[see Eq.~(\ref{KonEbeWel}) of Chap.~\ref{KapProSchrFel} and
Eqs.~(\ref{KonGreGes1}) and (\ref{KonGreGes2}) of Chap.~\ref{ProWecKap}] imply
the following conjugation properties for the S-ma\textbf{\-}trix elements in
momentum space:%
\begin{equation}
\overline{(S_{R})_{\mathbf{p\hspace{0.01in}p}^{\prime}}^{\pm}}=(S_{L}%
)_{\mathbf{p}^{\prime}\mathbf{p}}^{\pm},\qquad\overline{(S_{L}^{\ast
})_{\mathbf{p\hspace{0.01in}p}^{\prime}}^{\pm}}=(S_{R}^{\ast})_{\mathbf{p}%
^{\prime}\mathbf{p}}^{\pm}. \label{KonEigSMatImp}%
\end{equation}

A look at Eq.~(\ref{SMatImpDar1}) shows that the Green's function
$(G_{R})^{\pm}$ determines the S-ma\textbf{\-}trix $(S_{R})^{\pm}$. If we
represent $(G_{R})^{\pm}$ by the Born series in Eq.~(\ref{BorReiGL}) of
Chap.~\ref{ProWecKap}, we obtain the following perturbation expansion for the
S-ma\textbf{\-}trix $(S_{R})^{\pm}$ in momentum space:%
\begin{align}
(S_{R})_{\mathbf{p\hspace{0.01in}p}^{\prime}}^{\pm}=  &  \,\operatorname*{vol}%
\nolimits^{-1}\hspace{-0.01in}\delta_{q}^{3}((\ominus\hspace{0.01in}%
\kappa^{-1}\mathbf{p})\oplus\mathbf{p}^{\prime})+\text{i}^{\mp1}%
\hspace{-0.02in}\int\text{d}_{q}^{3}x\int\text{d}t\,(u^{\ast})_{\mathbf{p}%
}(x)\circledast V(\mathbf{x})\circledast u_{\hspace{0.01in}\mathbf{p}^{\prime
}}(x)\nonumber\\
&  \,+\text{i}^{\mp\hspace{0.01in}2}\hspace{-0.02in}\int\text{d}_{q}%
^{3}x\,\text{d}_{q}^{3}x^{\hspace{0.01in}\prime}\hspace{-0.02in}\int
\text{d}t\,\text{d}t^{\hspace{0.01in}\prime}\hspace{0.01in}(u^{\ast
})_{\mathbf{p}}(x)\circledast V(\mathbf{x})\nonumber\\
&  \qquad\qquad\circledast(K_{R})^{\pm}(x;x^{\hspace{0.01in}\prime
})\circledast V(\mathbf{x}^{\hspace{0.01in}\prime})\circledast u_{\hspace
{0.01in}\mathbf{p}^{\prime}}(x^{\hspace{0.01in}\prime})+\ldots
\label{StoEntSMatRImp}%
\end{align}
In deriving this formula, we have made use of the following identities:%
\begin{align}
u_{\hspace{0.01in}\mathbf{p}}(\mathbf{x},t)  &  =\lim_{t^{\prime}%
\rightarrow\hspace{0.01in}\mp\infty}\int\text{d}_{q}^{3}x^{\hspace
{0.01in}\prime}\hspace{0.01in}(K_{R})^{\pm}(x;x^{\hspace{0.01in}\prime
})\circledast u_{\hspace{0.01in}\mathbf{p}}(x^{\hspace{0.01in}\prime
}),\nonumber\\
(u^{\ast})_{\mathbf{p}}(\mathbf{x}^{\hspace{0.01in}\prime},t^{\hspace
{0.01in}\prime})  &  =\lim_{t\hspace{0.01in}\rightarrow\hspace{0.01in}%
\pm\infty}\int\text{d}_{q}^{3}x\,(u^{\ast})_{\mathbf{p}}(x)\circledast
(K_{R})^{\pm}(x;x^{\hspace{0.01in}\prime}).
\end{align}
Moreover, we have taken into account the completeness of $q$-de\textbf{\-}%
formed plane waves [see Eq.~(\ref{VolRelZeiWelDreDim1}) in
Chap.~\ref{KapProSchrFel}]. There are similar expansions for the other
$q$-ver\textbf{\-}sions of the S-ma\textbf{\-}trix. For example, if we use the
first identity in Eq.~(\ref{IdeSMatLR}) together with $(K_{L}^{\hspace
{0.01in}\ast})^{\pm}=(K_{R})^{\pm}$, we find:%
\begin{align}
(S_{L}^{\ast})_{\mathbf{p\hspace{0.01in}p}^{\prime}}^{\pm}=  &
\,\operatorname*{vol}\nolimits^{-1}\hspace{-0.01in}\delta_{q}^{3}%
((\ominus\hspace{0.01in}\kappa^{-1}\mathbf{p})\oplus\mathbf{p}^{\prime
})+\text{i}^{\pm1}\hspace{-0.02in}\int\text{d}_{q}^{3}x\int\text{d}%
t\,(u^{\ast})_{\mathbf{p}}(x)\circledast V(\mathbf{x})\circledast
u_{\hspace{0.01in}\mathbf{p}^{\prime}}(x)\nonumber\\
&  \,+\text{i}^{\pm\hspace{0.01in}2}\hspace{-0.02in}\int\text{d}_{q}%
^{3}x\,\text{d}_{q}^{3}x^{\hspace{0.01in}\prime}\hspace{-0.02in}\int
\text{d}t\,\text{d}t^{\hspace{0.01in}\prime}\hspace{0.01in}(u^{\ast
})_{\mathbf{p}}(x)\circledast V(\mathbf{x})\nonumber\\
&  \qquad\qquad\circledast(K_{L}^{\hspace{0.01in}\ast})^{\pm}(x;x^{\hspace
{0.01in}\prime})\circledast V(\mathbf{x}^{\hspace{0.01in}\prime})\circledast
u_{\hspace{0.01in}\mathbf{p}^{\prime}}(x^{\hspace{0.01in}\prime})+\ldots
\end{align}

\subsection{Unitarity and conservation of probability\label{KapUniCon}}

We clarify in which way the $q$-de\-formed S-ma\-trices are unitary. Using the
results in Eqs.~(\ref{SMatImpDar1}) and (\ref{SMatImpDar4}), we can calculate
the following product of two S-ma\-trices:%
\begin{gather}
\int\text{d}_{q}^{3}p^{\prime\prime}\,(S_{L}^{\ast})_{\mathbf{p}^{\prime
}\mathbf{p}^{\prime\prime}}^{\pm}\hspace{-0.01in}\circledast(S_{R}%
)_{\mathbf{p}^{\prime\prime}\mathbf{p}}^{\pm}=\nonumber\\[0.04in]
=\lim_{\substack{t,t^{\prime}\rightarrow\hspace{0.01in}\mp\infty
\\t^{\prime\prime}\hspace{-0.01in},\hspace{0.01in}t^{\prime\prime\prime
}\rightarrow\hspace{0.01in}\pm\infty}}\int\text{d}_{q}^{3}p^{\prime\prime
}\hspace{-0.02in}\int\text{d}_{q}^{3}x^{\prime}\hspace{0.01in}\text{d}_{q}%
^{3}x^{\hspace{0.01in}\prime\prime}(u^{\ast})_{\mathbf{p}^{\prime}}(x^{\prime
})\circledast(G_{L}^{\hspace{0.01in}\ast})^{\pm}(x^{\hspace{0.01in}\prime
}\hspace{-0.01in};x^{\hspace{0.01in}\prime\prime})\circledast u_{\hspace
{0.01in}\mathbf{p}^{\prime\prime}}(x^{\hspace{0.01in}\prime\prime})\nonumber\\
\circledast\int\text{d}_{q}^{3}x^{\hspace{0.01in}\prime\prime\prime}%
\hspace{0.01in}\text{d}_{q}^{3}x\,(u^{\ast})_{\mathbf{p}^{\prime\prime}%
}(x^{\hspace{0.01in}\prime\prime\prime})\circledast(G_{R})^{\pm}%
(x^{\hspace{0.01in}\prime\prime\prime}\hspace{-0.01in};x)\circledast
u_{\hspace{0.01in}\mathbf{p}}(x). \label{SMatProZwi1}%
\end{gather}
With the identities%
\begin{gather}
\lim_{t^{\prime\prime}\hspace{-0.01in},\hspace{0.01in}t^{\prime\prime\prime
}\rightarrow\hspace{0.01in}\pm\infty}\int\text{d}_{q}^{3}p^{\prime\prime
}\hspace{0.01in}u_{\hspace{0.01in}\mathbf{p}^{\prime\prime}}(\mathbf{x}%
^{\hspace{0.01in}\prime\prime}\hspace{-0.01in},t^{\hspace{0.01in}\prime\prime
})\circledast(u^{\ast})_{\mathbf{p}^{\prime\prime}}(\mathbf{x}^{\hspace
{0.01in}\prime\prime\prime}\hspace{-0.01in},t^{\hspace{0.01in}\prime
\prime\prime})=\nonumber\\
=\int\text{d}_{q}^{3}p^{\prime\prime}\hspace{0.01in}u_{\hspace{0.01in}%
\mathbf{p}^{\prime\prime}}(\mathbf{x}^{\hspace{0.01in}\prime\prime}%
\hspace{-0.01in},t^{\hspace{0.01in}\prime\prime})\circledast(u^{\ast
})_{\mathbf{p}^{\prime\prime}}(\mathbf{x}^{\hspace{0.01in}\prime\prime\prime
}\hspace{-0.01in},t^{\hspace{0.01in}\prime\prime})=\operatorname*{vol}%
\nolimits^{-1}\hspace{-0.01in}\delta_{q}^{3}(\mathbf{x}^{\hspace{0.01in}%
\prime\prime}\oplus(\ominus\hspace{0.01in}\kappa^{-1}\mathbf{x}^{\hspace
{0.01in}\prime\prime\prime})),
\end{gather}
we can carry out the integral with respect to $x^{\hspace{0.01in}\prime
\prime\prime}$ on the right-hand side of Eq.~(\ref{SMatProZwi1}):%
\begin{gather}
\int\text{d}_{q}^{3}p^{\prime\prime}\hspace{0.01in}(S_{L}^{\ast}%
)_{\mathbf{p}^{\prime}\mathbf{p}^{\prime\prime}}^{\pm}\hspace{-0.01in}%
\circledast(S_{R})_{\mathbf{p}^{\prime\prime}\mathbf{p}}^{\pm}=\nonumber\\
=\lim_{t,t^{\prime}\rightarrow\hspace{0.01in}\mp\infty}\hspace{0.01in}%
\lim_{t^{\prime\prime}\rightarrow\hspace{0.01in}\pm\infty}\int\text{d}_{q}%
^{3}x^{\hspace{0.01in}\prime}\hspace{0.01in}\text{d}_{q}^{3}x^{\hspace
{0.01in}\prime\prime}\text{d}_{q}^{3}x\,(u^{\ast})_{\mathbf{p}^{\prime}%
}(x^{\hspace{0.01in}\prime})\circledast(G_{L}^{\hspace{0.01in}\ast})^{\pm
}(x^{\hspace{0.01in}\prime}\hspace{-0.01in};x^{\hspace{0.01in}\prime\prime
})\nonumber\\
\circledast(G_{R})^{\pm}(x^{\hspace{0.01in}\prime\prime}\hspace{-0.01in}%
;x)\circledast u_{\hspace{0.01in}\mathbf{p}}(x). \label{ZwiRecMatProS1}%
\end{gather}
Due to Eqs.~(\ref{KomGreDel1}) and (\ref{ZusGreAvaRet}) of
Chap.~\ref{ProWecKap}, we have:%
\begin{equation}
\int\text{d}_{q}^{3}x^{\hspace{0.01in}\prime\prime}\hspace{0.01in}%
(G_{L}^{\hspace{0.01in}\ast})^{\pm}(x^{\hspace{0.01in}\prime}\hspace
{-0.01in};x^{\hspace{0.01in}\prime\prime})\circledast(G_{R})^{\pm}%
(x^{\hspace{0.01in}\prime\prime}\hspace{-0.01in};x)=\operatorname*{vol}%
\nolimits^{-1}\hspace{-0.01in}\delta_{q}^{3}(\mathbf{x}^{\hspace{0.01in}%
\prime}\oplus(\ominus\hspace{0.01in}\kappa^{-1}\mathbf{x})).
\label{KomIdeGreRSteGL}%
\end{equation}
If we apply this identity to the right-hand side of Eq.~(\ref{ZwiRecMatProS1}%
), we can also carry out the integral with respect to $x^{\hspace
{0.01in}\prime}$:%
\begin{align}
&  \int\text{d}_{q}^{3}p^{\prime\prime}\hspace{0.01in}(S_{L}^{\ast
})_{\mathbf{p}^{\prime}\mathbf{p}^{\prime\prime}}^{\pm}\hspace{-0.01in}%
\circledast(S_{R})_{\mathbf{p}^{\prime\prime}\mathbf{p}}^{\pm}=\nonumber\\
&  \qquad=\lim_{t,t^{\prime}\rightarrow\hspace{0.01in}\mp\infty}\int
\text{d}_{q}^{3}x^{\hspace{0.01in}\prime}\hspace{0.01in}\text{d}_{q}%
^{3}x\,(u^{\ast})_{\mathbf{p}^{\prime}}(x^{\hspace{0.01in}\prime}%
)\circledast\operatorname*{vol}\nolimits^{-1}\hspace{-0.01in}\delta_{q}%
^{3}(\mathbf{x}^{\hspace{0.01in}\prime}\oplus(\ominus\hspace{0.01in}%
\kappa^{-1}\mathbf{x}))\circledast u_{\hspace{0.01in}\mathbf{p}}(x)\nonumber\\
&  \qquad=\lim_{t\hspace{0.01in}\rightarrow\hspace{0.01in}\mp\infty}%
\int\text{d}_{q}^{3}x\,(u^{\ast})_{\mathbf{p}^{\prime}}(\mathbf{x}%
^{\hspace{0.01in}\prime}\hspace{-0.01in},t)\circledast u_{\hspace
{0.01in}\mathbf{p}}(\mathbf{x},t)=\operatorname*{vol}\nolimits^{-1}%
\hspace{-0.01in}\delta_{q}^{3}((\ominus\hspace{0.01in}\kappa^{-1}%
\mathbf{p}^{\prime})\oplus\mathbf{p}). \label{UniBedSMat1}%
\end{align}
By similar reasoning, we obtain:%
\begin{equation}
\int\text{d}_{q}^{3}p^{\prime\prime}\hspace{0.01in}(S_{L})_{\mathbf{p}%
^{\prime}\mathbf{p}^{\prime\prime}}^{\pm}\hspace{-0.01in}\circledast
(S_{R}^{\ast})_{\mathbf{p}^{\prime\prime}\mathbf{p}}^{\pm}=\operatorname*{vol}%
\nolimits^{-1}\hspace{-0.01in}\delta_{q}^{3}(\hspace{0.01in}\mathbf{p}%
^{\prime}\oplus(\ominus\hspace{0.01in}\kappa^{-1}\mathbf{p})).
\label{UniBedSMat2}%
\end{equation}
Since the $q$-de\-formed plane waves form a complete system of functions, we
can interpret the identities in Eqs.~(\ref{UniBedSMat1}) and
(\ref{UniBedSMat2}) as $q$-ver\-sions of the\textbf{ }\textit{unitarity
condition of the S-ma\-trix}.

The normalizations of the wave functions do not change by scattering due to
the unitarity of the S-ma\-trix. We can show this fact by the following
calculation:%
\begin{align}
&  \int\text{d}_{q}^{3}x\,(\psi_{L}^{\ast})^{\pm}(\mathbf{x},t)\circledast
(\psi_{R})^{\pm}(\mathbf{x},t)=\nonumber\\
&  =\lim_{t^{\prime}\hspace{-0.01in},\hspace{0.01in}t^{\prime\prime
}\rightarrow\hspace{0.01in}\mp\infty}\int\text{d}_{q}^{3}x^{\hspace
{0.01in}\prime\prime}\hspace{0.01in}\text{d}_{q}^{3}x^{\hspace{0.01in}\prime
}\hspace{0.01in}\text{d}_{q}^{3}x\,\phi_{L}^{\ast}(x^{\hspace{0.01in}%
\prime\prime})\circledast(G_{L}^{\hspace{0.01in}\ast})^{\pm}(x^{\hspace
{0.01in}\prime\prime}\hspace{-0.01in};x)\circledast(G_{R})^{\pm}%
(x;x^{\hspace{0.01in}\prime})\circledast\phi_{R}(x^{\hspace{0.01in}\prime
})\nonumber\\
&  =\operatorname*{vol}\nolimits^{-1}\hspace{-0.02in}\lim_{t^{\prime}%
\hspace{-0.01in},\hspace{0.01in}t^{\prime\prime}\rightarrow\hspace{0.01in}%
\mp\infty}\int\text{d}_{q}^{3}x^{\hspace{0.01in}\prime\prime}\hspace
{0.01in}\text{d}_{q}^{3}x^{\hspace{0.01in}\prime}\hspace{0.01in}\phi_{L}%
^{\ast}(x^{\hspace{0.01in}\prime\prime})\circledast\delta_{q}^{3}%
(\mathbf{x}^{\hspace{0.01in}\prime\prime}\hspace{-0.01in}\oplus(\ominus
\hspace{0.01in}\kappa^{-1}\mathbf{x}^{\hspace{0.01in}\prime}))\circledast
\phi_{R}(x^{\hspace{0.01in}\prime})\nonumber\\
&  =\lim_{t^{\prime}\rightarrow\hspace{0.01in}\mp\infty}\int\text{d}_{q}%
^{3}x^{\hspace{0.01in}\prime}\hspace{0.01in}\phi_{L}^{\ast}(\mathbf{x}%
^{\hspace{0.01in}\prime}\hspace{-0.01in},t^{\hspace{0.01in}\prime}%
)\circledast\phi_{R}(\mathbf{x}^{\hspace{0.01in}\prime}\hspace{-0.01in}%
,t^{\hspace{0.01in}\prime})\nonumber\\
&  =\int\text{d}_{q}^{3}x\,\phi_{L}^{\ast}(\mathbf{x},t)\circledast\phi
_{R}(\mathbf{x},t). \label{ErhNorGreBew}%
\end{align}
First, we have applied the identities in Eqs.~(\ref{DefGreGesInh1}) and
(\ref{DefGreGesInh2}) of Chap.~\ref{ProWecKap}. The next step follows from
Eq.~(\ref{KomIdeGreRSteGL}). Due to the $q$-de\-formed delta function, we
could carry out the integral with respect to $x^{\prime\prime}$. Moreover, we
have identified the time variables $t^{\prime}$ and $t^{\prime\prime}$ since
they approach the same limit. The last identity holds since the normalization
of the wave functions for a free particle is time-in\-de\-pen\-dent
\cite{Wachter:2020A}. We can verify the following identity by similar
reasoning:%
\begin{equation}
\int\text{d}_{q}^{3}x\,(\psi_{L})^{\pm}(\mathbf{x},t)\circledast(\psi
_{R}^{\ast})^{\pm}(\mathbf{x},t)=\int\text{d}_{q}^{3}x\,\phi_{L}%
(\mathbf{x},t)\circledast\phi_{R}^{\ast}(\mathbf{x},t).
\end{equation}

\subsection{Scattering of a scalar particle\label{KapStrKleGorFelN}}

We can also employ our reasonings in the case of a $q$-de\-formed scalar
particle. To this end, we extend $q$-ver\-sions of the free Klein-Gordon
equation [see Eqs.~(\ref{KleGorGleLin}), (\ref{KleGorGleRec}), and
(\ref{KleGorGleLin2}) in Chap.~\ref{KapProSchrFel}] by an interaction term
that depends on a real scattering potential $V$, i.~e.%
\begin{align}
c^{-2}\partial_{t}^{\hspace{0.01in}2}\triangleright\hspace{-0.01in}\phi
_{R}-\hspace{-0.01in}\nabla_{q}^{\hspace{0.01in}2}\hspace{0.01in}%
\triangleright\phi_{R}+(m\hspace{0.01in}c_{0})^{2}\phi_{R}  &  =-V\circledast
\phi_{R},\nonumber\\
c^{-2}\partial_{t}^{\hspace{0.01in}2}\hspace{0.01in}\,\bar{\triangleright
}\,\phi_{R}^{\ast}-\hspace{-0.01in}\nabla_{q}^{\hspace{0.01in}2}%
\,\hspace{0.01in}\bar{\triangleright}\,\phi_{R}^{\ast}+(m\hspace{0.01in}%
c_{0})^{2}\phi_{R}^{\ast}  &  =-V\circledast\phi_{R}^{\ast},
\label{KleGorGleLinInh}%
\end{align}
or%
\begin{align}
\phi_{L}\,\bar{\triangleleft}\,\hspace{0.01in}\partial_{t}^{\hspace{0.01in}%
2}c^{-2}\hspace{-0.01in}-\phi_{L}\,\bar{\triangleleft}\,\nabla_{q}%
^{\hspace{0.01in}2}+\phi_{L}\hspace{0.01in}(m\hspace{0.01in}c_{0})^{2}  &
=-\hspace{0.01in}\phi_{L}\circledast V,\nonumber\\
\phi_{L}^{\ast}\hspace{-0.01in}\triangleleft\partial_{t}^{\hspace{0.01in}%
2}c^{-2}\hspace{-0.01in}-\phi_{L}^{\ast}\triangleleft\nabla_{q}^{\hspace
{0.01in}2}+\phi_{L}^{\ast}\hspace{0.01in}(m\hspace{0.01in}c_{0})^{2}  &
=-\hspace{0.01in}\phi_{L}^{\ast}\circledast V. \label{KleGorGleRecInh}%
\end{align}

We can get Lipp\-mann-Schwin\-ger equations for the solutions to the above
$q$-de\-formed Klein-Gordon equations by using the propagators for the
$q$-ver\-sions of the free Klein-Gordon equation. For example, if we replace
the inhomogeneity $\varrho$ in Eq.~(\ref{LoeInhKleGorGle}) of Chap.
\ref{KapProSchrFel} with the interaction term $-V\circledast\phi_{R}$ or
$-\hspace{0.01in}\phi_{L}\circledast V$, we find:\footnote{We denote the
solutions to the $q$-versions of the free Klein-Gordon equations by
$\varphi_{L}$ or $\varphi_{R}$.}%
\begin{align}
\phi_{R}(x)  &  =\varphi_{R}(x)-\text{i\hspace{-0.01in}}\int\text{d}_{q}%
^{3}x^{\hspace{0.01in}\prime}\hspace{0.01in}\text{d}t^{\hspace{0.01in}\prime
}\hspace{0.01in}\Delta_{R}(x\hspace{0.01in};x^{\hspace{0.01in}\prime
})\circledast V(x^{\hspace{0.01in}\prime})\circledast\phi_{R}(x^{\hspace
{0.01in}\prime}),\nonumber\\
\phi_{L}(x)  &  =\varphi_{L}(x)+\text{i\hspace{-0.01in}}\int\text{d}_{q}%
^{3}x^{\hspace{0.01in}\prime}\hspace{0.01in}\text{d}t^{\hspace{0.01in}\prime
}\hspace{0.01in}\phi_{L}(x^{\hspace{0.01in}\prime})\circledast V(x^{\hspace
{0.01in}\prime})\circledast\Delta_{L}(x^{\hspace{0.01in}\prime}\hspace
{-0.01in};x). \label{LipSchwGleKleGorFelPhiL}%
\end{align}
By similar reasoning, we get:%
\begin{align}
\phi_{R}^{\ast}(x)  &  =\varphi_{R}^{\ast}(x)-\text{i\hspace{-0.01in}}%
\int\text{d}_{q}^{3}x^{\hspace{0.01in}\prime}\hspace{0.01in}\text{d}%
t^{\hspace{0.01in}\prime}\hspace{0.01in}\Delta_{R}^{\ast}(x\hspace
{0.01in};x^{\hspace{0.01in}\prime})\circledast V(x^{\hspace{0.01in}\prime
})\circledast\phi_{R}^{\ast}(x^{\hspace{0.01in}\prime}),\nonumber\\
\phi_{L}^{\ast}(x)  &  =\varphi_{L}^{\ast}(x)+\text{i\hspace{-0.01in}}%
\int\text{d}_{q}^{3}x^{\hspace{0.01in}\prime}\hspace{0.01in}\text{d}%
t^{\hspace{0.01in}\prime}\hspace{0.01in}\phi_{L}^{\ast}(x^{\hspace
{0.01in}\prime})\circledast V(x^{\hspace{0.01in}\prime})\circledast\Delta
_{L}^{\ast}(x^{\hspace{0.01in}\prime}\hspace{-0.01in};x).
\end{align}

Next, we introduce S-ma\textbf{\-}trix elements for the scattering of a
$q$-de\-formed scalar particle. Similarly to the definitions in
Eqs.~(\ref{SMatDef1})-(\ref{SMatDef4}) of Chap.~\ref{KapSMatDef}, we have%
\begin{gather}
(S_{R})^{\pm}(\varphi,\phi)=\nonumber\\
=\lim_{t\hspace{0.01in}\rightarrow\hspace{0.01in}\pm\infty}\text{i\hspace
{0.01in}}c^{-2}\hspace{-0.02in}\int\text{d}_{q}^{3}x\,\big[(\varphi_{L}^{\ast
})^{\pm}(x)\triangleleft\partial_{t}\circledast\phi_{R}(x)+(\varphi_{L}^{\ast
})^{\pm}(x)\circledast\partial_{t}\triangleright\hspace{-0.01in}\phi
_{R}(x)\big],\\[0.13in]
(S_{L})^{\pm}(\phi,\varphi)=\nonumber\\
=\lim_{t\hspace{0.01in}\rightarrow\hspace{0.01in}\pm\infty}\text{i\hspace
{0.01in}}c^{-2}\hspace{-0.02in}\int\text{d}_{q}^{3}x\,\big[\phi_{L}%
(x)\triangleleft\partial_{t}\circledast\hspace{-0.01in}(\varphi_{R}^{\ast
})^{\pm}(x)+\phi_{L}(x)\circledast\partial_{t}\triangleright\hspace
{-0.01in}(\varphi_{L}^{\ast})^{\pm}(x)\big], \label{SLKleGorFel}%
\end{gather}
and%
\begin{gather}
(S_{R}^{\ast})^{\pm}(\varphi,\phi)=\nonumber\\
=\lim_{t\hspace{0.01in}\rightarrow\hspace{0.01in}\pm\infty}\text{i\hspace
{0.01in}}c^{-2}\hspace{-0.02in}\int\text{d}_{q}^{3}x\,\big[(\varphi_{L})^{\pm
}(x)\triangleleft\partial_{t}\circledast\phi_{R}^{\ast}(x)+(\varphi_{L})^{\pm
}(x)\circledast\partial_{t}\triangleright\hspace{-0.01in}\phi_{R}^{\ast
}(x)\big],\\[0.13in]
(S_{L}^{\ast})^{\pm}(\phi,\varphi)=\nonumber\\
=\lim_{t\hspace{0.01in}\rightarrow\hspace{0.01in}\pm\infty}\text{i\hspace
{0.01in}}c^{-2}\hspace{-0.02in}\int\text{d}_{q}^{3}x\,\big[\phi_{L}^{\ast
}(x)\triangleleft\partial_{t}\circledast\hspace{-0.01in}(\varphi_{R})^{\pm
}(x)+\phi_{L}^{\ast}(x)\circledast\partial_{t}\triangleright\hspace
{-0.01in}(\varphi_{R})^{\pm}(x)\big].
\end{gather}
The plus or minus sign indicates whether we consider particle states with
positive or negative energy after scattering \cite{Bjorken1Engl}.

We consider a situation in which a free scalar particle starts with momentum
$\mathbf{p}$. After it has taken part in an interaction, the particle will
reappear as a free particle with momentum $\mathbf{p}^{\prime}$. For this
reason, the wave function for the final free particle is given by [see
Eq.~(\ref{EbeWelKGFSte}) in Chap. \ref{KapProSchrFel}]:%
\begin{equation}
(\varphi_{L}^{\ast})^{+}(\mathbf{x},t)=(\varphi^{\ast})_{\mathbf{p}^{\prime}%
}^{[+]}(\mathbf{x},t)=\frac{c}{\sqrt{2}}\,E_{\mathbf{p}^{\prime}}%
^{\hspace{0.01in}-1/2}\hspace{-0.01in}\circledast\exp(\text{i\hspace{0.01in}%
}tE_{\mathbf{p}})\circledast(u^{\ast})_{\mathbf{p}}(\mathbf{x}).
\label{FinWav}%
\end{equation}
Since the wave function $\phi_{R}(\mathbf{x},t)$ for the interacting particle
evolves from a state with positive energy and momentum $\mathbf{p}$ in the
distant past, it holds%
\begin{equation}
\lim_{t\hspace{0.01in}\rightarrow\hspace{0.01in}-\infty}\big[\phi
_{R}(\mathbf{x},t)-\varphi_{\mathbf{p}}^{[+]}(\mathbf{x},t)\big]=0
\label{RanBedPhiL}%
\end{equation}
with [see Eq.~(\ref{EbeWelKGF}) in Chap.~\ref{KapProSchrFel}]%
\begin{equation}
\varphi_{\mathbf{p}}^{[+]}(\mathbf{x},t)=\frac{c}{\sqrt{2}}\,u_{\hspace
{0.01in}\mathbf{p}}(\mathbf{x})\circledast E_{\mathbf{p}}^{\hspace
{0.01in}-1/2}\hspace{-0.01in}\circledast\exp(-\text{i\hspace{0.01in}%
}tE_{\mathbf{p}}).
\end{equation}
The wave function $\phi_{R}(\mathbf{x},t)$ is a solution to the first
Klein-Gor\-don equation in (\ref{KleGorGleLinInh}) and satisfies the boundary
condition in Eq.~(\ref{RanBedPhiL}). For this reason, $\phi_{R}(\mathbf{x},t)$
has to be a solution to the first Lipp\-mann-Schwin\-ger equation in
Eq.~(\ref{LipSchwGleKleGorFelPhiL}):%
\begin{equation}
\phi_{R}(\mathbf{x},t)=\varphi_{\mathbf{p}}^{[+]}(\mathbf{x},t)-\text{i\hspace
{-0.01in}}\int\text{d}^{3}x^{\hspace{0.01in}\prime}\hspace{0.01in}%
\text{d}t^{\hspace{0.01in}\prime}\hspace{0.01in}\Delta_{R}(x\hspace
{0.01in};x^{\hspace{0.01in}\prime})\circledast V(\mathbf{x}^{\hspace
{0.01in}\prime})\circledast\phi_{R}(x^{\hspace{0.01in}\prime}).
\label{IteLoePhiL}%
\end{equation}

If we plug the expressions of Eqs.~(\ref{FinWav}) and (\ref{IteLoePhiL}) into
the first formula of Eq.~(\ref{SLKleGorFel}), we finally obtain:%
\begin{gather}
(S_{R})_{\mathbf{p}^{\prime}\mathbf{p}}^{+}=\,\lim_{t\hspace{0.01in}%
\rightarrow\hspace{0.01in}+\infty}\text{i}c^{-2}\hspace{-0.02in}\int
\text{d}_{q}^{3}x\,\Big [(\varphi^{\ast})_{\mathbf{p}^{\prime}}^{[+]}%
\triangleleft\partial_{t}\circledast\varphi_{\mathbf{p}}^{[+]}+(\varphi^{\ast
})_{\mathbf{p}^{\prime}}^{[+]}\circledast\partial_{t}\triangleright
\hspace{-0.01in}\varphi_{\mathbf{p}}^{[+]}\Big]\nonumber\\
+\lim_{t\hspace{0.01in}\rightarrow\hspace{0.01in}+\infty}c^{-2}\hspace
{-0.02in}\int\text{d}_{q}^{3}x\int\text{d}^{3}x^{\hspace{0.01in}\prime}%
\hspace{0.01in}\text{d}t^{\hspace{0.01in}\prime}\hspace{0.01in}\Big [(\varphi
^{\ast})_{\mathbf{p}^{\prime}}^{[+]}(x)\triangleleft\partial_{t}%
\circledast\hspace{-0.01in}\Delta_{R}(x\hspace{0.01in};x^{\hspace
{0.01in}\prime})\circledast V(\mathbf{x}^{\hspace{0.01in}\prime}%
)\circledast\phi_{R}(x^{\hspace{0.01in}\prime})\nonumber\\
+\,(\varphi^{\ast})_{\mathbf{p}^{\prime}}^{[+]}(x)\circledast\partial
_{t}\triangleright\hspace{-0.01in}\Delta_{R}(x\hspace{0.01in};x^{\hspace
{0.01in}\prime})\circledast V(\mathbf{x}^{\hspace{0.01in}\prime}%
)\circledast\phi_{R}(x^{\hspace{0.01in}\prime})\Big].
\label{SMatKleGorTeiTeiStr}%
\end{gather}
The propagator $\Delta_{R}$ is subject to the following identity
\cite{Wachter:2021B}:%
\begin{align}
&  \text{i}c^{-2}\hspace{-0.02in}\int\text{d}_{q}^{3}x\,\varphi_{L}^{\ast
}(\mathbf{x},t)\triangleleft\partial_{t}\circledast\hspace{-0.01in}\Delta
_{R}(\mathbf{x},t\hspace{0.01in};\mathbf{x}^{\hspace{0.01in}\prime}%
\hspace{-0.01in},t^{\hspace{0.01in}\prime})\nonumber\\
&  +\text{i}c^{-2}\hspace{-0.02in}\int\text{d}_{q}^{3}x\,\varphi_{L}^{\ast
}(\mathbf{x},t)\circledast\partial_{t}\triangleright\hspace{-0.01in}\Delta
_{R}(\mathbf{x},t\hspace{0.01in};\mathbf{x}^{\hspace{0.01in}\prime}%
\hspace{-0.01in},t^{\hspace{0.01in}\prime})=\nonumber\\
&  \qquad\qquad=\theta(t-t^{\hspace{0.01in}\prime})\,(\varphi_{L}^{\ast
})^{[+]}(\mathbf{x}^{\hspace{0.01in}\prime}\hspace{-0.01in},t^{\hspace
{0.01in}\prime})-\theta(t^{\hspace{0.01in}\prime}\hspace{-0.02in}%
-t)\,(\varphi_{L}^{\ast})^{[-]}(\mathbf{x}^{\hspace{0.01in}\prime}%
\hspace{-0.01in},t^{\hspace{0.01in}\prime}).
\end{align}
For this reason, we have:%
\begin{align}
&  \text{i}c^{-2}\hspace{-0.02in}\int\text{d}_{q}^{3}x\,\big[(\varphi^{\ast
})_{\mathbf{p}^{\prime}}^{[+]}(x)\triangleleft\partial_{t}\circledast
\hspace{-0.01in}\Delta_{R}(x\hspace{0.01in};x^{\hspace{0.01in}\prime
})\circledast V(\mathbf{x}^{\hspace{0.01in}\prime})\circledast\phi
_{R}(x^{\prime})\nonumber\\
&  +\text{i}c^{-2}\hspace{-0.02in}\int\text{d}_{q}^{3}x\,(\varphi^{\ast
})_{\mathbf{p}^{\prime}}^{[+]}(x)\circledast\partial_{t}\triangleright
\Delta_{R}(x\hspace{0.01in};x^{\prime})\circledast V(\mathbf{x}^{\hspace
{0.01in}\prime})\circledast\phi_{R}(x^{\hspace{0.01in}\prime}%
)\big]=\nonumber\\
&  \qquad\qquad\qquad\qquad=\theta(t-t^{\hspace{0.01in}\prime})\,(\varphi
^{\ast})_{\mathbf{p}^{\prime}}^{[+]}(x^{\hspace{0.01in}\prime})\circledast
V(\mathbf{x}^{\hspace{0.01in}\prime})\circledast\phi_{R}(x^{\hspace
{0.01in}\prime}).
\end{align}
Using this result and the orthogonality relation in
Eq.~(\ref{OrtRelKleGorPhiSte}) of Chap.~\ref{KapProSchrFel}, we can write the
S-ma\textbf{\-}trix in Eq.~(\ref{SMatKleGorTeiTeiStr}) as follows
\cite{Bjorken1Engl}:%
\begin{equation}
(S_{R})_{\mathbf{p}^{\prime}\mathbf{p}}^{+}=\operatorname*{vol}\nolimits^{-1}%
\hspace{-0.01in}\delta_{q}^{3}((\ominus\hspace{0.01in}\mathbf{p}^{\prime
})\oplus\mathbf{p})-\text{i\hspace{-0.01in}}\int\text{d}^{3}x^{\hspace
{0.01in}\prime}\hspace{0.01in}\text{d}t^{\hspace{0.01in}\prime}\hspace
{0.01in}(\varphi^{\ast})_{\mathbf{p}^{\prime}}^{[+]}(x^{\hspace{0.01in}\prime
})\circledast V(\mathbf{x}^{\hspace{0.01in}\prime})\circledast\phi
_{R}(x^{\hspace{0.01in}\prime}). \label{IteGleSLKleGor}%
\end{equation}
There are no limits in the above formula since the orthogonality relations
are time-in\textbf{\-}de\textbf{\-}pen\textbf{\-}dent. Moreover, the following
identity holds:%
\begin{equation}
\lim_{t\hspace{0.01in}\rightarrow\hspace{0.01in}\infty}\theta(t-t^{\hspace
{0.01in}\prime})=1.
\end{equation}

We can solve the Lipp\textbf{\-}mann-Schwin\textbf{\-}ger equation for\ the
wave function $\phi_{R}(x)$ by iteration [see Eq.~(\ref{IteLoePhiL})]:%
\begin{align}
\phi_{R}(x)=  &  \hspace{0.03in}\varphi_{\mathbf{p}}^{[+]}(x)-\text{i\hspace
{-0.01in}}\int\text{d}^{3}x^{\hspace{0.01in}\prime}\hspace{0.01in}%
\text{d}t^{\hspace{0.01in}\prime}\hspace{0.01in}\Delta_{R}(x;x^{\hspace
{0.01in}\prime})\circledast V(\mathbf{x}^{\hspace{0.01in}\prime}%
)\circledast\varphi_{\mathbf{p}}^{[+]}(x^{\prime})\nonumber\\
&  +(-\text{i})^{2}\hspace{-0.01in}\int\text{d}^{3}x^{\prime}\,\text{d}%
t^{\hspace{0.01in}\prime}\hspace{-0.01in}\int\text{d}^{3}x^{\hspace
{0.01in}\prime\prime}\,\text{d}t^{\hspace{0.01in}\prime\prime}\,\Delta
_{R}(x;x^{\hspace{0.01in}\prime})\circledast V(\mathbf{x}^{\hspace
{0.01in}\prime})\nonumber\\
&  \qquad\qquad\qquad\circledast\Delta_{R}(x^{\hspace{0.01in}\prime}%
\hspace{-0.01in};x^{\hspace{0.01in}\prime\prime})\circledast V(\mathbf{x}%
^{\hspace{0.01in}\prime\prime})\circledast\varphi_{\mathbf{p}}^{[+]}%
(x^{\hspace{0.01in}\prime\prime})+\ldots
\end{align}
By substituting this result into the right-hand side of
Eq.~(\ref{IteGleSLKleGor}), we get:%
\begin{align}
(S_{R})_{\mathbf{p}^{\prime}\mathbf{p}}^{+}=  &  \,\operatorname*{vol}%
\nolimits^{-1}\hspace{-0.01in}\delta_{q}^{3}((\ominus\hspace{0.01in}%
\mathbf{p}^{\prime})\oplus\mathbf{p})-\text{i\hspace{-0.01in}}\int\text{d}%
^{3}x^{\hspace{0.01in}\prime}\hspace{0.01in}\text{d}t^{\hspace{0.01in}\prime
}\hspace{0.01in}(\varphi^{\ast})_{\mathbf{p}^{\prime}}^{[+]}(x^{\hspace
{0.01in}\prime})\circledast V(\mathbf{x}^{\hspace{0.01in}\prime}%
)\circledast\varphi_{\mathbf{p}}^{[+]}(x^{\hspace{0.01in}\prime})\nonumber\\
&  \,-\int\text{d}^{3}x^{\hspace{0.01in}\prime}\hspace{0.01in}\text{d}%
t^{\hspace{0.01in}\prime}\hspace{-0.01in}\int\text{d}^{3}x^{\hspace
{0.01in}\prime\prime}\hspace{0.01in}\text{d}t^{\hspace{0.01in}\prime\prime
}\hspace{0.01in}(\varphi^{\ast})_{\mathbf{p}^{\prime}}^{[+]}(x^{\hspace
{0.01in}\prime})\circledast V(\mathbf{x}^{\hspace{0.01in}\prime})\nonumber\\
&  \qquad\qquad\qquad\circledast\Delta_{R}(x^{\hspace{0.01in}\prime}%
\hspace{-0.01in};x^{\hspace{0.01in}\prime\prime})\circledast V(\mathbf{x}%
^{\hspace{0.01in}\prime\prime})\circledast\varphi_{\mathbf{p}}^{[+]}%
(x^{\hspace{0.01in}\prime\prime})+\ldots\label{SMatStrTeiTei}%
\end{align}

We can modify the above considerations to apply them to antiparticle
scattering, pair production, and annihilation.

\section{Interaction picture\label{IntPic}}

In Ref.~\cite{Wachter:2020A}, we have discussed two formalisms of quantum
dynamics on the $q$-de\textbf{\-}formed Euclidean space: the Schr\"{o}dinger
picture and the Heisenberg picture. Whenever the Hamiltonian operator of a
quantum system splits into a time-in\textbf{\-}de\textbf{\-}pen\textbf{\-}dent
part $H_{0}$ and a time-de\textbf{\-}pen\textbf{\-}dent interaction $V(t)$,
the \textit{interaction picture} is often more helpful \cite{Sakurai:1994}.
The interaction picture also applies to scattering processes in the $q$-de\-formed Euclidean space.

We can assume that a particle before scattering is non-in\textbf{\-}%
ter\textbf{\-}act\textbf{\-}ing. Moreover, we can expand its wave function in
terms of $q$-de\textbf{\-}formed momentum eigenfunctions \cite{Wachter:2019A}.
The corresponding expansion coefficients determine the probability for the
particle being in a particular momentum eigenstate. After the interaction, we
can again expand the wave function in terms of $q$-de\textbf{\-}formed
momentum eigenfunctions. The expansion coefficients will have changed due to
the scattering.

For the wave functions describing the scattered $q$-de\textbf{\-}formed
particle in the interaction picture, we have the expansions\footnote{We write
wave functions of the interaction picture in upper case Greek letters.}%
\begin{align}
\Psi_{R}(\mathbf{x},t)  &  =\int\text{d}_{q}^{3}p\,u_{\hspace{0.01in}%
\mathbf{p}}(\mathbf{x})\circledast C_{\mathbf{p}}(t),\nonumber\\
\Psi_{L}(\mathbf{x},t)  &  =\int\text{d}_{q}^{3}p\,C^{\hspace{0.01in}%
\mathbf{p}}(t)\circledast u^{\mathbf{p}}(\mathbf{x}), \label{EntWecBil1}%
\end{align}
and%
\begin{align}
\Psi_{L}^{\ast}(\mathbf{x},t)  &  =\int\text{d}_{q}^{3}p\,(C^{\ast
})_{\mathbf{p}}(t)\circledast(u^{\ast})_{\mathbf{p}}(\mathbf{x}%
),\nonumber\\[0.1in]
\Psi_{R}^{\ast}(\mathbf{x},t)  &  =\int\text{d}_{q}^{3}p\,(u^{\ast
})^{\mathbf{p}}(\mathbf{x})\circledast(C^{\ast})^{\mathbf{p}}(t)
\label{EntWecBil2}%
\end{align}
with%
\begin{align}
u_{\hspace{0.01in}\mathbf{p}}(\mathbf{x})  &  =u_{\hspace{0.01in}\mathbf{p}%
}(\mathbf{x},t=0), & (u^{\ast})_{\mathbf{p}}(\mathbf{x})  &  =(u^{\ast
})_{\mathbf{p}}(\mathbf{x},t=0),\nonumber\\
u^{\mathbf{p}}(\mathbf{x})  &  =u^{\mathbf{p}}(\mathbf{x},t=0), & (u^{\ast
})^{\mathbf{p}}(\mathbf{x})  &  =(u^{\ast})^{\mathbf{p}}(\mathbf{x},t=0).
\label{MomEigFktWdh}%
\end{align}
The potential $V(t)$ alone determines the time dependence of the expansion
coefficients $C_{\mathbf{p}}(t)$, $C^{\hspace{0.01in}\mathbf{p}}(t)$,
$(C^{\ast})_{\mathbf{p}}(t)$, and $(C^{\ast})^{\mathbf{p}}(t)$.

We can regain the wave functions of the Schr\"{o}dinger picture from the
expansions in Eq.~(\ref{EntWecBil1}) or Eq.~(\ref{EntWecBil2}). To this end,
we only need to replace the $q$-de\-formed momentum eigenfunctions with the
corresponding plane wave solutions of the $q$-de\-formed free Schr\"{o}dinger
equations, i.~e.%
\begin{align}
\psi_{R}(\mathbf{x},t)  &  =\exp(-\text{i\hspace{0.01in}}tH_{0})\triangleright
\Psi_{R}(\mathbf{x},t)=\int\text{d}_{q}^{3}p\,u_{\hspace{0.01in}\mathbf{p}%
}(\mathbf{x},t)\circledast C_{\mathbf{p}}(t),\nonumber\\[0.1in]
\psi_{R}^{\ast}(\mathbf{x},t)  &  =\exp(-\text{i\hspace{0.01in}}tH_{0}%
)\,\bar{\triangleright}\,\Psi_{R}^{\ast}(\mathbf{x},t)=\int\text{d}_{q}%
^{3}p\,(u^{\ast})^{\mathbf{p}}(x,t)\circledast(C^{\ast})^{\mathbf{p}}(t),
\label{ZusSchWecBil1}%
\end{align}
and%
\begin{align}
\psi_{L}^{\ast}(\mathbf{x},t)  &  =\Psi_{L}^{\ast}(\mathbf{x},t)\triangleleft
\exp(\text{i}H_{0}\text{\hspace{0.01in}}t)=\int\text{d}_{q}^{3}p\,(C^{\ast
})_{\mathbf{p}}(t)\circledast(u^{\ast})_{\mathbf{p}}(x,t),\nonumber\\[0.1in]
\psi_{L}(\mathbf{x},t)  &  =\Psi_{L}(\mathbf{x},t)\,\bar{\triangleleft}%
\,\exp(\text{i}H_{0}\text{\hspace{0.01in}}t)=\int\text{d}_{q}^{3}%
p\,C^{\hspace{0.01in}\mathbf{p}}(t)\circledast u^{\mathbf{p}}(\mathbf{x},t).
\label{ZusSchWecBil2}%
\end{align}

The wave functions of the Schr\"{o}dinger picture are solutions to
Schr\"{o}dinger equations with the complete Hamiltonian opera\-tor $H$:%
\begin{align}
\text{i\hspace{0.01in}}\partial_{t}\triangleright\hspace{-0.01in}\psi
_{R}(\mathbf{x},t)  &  =H\triangleright\psi_{R}(\mathbf{x},t), & \psi
_{L}^{\ast}(\mathbf{x},t)\triangleleft\partial_{t}\hspace{0.01in}\text{i}  &
=\psi_{L}^{\ast}(\mathbf{x},t)\triangleleft H,\nonumber\\
\text{i\hspace{0.01in}}\partial_{t}\,\bar{\triangleright}\,\psi_{R}^{\ast
}(\mathbf{x},t)  &  =H\,\bar{\triangleright}\,\psi_{R}^{\ast}(\mathbf{x},t), &
\psi_{L}(\mathbf{x},t)\,\bar{\triangleleft}\,\partial_{t}\hspace
{0.01in}\text{i}  &  =\psi_{L}(\mathbf{x},t)\,\bar{\triangleleft}\,H.
\label{SchroGlePsiLWec}%
\end{align}
The wave functions of the interaction picture, however, satisfy the equations%
\begin{align}
\text{i\hspace{0.01in}}\partial_{t}\triangleright\hspace{-0.01in}\Psi
_{R}(\mathbf{x},t)  &  =V_{I}\triangleright\Psi_{R}(\mathbf{x},t), & \Psi
_{L}^{\ast}(\mathbf{x},t)\triangleleft\partial_{t}\hspace{0.01in}\text{i}  &
=\Psi_{L}^{\ast}(\mathbf{x},t)\triangleleft V_{I},\nonumber\\
\text{i\hspace{0.01in}}\partial_{t}\,\bar{\triangleright}\,\Psi_{R}^{\ast
}(\mathbf{x},t)  &  =V_{I}\,\bar{\triangleright}\,\Psi_{R}^{\ast}%
(\mathbf{x},t), & \Psi_{L}(\mathbf{x},t)\,\bar{\triangleleft}\,\partial
_{t}\hspace{0.01in}\text{i}  &  =\Psi_{L}(\mathbf{x},t)\,\bar{\triangleleft
}\,V_{I}, \label{SchGleWecBil1}%
\end{align}
where $V_{I}$ is understood to be the potential in the interaction picture:%
\begin{equation}
V_{I}=\exp(\text{i\hspace{0.01in}}tH_{0})\,V\exp(-\text{i\hspace{0.01in}%
}tH_{0}).
\end{equation}
We can prove the identities in Eq.~(\ref{SchGleWecBil1}) in the following way:%
\begin{align}
\text{i\hspace{0.01in}}\partial_{t}\triangleright\hspace{-0.01in}\Psi
_{R}(\mathbf{x},t)  &  =\text{i\hspace{0.01in}}\partial_{t}\triangleright
\lbrack\hspace{0.01in}\exp(\text{i\hspace{0.01in}}tH_{0})\triangleright
\psi_{R}(\mathbf{x},t)]\nonumber\\
&  =-H_{0}\exp(\text{i\hspace{0.01in}}tH_{0})\triangleright\psi_{R}%
(\mathbf{x},t)+\exp(\text{i\hspace{0.01in}}tH_{0})\hspace{0.01in}%
H\triangleright\psi_{R}(\mathbf{x},t)\nonumber\\
&  =-H_{0}\triangleright\hspace{-0.01in}\Psi_{R}(\mathbf{x},t)+\exp
(\text{i\hspace{0.01in}}tH_{0})\hspace{0.01in}H\exp(-\text{i\hspace{0.01in}%
}tH_{0})\triangleright\hspace{-0.01in}\Psi_{R}(\mathbf{x},t)\nonumber\\
&  =\exp(\text{i\hspace{0.01in}}tH_{0})\,V\exp(-\text{i\hspace{0.01in}}%
tH_{0})\triangleright\hspace{-0.01in}\Psi_{R}(\mathbf{x},t).
\end{align}
First, we have expressed the wave function of the interaction picture by that
of the Schr\"{o}dinger picture. After applying the product rule for the time
derivative, we use the Schr\"{o}dinger equation for $\psi_{R}$ [cf.
Eq.~(\ref{SchroGlePsiLWec})]. In the penultimate step, we switch from the
Schr\"{o}dinger picture to the interaction picture. The final step follows from
the decomposition $H=H_{0}+V$.

To get solutions to the equations in (\ref{SchGleWecBil1}), we introduce
time-evo\-lu\-tion operators defined by%
\begin{align}
\Psi_{R}(\mathbf{x},t)  &  =(\mathcal{U}_{R})_{I}(t,t_{0})\triangleright
\hspace{-0.01in}\Psi_{R}(\mathbf{x},t_{0}),\nonumber\\
\Psi_{R}^{\ast}(\mathbf{x},t)  &  =(\mathcal{U}_{R}^{\ast})_{I}(t,t_{0}%
)\,\bar{\triangleright}\,\Psi_{R}^{\ast}(\mathbf{x},t_{0}),
\label{OpeGleZeiOpeWec1}%
\end{align}
or%
\begin{align}
\Psi_{L}^{\ast}(\mathbf{x},t)  &  =\Psi_{L}^{\ast}(\mathbf{x},t_{0}%
)\triangleleft(\mathcal{U}_{L}^{\ast})_{I}(t_{0},t),\nonumber\\
\Psi_{L}(\mathbf{x},t)  &  =\Psi_{L}(\mathbf{x},t_{0})\,\bar{\triangleleft
}\,\hspace{0.01in}(\mathcal{U}_{L})_{I}(t_{0},t). \label{OpeGleZeiOpeWec2}%
\end{align}
Due to these definitions and the identities in\ Eq.~(\ref{SchGleWecBil1}), the
time-evo\-lu\-tion operators satisfy the equations%
\begin{align}
\text{i}\hspace{0.01in}\partial_{t}\triangleright\hspace{-0.01in}%
(\mathcal{U}_{R})_{I}(t,t_{0})  &  =V_{I}(t)\hspace{0.01in}(\mathcal{U}%
_{R})_{I}(t,t_{0}),\nonumber\\
\text{i}\hspace{0.01in}\partial_{t}\,\bar{\triangleright}\,(\mathcal{U}%
_{R}^{\ast})_{I}(t,t_{0})  &  =V_{I}(t)\hspace{0.01in}(\mathcal{U}_{R}^{\ast
})_{I}(t,t_{0}), \label{DifGleZeiEntOpeWec1}%
\end{align}
or%
\begin{align}
(\mathcal{U}_{L}^{\ast})_{I}(t_{0},t)\triangleleft\partial_{t}\hspace
{0.01in}\text{i}  &  =(\mathcal{U}_{L}^{\ast})_{I}(t_{0},t)\hspace
{0.01in}V_{I}(t),\nonumber\\
(\mathcal{U}_{L})_{I}(t_{0},t)\,\bar{\triangleleft}\,\hspace{0.01in}%
\partial_{t}\hspace{0.01in}\text{i}  &  =(\mathcal{U}_{L})_{I}(t_{0}%
,t)\hspace{0.01in}V_{I}(t). \label{DifGleZeiEntOpeWec2}%
\end{align}

We require that the solutions to the equations in (\ref{OpeGleZeiOpeWec1}) or
(\ref{OpeGleZeiOpeWec2}) be subject to the following conditions:%
\begin{equation}
(\mathcal{U}_{R})_{I}(t,t)=(\mathcal{U}_{R}^{\ast})_{I}(t,t)=(\mathcal{U}%
_{L})_{I}(t,t)=(\mathcal{U}_{L}^{\ast})_{I}(t,t)=1. \label{RanBedZeiEntOpeDir}%
\end{equation}
In this case, the differential equations in Eq.~(\ref{DifGleZeiEntOpeWec1}) or
Eq.~(\ref{DifGleZeiEntOpeWec2}) are equivalent to the integral equations%
\begin{align}
(\mathcal{U}_{R})_{I}(t,t_{0})  &  =1-\text{i\hspace{-0.01in}}\int_{t_{0}}%
^{t}\text{d}t^{\hspace{0.01in}\prime}\hspace{0.01in}V_{I}(t^{\hspace
{0.01in}\prime})\hspace{0.01in}(\mathcal{U}_{R})_{I}(t^{\hspace{0.01in}\prime
}\hspace{-0.01in},t_{0}),\nonumber\\
(\mathcal{U}_{R}^{\ast})_{I}(t,t_{0})  &  =1-\text{i\hspace{-0.01in}}%
\int_{t_{0}}^{t}\text{d}t^{\hspace{0.01in}\prime}\hspace{0.01in}%
V_{I}(t^{\hspace{0.01in}\prime})\hspace{0.01in}(\mathcal{U}_{R}^{\ast}%
)_{I}(t^{\hspace{0.01in}\prime}\hspace{-0.01in},t_{0}),
\end{align}
or%
\begin{align}
(\mathcal{U}_{L}^{\ast})_{I}(t_{0},t)  &  =1+\text{i\hspace{-0.01in}}%
\int_{t_{0}}^{t}\text{d}t^{\hspace{0.01in}\prime}\hspace{0.01in}%
(\mathcal{U}_{L}^{\ast})_{I}(t_{0},t^{\hspace{0.01in}\prime})\hspace
{0.01in}V_{I}(t^{\hspace{0.01in}\prime}),\nonumber\\
(\mathcal{U}_{L})_{I}(t_{0},t)  &  =1+\text{i\hspace{-0.01in}}\int_{t_{0}}%
^{t}\text{d}t^{\hspace{0.01in}\prime}\hspace{0.01in}(\mathcal{U}_{L}%
)_{I}(t_{0},t^{\hspace{0.01in}\prime})\hspace{0.01in}V_{I}(t^{\hspace
{0.01in}\prime}).
\end{align}
Solving these integral equations by iteration, we obtain $q$%
\textit{-ver\-sions of the Dyson series} i.~e.%
\begin{align}
(\mathcal{U}_{R})_{I}(t,t_{0})  &  =1+\sum_{n\hspace{0.01in}=\hspace{0.01in}%
1}^{\infty}\text{i}^{-n}\hspace{-0.01in}\int\limits_{t_{0}}^{t}\text{d}%
t_{1}\int\limits_{t_{0}}^{t_{1}}\text{d}t_{2}\ldots\hspace{-0.01in}%
\int\limits_{t_{0}}^{t_{n-1}}\text{d}t_{n}\,V_{I}(t_{1})\ldots V_{I}%
(t_{n}),\nonumber\\
(\mathcal{U}_{R}^{\ast})_{I}(t,t_{0})  &  =1+\sum_{n\hspace{0.01in}%
=\hspace{0.01in}1}^{\infty}\text{i}^{-n}\hspace{-0.01in}\int\limits_{t_{0}%
}^{t}\text{d}t_{1}\int\limits_{t_{0}}^{t_{1}}\text{d}t_{2}\ldots
\hspace{-0.01in}\int\limits_{t_{0}}^{t_{n-1}}\text{d}t_{n}\,V_{I}(t_{1})\ldots
V_{I}(t_{n}),
\end{align}
or%
\begin{align}
(\mathcal{U}_{L})_{I}(t_{0},t)  &  =1+\sum_{n\hspace{0.01in}=\hspace{0.01in}%
1}^{\infty}\text{i}^{n}\hspace{-0.01in}\int\limits_{t_{0}}^{t}\text{d}%
t_{1}\int\limits_{t_{0}}^{t_{1}}\text{d}t_{2}\ldots\hspace{-0.01in}%
\int\limits_{t_{0}}^{t_{n-1}}\text{d}t_{n}\,V_{I}(t_{n})\ldots V_{I}%
(t_{1}),\nonumber\\
(\mathcal{U}_{L}^{\ast})_{I}(t_{0},t)  &  =1+\sum_{n\hspace{0.01in}%
=\hspace{0.01in}1}^{\infty}\text{i}^{n}\hspace{-0.01in}\int\limits_{t_{0}}%
^{t}\text{d}t_{1}\int\limits_{t_{0}}^{t_{1}}\text{d}t_{2}\ldots\hspace
{-0.01in}\int\limits_{t_{0}}^{t_{n-1}}\text{d}t_{n}\,V_{I}(t_{n})\ldots
V_{I}(t_{1}).
\end{align}
Note that the product of the potentials is time-or\-dered. For this reason, we
can write the above expansions as follows \cite{Bjorken1Engl}:%
\begin{align}
(\mathcal{U}_{R})_{I}(t,t_{0})  &  =1+\sum_{n\hspace{0.01in}=\hspace{0.01in}%
1}^{\infty}\frac{\text{i}^{-n}}{n!}\int\limits_{t_{0}}^{t}\text{d}t_{1}%
\int\limits_{t_{0}}^{t}\text{d}t_{2}\ldots\hspace{-0.01in}\int\limits_{t_{0}%
}^{t}\text{d}t_{n}\,\hat{T}^{+}\big[V_{I}(t_{1})\ldots V_{I}(t_{n}%
)\big],\nonumber\\
(\mathcal{U}_{L}^{\ast})_{I}(t_{0},t)  &  =1+\sum_{n\hspace{0.01in}%
=\hspace{0.01in}1}^{\infty}\frac{\text{i}^{n}}{n!}\int\limits_{t_{0}}%
^{t}\text{d}t_{1}\int\limits_{t_{0}}^{t}\text{d}t_{2}\ldots\hspace
{-0.01in}\int\limits_{t_{0}}^{t}\text{d}t_{n}\,\hat{T}^{-}\big[V_{I}%
(t_{n})\ldots V_{I}(t_{1})\big]. \label{DirZeiEntOpeEin}%
\end{align}
The time-or\-dered products are defined by%
\begin{align}
\hat{T}^{+}\big[V_{I}(t_{1})\ldots V_{I}(t_{n})\big]  &  =V_{I}(t_{i_{1}%
})\hspace{0.01in}V_{I}(t_{i_{2}})\ldots V_{I}(t_{i_{n}}),\nonumber\\
\hat{T}^{-}\big[V_{I}(t_{1})\ldots V_{I}(t_{n})\big]  &  =V_{I}(t_{i_{n}%
})\ldots V_{I}(t_{i_{2}})\hspace{0.01in}V_{I}(t_{i_{1}})
\label{DysZeiEntOpePM}%
\end{align}
if $\{i_{1},i_{2},\ldots,i_{n}\}$ denotes a permutation of the natural numbers
from $1$ to $n$, such that $t_{i_{1}}\geq t_{i_{2}}\geq\ldots\geq t_{i_{n}}$.

The $q$-de\-formed momentum eigenfunctions in Eq.~(\ref{MomEigFktWdh}) form a
complete and orthogonal system [cf. Eq.~(\ref{OrtRelEbeWel1Schr}) and
Eq.~(\ref{VolRelZeiWelDreDim1}) in Chap.~\ref{KapProSchrFel}]. The
coefficients for the expansions in Eqs.~(\ref{EntWecBil1}) and
(\ref{EntWecBil2}) can be calculated by%
\begin{align}
C_{\mathbf{p}}(t)  &  =\int\text{d}_{q}^{3}x\,(u^{\ast})_{\mathbf{p}%
}(\mathbf{x})\circledast\Psi_{R}(\mathbf{x},t),\nonumber\\
C^{\hspace{0.01in}\mathbf{p}}(t)  &  =\int\text{d}_{q}^{3}x\,\Psi
_{L}(\mathbf{x},t)\circledast(u^{\ast})^{\mathbf{p}}(\mathbf{x}),
\end{align}
and%
\begin{align}
(C^{\ast})_{\mathbf{p}}(t)  &  =\int\text{d}_{q}^{3}x\,\Psi_{L}^{\ast
}(\mathbf{x},t)\circledast u_{\hspace{0.01in}\mathbf{p}}(\mathbf{x}%
),\nonumber\\
(C^{\ast})^{\mathbf{p}}(t)  &  =\int\text{d}_{q}^{3}x\,u^{\mathbf{p}%
}(\mathbf{x})\circledast\Psi_{R}^{\ast}(\mathbf{x},t).
\end{align}
These coefficients determine the probability for a particle to be in a
particular momentum eigenstate after its interaction. In a scattering
experiment, they are related to the S-ma\-trix.

Remember that only the potential $V$ determines the time evolution of the wave
function in the interaction picture. For this reason, the wave function of a
free particle does not depend on time in the interaction picture. It follows
from Eqs.~(\ref{ZusSchWecBil1}) and (\ref{ZusSchWecBil2}) that the wave
function of a free particle at time $t=0$ in the Schr\"{o}dinger picture is
the same as in the interaction picture:%
\begin{align}
\Phi_{R}(\mathbf{x})  &  =\phi_{R}(\mathbf{x},t=0), & \Phi_{R}^{\ast
}(\mathbf{x})  &  =\phi_{R}^{\ast}(\mathbf{x},t=0),\nonumber\\
\Phi_{L}(\mathbf{x})  &  =\phi_{L}(\mathbf{x},t=0), & \Phi_{L}^{\ast
}(\mathbf{x})  &  =\phi_{L}^{\ast}(\mathbf{x},t=0).
\end{align}

In a scattering experiment, a particle is free in the distant past or distant
future. For this reason, the wave functions for a scattered particle in the
interaction picture satisfy the conditions%
\begin{align}
\lim_{t\hspace{0.01in}\rightarrow\hspace{0.01in}\mp\infty}(\Psi_{R})^{\pm
}(\mathbf{x},t)  &  =\Phi_{R}(\mathbf{x}),\nonumber\\
\lim_{t\hspace{0.01in}\rightarrow\hspace{0.01in}\mp\infty}(\Psi_{L})^{\pm
}(\mathbf{x},t)  &  =\Phi_{L}(\mathbf{x}),
\end{align}
and%
\begin{align}
\lim_{t\hspace{0.01in}\rightarrow\hspace{0.01in}\mp\infty}(\Psi_{R}^{\ast
})^{\pm}(\mathbf{x},t)  &  =\Phi_{R}^{\ast}(\mathbf{x}),\nonumber\\
\lim_{t\hspace{0.01in}\rightarrow\hspace{0.01in}\mp\infty}(\Psi_{L}^{\ast
})^{\pm}(\mathbf{x},t)  &  =\Phi_{L}^{\ast}(\mathbf{x}).
\end{align}
If we take into account Eq.~(\ref{OpeGleZeiOpeWec1}) or
Eq.~(\ref{OpeGleZeiOpeWec2}), the above conditions imply%
\begin{align}
(\Psi_{R})^{+}(\mathbf{x},t)  &  =(\mathcal{U}_{R})_{I}(t,-\infty
)\triangleright\Phi_{R}(\mathbf{x}),\nonumber\\
(\Psi_{R}^{\ast})^{+}(\mathbf{x},t)  &  =(\mathcal{U}_{R}^{\ast}%
)_{I}(t,-\infty)\,\bar{\triangleright}\,\Phi_{R}^{\ast}(\mathbf{x}),
\end{align}
or%
\begin{align}
(\Psi_{L})^{+}(\mathbf{x},t)  &  =\Phi_{L}(\mathbf{x})\,\bar{\triangleleft
}\,(\mathcal{U}_{L})_{I}(-\infty,t),\nonumber\\
(\Psi_{L}^{\ast})^{+}(\mathbf{x},t)  &  =\Phi_{L}^{\ast}(\mathbf{x}%
)\triangleleft(\mathcal{U}_{L}^{\ast})_{I}(-\infty,t).
\end{align}
Thus, the matrix representations of the time evolution operator in the
interaction picture determine the S-ma\textbf{\-}trices for a scattering
experiment \cite{Weinberg:2000}, i.~e.%
\begin{align}
(S_{R})^{+}(\Phi,\Psi)  &  =\lim_{t\hspace{0.01in}\rightarrow\hspace
{0.01in}\infty}\int\text{d}_{q}^{3}x\,\Phi_{L}^{\ast}(\mathbf{x}%
)\circledast(\Psi_{R})^{+}(\mathbf{x},t)\nonumber\\
&  =\int\text{d}_{q}^{3}x\,\Phi_{L}^{\ast}(\mathbf{x})\circledast
(\mathcal{U}_{R})_{I}(\infty,-\infty)\triangleright\Phi_{R}(\mathbf{x}%
),\\[0.1in]
(S_{L})^{+}(\Phi,\Psi)  &  =\lim_{t\hspace{0.01in}\rightarrow\hspace
{0.01in}\infty}\int\text{d}_{q}^{3}x\,(\Psi_{L})^{+}(\mathbf{x},t)\circledast
\Phi_{R}^{\ast}(\mathbf{x})\nonumber\\
&  =\int\text{d}_{q}^{3}x\,\Phi_{L}(\mathbf{x})\,\bar{\triangleleft
}\,(\mathcal{U}_{L})_{I}(-\infty,\infty)\circledast\Phi_{R}^{\ast}%
(\mathbf{x}),
\end{align}
and%
\begin{align}
(S_{R}^{\ast})^{+}(\Phi,\Psi)  &  =\lim_{t\hspace{0.01in}\rightarrow
\hspace{0.01in}\infty}\int\text{d}_{q}^{3}x\,\Phi_{L}(\mathbf{x}%
)\circledast(\Psi_{R}^{\ast})^{+}(\mathbf{x},t)\nonumber\\
&  =\int\text{d}_{q}^{3}x\,\Phi_{L}(\mathbf{x})\circledast(\mathcal{U}%
_{R}^{\ast})_{I}(\infty,-\infty)\,\bar{\triangleright}\,\Phi_{R}^{\ast
}(\mathbf{x}),\\[0.1in]
(S_{L}^{\ast})^{+}(\Phi,\Psi)  &  =\lim_{t\hspace{0.01in}\rightarrow
\hspace{0.01in}\infty}\int\text{d}_{q}^{3}x\,(\Psi_{L}^{\ast})^{+}%
(\mathbf{x},t)\circledast\Phi_{R}(\mathbf{x})\nonumber\\
&  =\int\text{d}_{q}^{3}x\,\Phi_{L}^{\ast}(\mathbf{x})\triangleleft
(\mathcal{U}_{L}^{\ast})_{I}(-\infty,\infty)\circledast\Phi_{L}(\mathbf{x}).
\end{align}

\section{Time-dependent perturbation theory\label{KapZeiStoeThe}}

We describe how a time-de\-pen\-dent perturbation causes transitions between
the $q$-de\-formed momentum eigenstates of a nonrelativistic particle. For
this purpose, we assume that the particle moves in the presence of a weak
potential $V(\mathbf{x},t)$. The corresponding Schr\"{o}dinger equations read
as [cf. Eqs.~(\ref{SchGleStr1}) and (\ref{SchGleStr2}) in
Chap.~\ref{ProWecKap}]%
\begin{align}
\text{i}\partial_{t}\triangleright\psi_{R}(\mathbf{x},t)-H_{0}\triangleright
\psi_{R}(\mathbf{x},t)  &  =V(\mathbf{x},t)\circledast\psi_{R}(\mathbf{x}%
,t),\nonumber\\
\text{i}\partial_{t}\,\bar{\triangleright}\,\psi_{R}^{\ast}(\mathbf{x}%
,t)-H_{0}\,\bar{\triangleright}\,\psi_{R}^{\ast}(\mathbf{x},t)  &
=V(\mathbf{x},t)\circledast\psi_{R}^{\ast}(\mathbf{x},t), \label{SchGleVWdh1}%
\end{align}
or%
\begin{align}
\psi_{L}(\mathbf{x},t)\,\bar{\triangleleft}\,\partial_{t}\text{i}-\psi
_{L}(\mathbf{x},t)\,\bar{\triangleleft}\,H_{0}  &  =\psi_{L}(\mathbf{x}%
,t)\circledast V(\mathbf{x},t),\nonumber\\
\psi_{L}^{\ast}(\mathbf{x},t)\triangleleft\partial_{t}\text{i}-\psi_{L}^{\ast
}(\mathbf{x},t)\triangleleft H_{0}  &  =\psi_{L}^{\ast}(\mathbf{x}%
,t)\circledast V(\mathbf{x},t). \label{SchGleVWdh2}%
\end{align}
We can expand their solutions in terms of $q$-de\-formed plane waves as the
latter form a complete and orthogonal system [cf. Eq.~(\ref{OrtRelEbeWel1Schr}%
) and Eq.~(\ref{VolRelZeiWelDreDim1}) in Chap.~\ref{KapProSchrFel}].
Concretely, we have%
\begin{align}
\psi_{R}(\mathbf{x},t)  &  =\int\text{d}_{q}^{3}p\,u_{\hspace{0.01in}%
\mathbf{p}}(\mathbf{x},t)\circledast c_{\hspace{0.01in}\mathbf{p}%
}(t),\nonumber\\[0.05in]
\psi_{L}(\mathbf{x},t)  &  =\int\text{d}_{q}^{3}p\,c^{\hspace{0.01in}%
\mathbf{p}}(t)\circledast u^{\mathbf{p}}(\mathbf{x},t),
\end{align}
or%
\begin{align}
\psi_{R}^{\ast}(\mathbf{x},t)  &  =\int\text{d}_{q}^{3}p\,(u^{\ast
})^{\mathbf{p}}(\mathbf{x},t)\circledast(c^{\ast})^{\mathbf{p}}%
(t),\nonumber\\[0.05in]
\psi_{L}^{\ast}(\mathbf{x},t)  &  =\int\text{d}_{q}^{3}p\,(c^{\ast
})_{\mathbf{p}}(t)\circledast(u^{\ast})_{\mathbf{p}}(\mathbf{x},t).
\end{align}
We plug these expansions into the Schr\"{o}dinger equations (\ref{SchGleVWdh1}%
) or (\ref{SchGleVWdh2}). If we take into account the identities%
\begin{align}
\text{i}\hspace{0.01in}\partial_{t}\triangleright u_{\hspace{0.01in}%
\mathbf{p}}(\mathbf{x},t)  &  =H_{0}\triangleright u_{\hspace{0.01in}%
\mathbf{p}}(\mathbf{x},t),\nonumber\\
u^{\mathbf{p}}(\mathbf{x},t)\,\bar{\triangleleft}\,\partial_{t}\hspace
{0.01in}\text{i}  &  =u^{\mathbf{p}}(\mathbf{x},t)\,\bar{\triangleleft}%
\,H_{0},
\end{align}
or%
\begin{align}
\text{i}\hspace{0.01in}\partial_{t}\,\bar{\triangleright}\,(u^{\ast
})^{\mathbf{p}}(\mathbf{x},t)  &  =H_{0}\,\bar{\triangleright}\,(u^{\ast
})^{\mathbf{p}}(\mathbf{x},t),\nonumber\\
(u^{\ast})_{\mathbf{p}}(\mathbf{x},t)\triangleleft\partial_{t}\hspace
{0.01in}\text{i}  &  =(u^{\ast})_{\mathbf{p}}(\mathbf{x},t)\triangleleft
H_{0},
\end{align}
we obtain, for example:%
\begin{equation}
\int\text{d}_{q}^{3}p\,\big [\hspace{0.01in}u_{\hspace{0.01in}\mathbf{p}%
}(\mathbf{x},t)\circledast\text{i}\hspace{0.01in}\partial_{t}\triangleright
c_{\hspace{0.01in}\mathbf{p}}(t)-V(\mathbf{x},t)\circledast u_{\hspace
{0.01in}\mathbf{p}}(\mathbf{x},t)\circledast c_{\hspace{0.01in}\mathbf{p}%
}(t)\big]=0.
\end{equation}
We multiply this equation by $(u^{\ast})_{\mathbf{p}^{\prime}}(\mathbf{x},t)$
from the left and integrate over all space. Due to the orthogonality of
$q$-de\-formed plane waves, we get
\begin{equation}
\text{i}\hspace{0.01in}\partial_{t}\triangleright c_{\hspace{0.01in}%
\mathbf{p}^{\prime}}(t)=\int\text{d}_{q}^{3}p\,\operatorname*{e}%
\nolimits^{\text{i}t\varepsilon_{\mathbf{p}^{\prime}}}\hspace{-0.02in}%
\circledast\,V_{\mathbf{p}^{\prime}\mathbf{p}}(t)\circledast\operatorname*{e}%
\nolimits^{-\text{i}t\varepsilon_{\mathbf{p}}}\circledast\,c_{\hspace
{0.01in}\mathbf{p}}(t) \label{DifGleZeiSto1}%
\end{equation}
with%
\begin{equation}
V_{\mathbf{p}^{\prime}\mathbf{p}}(t)=\int\text{d}_{q}^{3}x\,(u^{\ast
})_{\mathbf{p}^{\prime}}(\mathbf{x})\circledast V(\mathbf{x},t)\circledast
u_{\hspace{0.01in}\mathbf{p}}(\mathbf{x}).
\end{equation}
Similarly, we can show%
\begin{equation}
c^{\hspace{0.01in}\mathbf{p}^{\prime}}\hspace{-0.02in}(t)\,\bar{\triangleleft
}\,\partial_{t}\hspace{0.01in}\text{i}=\int\text{d}_{q}^{3}p\,c^{\hspace
{0.01in}\mathbf{p}}(t)\circledast\operatorname*{e}\nolimits^{\text{i}%
t\varepsilon_{\mathbf{p}}}\circledast\,V^{\hspace{0.01in}\mathbf{p\hspace
{0.01in}p}^{\prime}}\hspace{-0.02in}(t)\circledast\operatorname*{e}%
\nolimits^{-\text{i}t\varepsilon_{\mathbf{p}^{\prime}}} \label{DifGleZeiSto2}%
\end{equation}
and%
\begin{align}
\text{i}\hspace{0.01in}\partial_{t}\,\bar{\triangleright}\,(c^{\ast
})^{\mathbf{p}^{\prime}}\hspace{-0.02in}(t)  &  =\int\text{d}_{q}%
^{3}p\,\operatorname*{e}\nolimits^{\text{i}\varepsilon_{\mathbf{p}^{\prime}}%
t}\circledast\,V^{\hspace{0.01in}\mathbf{p}^{\prime}\mathbf{p}}(t)\circledast
\operatorname*{e}\nolimits^{-\text{i}\varepsilon_{\mathbf{p}}t}\circledast
\,(c^{\ast})^{\mathbf{p}}(t),\nonumber\\
(c^{\ast})_{\mathbf{p}^{\prime}}(t)\triangleleft\partial_{t}\hspace
{0.01in}\text{i}  &  =\int\text{d}_{q}^{3}p\,(c^{\ast})_{\mathbf{p}%
}(t)\circledast\operatorname*{e}\nolimits^{\text{i}t\varepsilon_{\mathbf{p}}%
}\circledast\hspace{0.01in}V_{\mathbf{p\hspace{0.01in}p}^{\prime}%
}(t)\circledast\operatorname*{e}\nolimits^{-\text{i}t\varepsilon
_{\mathbf{p}^{\prime}}} \label{DifGleZeiSto3}%
\end{align}
with%
\begin{equation}
V^{\hspace{0.01in}\mathbf{p}^{\prime}\mathbf{p}}(t)=\int\text{d}_{q}%
^{3}x\,u^{\mathbf{p}^{\prime}}\hspace{-0.02in}(\mathbf{x})\circledast
V(\mathbf{x},t)\circledast(u^{\ast})^{\mathbf{p}}(\mathbf{x}).
\end{equation}
Due to Eqs.~(\ref{DifGleZeiSto1}), (\ref{DifGleZeiSto2}), and
(\ref{DifGleZeiSto3}), the expansion coefficients satisfy the integral
equations%
\begin{align}
c_{\hspace{0.01in}\mathbf{p}^{\prime}}(t)  &  =c_{\hspace{0.01in}%
\mathbf{p}^{\prime}}(0)+\text{i}^{-1}\hspace{-0.02in}\int\limits_{0}%
^{t}\text{d}\tau\hspace{-0.02in}\int\text{d}_{q}^{3}p\,(V_{I})_{\mathbf{p}%
^{\prime}\mathbf{p}}(\tau)\circledast c_{\hspace{0.01in}\mathbf{p}}%
(\tau),\nonumber\\
c^{\hspace{0.01in}\mathbf{p}^{\prime}}\hspace{-0.02in}(t)  &  =c^{\hspace
{0.01in}\mathbf{p}^{\prime}}(0)+\text{i\hspace{-0.01in}}\int\limits_{0}%
^{t}\text{d}\tau\hspace{-0.02in}\int\text{d}_{q}^{3}p\,c^{\hspace
{0.01in}\mathbf{p}}(\tau)\circledast(V_{I})^{\mathbf{p\hspace{0.01in}%
p}^{\prime}}\hspace{-0.02in}(\tau),
\end{align}
and%
\begin{align}
(c^{\ast})^{\mathbf{p}^{\prime}}\hspace{-0.02in}(t)  &  =(c^{\ast
})^{\mathbf{p}^{\prime}}(0)+\text{i}^{-1}\hspace{-0.02in}\int\limits_{0}%
^{t}\text{d}\tau\hspace{-0.02in}\int\text{d}_{q}^{3}p\,(V_{I})^{\mathbf{p}%
^{\prime}\mathbf{p}}(\tau)\circledast(c^{\ast})^{\mathbf{p}}(\tau),\nonumber\\
(c^{\ast})_{\mathbf{p}^{\prime}}(t)  &  =(c^{\ast})_{\mathbf{p}^{\prime}%
}(0)+\text{i\hspace{-0.01in}}\int\limits_{0}^{t}\text{d}\tau\hspace
{-0.02in}\int\text{d}_{q}^{3}p\,(c^{\ast})_{\mathbf{p}}(\tau)\circledast
(V_{I})_{\mathbf{p\hspace{0.01in}p}^{\prime}}(\tau)
\end{align}
with%
\begin{align}
(V_{I})_{\mathbf{p}^{\prime}\mathbf{p}}(t)  &  =\operatorname*{e}%
\nolimits^{\text{i}t\varepsilon_{\mathbf{p}^{\prime}}}\hspace{-0.02in}%
\circledast\,V_{\mathbf{p}^{\prime}\mathbf{p}}(t)\circledast\operatorname*{e}%
\nolimits^{-\text{i}t\varepsilon_{\mathbf{p}}},\nonumber\\
(V_{I})^{\mathbf{p}^{\prime}\mathbf{p}}(t)  &  =\operatorname*{e}%
\nolimits^{\text{i}t\varepsilon_{\mathbf{p}^{\prime}}}\hspace{-0.02in}%
\circledast\,V^{\mathbf{p}^{\prime}\mathbf{p}}(t)\circledast\operatorname*{e}%
\nolimits^{-\text{i}t\varepsilon_{\mathbf{p}}}.
\end{align}
Solving the above integral equations by iteration, we obtain the perturbation
expansions%
\begin{align}
c_{\hspace{0.01in}\mathbf{p}}(t)=\,  &  c_{\hspace{0.01in}\mathbf{p}}%
(0)+\sum_{n\hspace{0.01in}=\hspace{0.01in}1}^{\infty}\text{i}^{-n}%
\hspace{-0.02in}\int\limits_{0\hspace{-0.02in}}^{t}\text{d}t_{n}\ldots
\hspace{-0.01in}\int\limits_{0}^{t_{2}}\text{d}t_{1}\int\text{d}p_{n}%
\ldots\hspace{0.01in}\text{d}p_{1}\,(V_{I})_{\mathbf{p\hspace{0.01in}p}%
_{n}\hspace{0.01in}}(t_{n})\nonumber\\
&  \qquad\qquad\qquad\circledast\ldots\circledast(V_{I})_{\mathbf{p}%
_{2}\hspace{0.01in}\mathbf{p}_{1}}(t_{1})\circledast c_{\hspace{0.01in}%
\mathbf{p}_{1}}(0),\\[0.1in]
c^{\hspace{0.01in}\mathbf{p}}(t)=\,  &  c^{\hspace{0.01in}\mathbf{p}}%
(0)+\sum_{n\hspace{0.01in}=\hspace{0.01in}1}^{\infty}\text{i}^{n}%
\hspace{-0.02in}\int\limits_{0}^{t}\text{d}t_{n}\ldots\hspace{-0.01in}%
\int\limits_{0}^{t_{2}}\text{d}t_{1}\int\text{d}p_{n}\ldots\hspace
{0.01in}\text{d}p_{1}\,c^{\hspace{0.01in}\mathbf{p}_{1}}(0)\nonumber\\
&  \qquad\qquad\qquad\circledast(V_{I})^{\mathbf{p}_{1}\mathbf{p}_{2}}%
(t_{1})\circledast\ldots\circledast(V_{I})^{\mathbf{p}_{n}\mathbf{p}}(t_{n}),
\end{align}
and%
\begin{align}
(c^{\ast})^{\mathbf{p}}(t)=\,  &  (c^{\ast})^{\mathbf{p}}(0)+\sum
_{n\hspace{0.01in}=\hspace{0.01in}1}^{\infty}\text{i}^{-n}\int\limits_{0}%
^{t}\text{d}t_{n}\ldots\hspace{-0.01in}\int\limits_{0}^{t_{2}}\text{d}%
t_{1}\int\text{d}p_{n}\ldots\hspace{0.01in}\text{d}p_{1}\,(V_{I}%
)^{\mathbf{p\hspace{0.01in}p}_{n}}(t_{n})\nonumber\\
&  \qquad\qquad\qquad\circledast\ldots\circledast(V_{I})^{\mathbf{p}%
_{2}\hspace{0.01in}\mathbf{p}_{1}}(t_{1})\circledast(c^{\ast})^{\mathbf{p}%
_{1}}(0),\\[0.1in]
(c^{\ast})_{\mathbf{p}}(t)=\,  &  (c^{\ast})_{\mathbf{p}}(0)+\sum
_{n\hspace{0.01in}=\hspace{0.01in}1}^{\infty}\text{i}^{n}\hspace{-0.02in}%
\int\limits_{0}^{t}\text{d}t_{n}\ldots\hspace{-0.01in}\int\limits_{0}^{t_{2}%
}\text{d}t_{1}\int\text{d}p_{n}\ldots\hspace{0.01in}\text{d}p_{1}\,(c^{\ast
})_{\mathbf{p}_{1}}(0)\nonumber\\
&  \qquad\qquad\qquad\circledast(V_{I})_{\mathbf{p}_{1}\mathbf{p}_{2}}%
(t_{1})\circledast\ldots\circledast(V_{I})_{\mathbf{p}_{n}\mathbf{p}}(t_{n}).
\end{align}

\appendix

\section{Star-products\label{KapQuaZeiEle}}

The three-di\-men\-sion\-al $q$-de\-formed Euclidean space $\mathbb{R}_{q}%
^{3}$ has the generators $X^{+}$, $X^{3}$, and $X^{-}$, subject to the
following commutation relations \cite{Lorek:1997eh}:%
\begin{align}
X^{3}X^{+}  &  =q^{2}X^{+}X^{3},\nonumber\\
X^{3}X^{-}  &  =q^{-2}X^{-}X^{3},\nonumber\\
X^{-}X^{+}  &  =X^{+}X^{-}+(q-q^{-1})\hspace{0.01in}X^{3}X^{3}.
\label{RelQuaEukDre}%
\end{align}
We can extend the algebra of $\mathbb{R}_{q}^{3}$ by a time element $X^{0}$,
which commutes with the generators $X^{+}$, $X^{3}$, and $X^{-}$
\cite{Wachter:2020A}:%
\begin{equation}
X^{0}X^{A}=X^{A}X^{0},\text{\qquad}A\in\{+,3,-\}. \label{ZusRelExtDreEukQUa}%
\end{equation}
In the following, we refer to the algebra spanned by the generators $X^{i}$
with $i\in\{0,+,3,-\}$ as $\mathbb{R}_{q}^{3,t}$.

There is a $q$-ana\-log of the three-di\-men\-sion\-al Euclidean metric
$g^{AB}$ with its inverse $g_{AB}$ \cite{Lorek:1997eh} (rows and columns are
arranged in the order $+,3,-$):%
\begin{equation}
g_{AB}=g^{AB}=\left(
\begin{array}
[c]{ccc}%
0 & 0 & -\hspace{0.01in}q\\
0 & 1 & 0\\
-\hspace{0.01in}q^{-1} & 0 & 0
\end{array}
\right)  . \label{DreDimMet}%
\end{equation}
We can use the $q$-de\-formed metric to raise and lower indices:%
\begin{equation}
X_{A}=g_{AB}\hspace{0.01in}X^{B},\qquad X^{A}=g^{AB}X_{B}. \label{HebSenInd}%
\end{equation}

The algebra $\mathbb{R}_{q}^{3,t}$ has a semilinear, involutive, and
anti-multiplicative mapping, which we call \textit{quantum space conjugation}.
If we indicate conjugate elements of a quantum space by a bar,\footnote{A bar
over a complex number indicates complex conjugation.} we can write the
properties of quantum space conjugation as follows ($\alpha,\beta\in
\mathbb{C}$ and $u,v\in\mathbb{R}_{q}^{3,t}$):%
\begin{equation}
\overline{\alpha\,u+\beta\,v}=\overline{\alpha}\,\overline{u}+\overline{\beta
}\,\overline{v},\quad\overline{\overline{u}}=u,\quad\overline{u\,v}%
=\overline{v}\,\overline{u}.
\end{equation}
The conjugation on the algebra $\mathbb{R}_{q}^{3,t}$ is compatible with the
commutation relations in Eqs.~(\ref{RelQuaEukDre}) and
(\ref{ZusRelExtDreEukQUa}) if the following applies \cite{Wachter:2020A}:%
\begin{equation}
\overline{X^{A}}=X_{A}=g_{AB}\hspace{0.01in}X^{B},\qquad\overline{X^{0}}%
=X_{0}. \label{ConSpaKoo}%
\end{equation}

We can only prove a physical theory if it predicts measurement results. The
problem, however, is: How can we associate the elements of the noncommutative
space $\mathbb{R}_{q}^{3,t}$ with real numbers? One solution to this problem
is to introduce a vector space isomorphism between the noncommutative algebra
$\mathbb{R}_{q}^{3,t}$ and a corresponding commutative coordinate algebra
$\mathbb{C}[\hspace{0.01in}x^{+},x^{3},x^{-},t\hspace{0.01in}]$.

We can write each element
$F\in$ $\mathbb{R}_{q}^{3,t}$ uniquely as a finite or infinite linear
combination of monomials with a given normal ordering
(\textit{Poincar\'{e}-Birkhoff-Witt property}):%
\begin{equation}
F=\sum\limits_{n_{+},\ldots,\hspace{0.01in}n_{0}}a_{\hspace{0.01in}n_{+}%
\ldots\hspace{0.01in}n_{0}}\,(X^{+})^{n_{+}}(X^{3})^{n_{3}}(X^{-})^{n_{-}%
}(X^{0})^{n_{0}},\quad\quad a_{\hspace{0.01in}n_{+}\ldots\hspace{0.01in}n_{0}%
}\in\mathbb{C}.
\end{equation}
For this reason, we can define a vector
space isomorphism%
\begin{equation}
\mathcal{W}:\mathbb{C}[\hspace{0.01in}x^{+},x^{3},x^{-},t\hspace
{0.01in}]\rightarrow\mathbb{R}_{q}^{3,t} \label{VecRauIsoInv}%
\end{equation}
with%
\begin{equation}
\mathcal{W}\left(  (x^{+})^{n_{+}}(x^{3})^{n_{3}}(x^{-})^{n_{-}}%
\hspace{0.01in}t^{\hspace{0.01in}n_{0}}\right)  =(X^{+})^{n_{+}}(X^{3}%
)^{n_{3}}(X^{-})^{n_{-}}(X^{0})^{n_{0}}. \label{StePro0}%
\end{equation}
In general, we have%
\begin{equation}
\mathbb{C}[\hspace{0.01in}x^{+},x^{3},x^{-},t\hspace{0.01in}]\ni f\mapsto
F\in\mathbb{R}_{q}^{3,t},
\end{equation}
where%
\begin{align}
f  &  =\sum\limits_{n_{+},\ldots,\hspace{0.01in}n_{0}}a_{\hspace{0.01in}%
n_{+}\ldots\hspace{0.01in}n_{0}}\,(x^{+})^{n_{+}}(x^{3})^{n_{3}}(x^{-}%
)^{n_{-}}\hspace{0.01in}t^{\hspace{0.01in}n_{0}},\nonumber\\
F  &  =\sum\limits_{n_{+},\ldots,\hspace{0.01in}n_{0}}a_{\hspace{0.01in}%
n_{+}\ldots\hspace{0.01in}n_{0}}\,(X^{+})^{n_{+}}(X^{3})^{n_{3}}(X^{-}%
)^{n_{-}}(X^{0})^{n_{0}}. \label{AusFfNorOrd}%
\end{align}
The vector space isomorphism $\mathcal{W}$ is nothing else than the
\textit{Moy\-al-Weyl mapping}, which gives an operator $F$ to a complex valued
function $f$
\cite{Bayen:1977ha,1997q.alg.....9040K,Madore:2000en,Moyal:1949sk}.

We can extend this vector space isomorphism to an algebra isomorphism if we
introduce a new product on the commutative coordinate algebra. This so-called
\textit{star-pro\-duct }symbolized by $\circledast$ satisfies the following
homomorphism condition:%
\begin{equation}
\mathcal{W}\left(  f\circledast g\right)  =\mathcal{W}\left(  f\right)
\cdot\mathcal{W}\left(  \hspace{0.01in}g\right)  . \label{HomBedWeyAbb}%
\end{equation}
Since the Moy\-al-Weyl mapping is invertible, we can write the star-prod\-uct
as follows:%
\begin{equation}
f\circledast g=\mathcal{W}^{\hspace{0.01in}-1}\big (\,\mathcal{W}\left(
f\right)  \cdot\mathcal{W}\left(  \hspace{0.01in}g\right)  \big ).
\label{ForStePro}%
\end{equation}

To get explicit formulas for calculating star-prod\-ucts, we first have to
write a noncommutative product of two nor\-mal-or\-dered monomials as a linear
combination of nor\-mal-or\-dered monomials again (see
Ref.~\cite{Wachter:2002A} for details):%
\begin{equation}
(X^{+})^{n_{+}}\ldots\hspace{0.01in}(X^{0})^{n_{0}}\cdot(X^{+})^{m_{+}}%
\ldots\hspace{0.01in}(X^{0})^{m_{0}}=\sum_{\underline{k}\hspace{0.01in}%
=\hspace{0.01in}0}B_{\underline{k}}^{\hspace{0.01in}\underline{n}%
,\underline{m}}\,(X^{+})^{k_{+}}\ldots\hspace{0.01in}(X^{0})^{k_{0}}.
\label{EntProMon}%
\end{equation}
We achieve this by using the commutation relations for the noncommutative
coordinates [cf. Eq.~(\ref{RelQuaEukDre})]. From the concrete form of the
expansion in\ Eq.~(\ref{EntProMon}), we can finally read off a formula to
calculate the star-prod\-uct of two power series in commutative space-time
coordinates ($\lambda=q-q^{-1}$):\footnote{The argument $\mathbf{x}$ indicates
a dependence on the spatial coordinates $x^{+}$, $x^{3}$, and $x^{-}$.}%
\begin{gather}
f(\mathbf{x},t)\circledast g(\mathbf{x},t)=\nonumber\\
\sum_{k\hspace{0.01in}=\hspace{0.01in}0}^{\infty}\lambda^{k}\hspace
{0.01in}\frac{(x^{3})^{2k}}{[[k]]_{q^{4}}!}\,q^{2(\hat{n}_{3}\hspace
{0.01in}\hat{n}_{+}^{\prime}+\,\hat{n}_{-}\hat{n}_{3}^{\prime})}%
D_{q^{4},\hspace{0.01in}x^{-}}^{k}f(\mathbf{x},t)\,D_{q^{4},\hspace
{0.01in}x^{\prime+}}^{k}g(\mathbf{x}^{\prime},t)\big|_{x^{\prime}%
\rightarrow\hspace{0.01in}x}. \label{StaProForExp}%
\end{gather}
The expression\ above depends on the operators%
\begin{equation}
\hat{n}_{A}=x^{A}\frac{\partial}{\partial x^{A}}%
\end{equation}
and the so-called Jackson derivatives \cite{Jackson:1910yd}:%
\begin{equation}
D_{q^{k},\hspace{0.01in}x}\,f=\frac{f(q^{k}x)-f(x)}{q^{k}x-x}.
\end{equation}
The $q$-numbers are given by%
\begin{equation}
\lbrack\lbrack a]]_{q}=\frac{1-q^{a}}{1-q}, \label{qNumDef}%
\end{equation}
and the $q$-factorials are defined in complete analogy to the undeformed case:%
\begin{equation}
\lbrack\lbrack n]]_{q}!=[[1]]_{q}\hspace{0.01in}[[2]]_{q}\ldots\lbrack
\lbrack\hspace{0.01in}n-1]]_{q}\hspace{0.01in}[[n]]_{q},\qquad\lbrack
\lbrack0]]_{q}!=1.
\end{equation}

The algebra isomorphism $\mathcal{W}^{-1}$ enables us to carry over the
conjugation for the quantum space algebra $\mathbb{R}_{q}^{3,t}$ to the
commutative coordinate algebra $\mathbb{C}[\hspace{0.01in}x^{+},x^{3}%
,x^{-},t\hspace{0.01in}]$. In other words, the mapping $\mathcal{W}%
^{\hspace{0.01in}-1}$ is a $\ast$-al\-ge\-bra homomorphism:%
\begin{equation}
\mathcal{W}(\hspace{0.01in}\overline{f}\hspace{0.01in})=\overline
{\mathcal{W}(f)}\qquad\Leftrightarrow\text{\qquad}\overline{f}=\mathcal{W}%
^{-1}\big (\hspace{0.01in}\overline{\mathcal{W}(f)}\hspace{0.01in}\big ).
\label{ConAlgIso}%
\end{equation}
This relationship implies the following property for the star-pro\-duct:%
\begin{equation}
\overline{f\circledast g}=\overline{g}\circledast\overline{f}.
\label{KonEigSteProFkt}%
\end{equation}

With $\bar{f}$, we designate the power series obtained from $f$ by quantum
space conjugation. It follows from Eq.~(\ref{ConSpaKoo})\ and
Eq.~(\ref{ConAlgIso}) that $\bar{f}$ takes the following form (if $\bar
{a}_{n_{+},n_{3},n_{-},n_{0}}$ stands for the complex conjugate of
$a_{n_{+},n_{3},n_{-},n_{0}}$) \cite{Wachter:2007A,Wachter:2020A}:%
\begin{align}
\overline{f(\mathbf{x},t)}  &  =\sum\nolimits_{\underline{n}}\bar{a}%
_{n_{+},n_{3},n_{-},n_{0}}\,(-\hspace{0.01in}q\hspace{0.01in}x^{-})^{n_{+}%
}(\hspace{0.01in}x^{3})^{n_{3}}(-\hspace{0.01in}q^{-1}x^{+})^{n_{-}}%
\hspace{0.01in}t^{\hspace{0.01in}n_{0}}\nonumber\\
&  =\sum\nolimits_{\underline{n}}(-\hspace{0.01in}q)^{n_{-}-\hspace
{0.02in}n_{+}}\hspace{0.01in}\bar{a}_{n_{-},n_{3},n_{+},n_{0}}\,(\hspace
{0.01in}x^{+})^{n_{+}}(\hspace{0.01in}x^{3})^{n_{3}}(\hspace{0.01in}%
x^{-})^{n_{-}}\hspace{0.01in}t^{\hspace{0.01in}n_{0}}\nonumber\\
&  =\bar{f}(\mathbf{x},t). \label{KonPotReiKom}%
\end{align}

\section{Partial derivatives and integrals\label{KapParDer}}

There are partial derivatives for $q$-de\-formed space-time coordinates
\cite{CarowWatamura:1990zp,Wess:1990vh}. These partial derivatives again form
a quantum space with the same algebraic structure as that of the
$q$-de\-formed space-time coordinates. Thus, the $q$-de\-formed partial
derivatives $\partial_{i}$ satisfy the same commutation relations as the
covariant coordinate generators $X_{i}$:%
\begin{gather}
\partial_{0}\hspace{0.01in}\partial_{+}=\hspace{0.01in}\partial_{+}%
\hspace{0.01in}\partial_{0},\quad\partial_{0}\hspace{0.01in}\partial
_{-}=\hspace{0.01in}\partial_{-}\hspace{0.01in}\partial_{0},\quad\partial
_{0}\hspace{0.01in}\partial_{3}=\partial_{3}\hspace{0.01in}\partial
_{0},\nonumber\\
\partial_{+}\hspace{0.01in}\partial_{3}=q^{2}\partial_{3}\hspace
{0.01in}\partial_{+},\quad\partial_{3}\hspace{0.01in}\partial_{-}%
=\hspace{0.01in}q^{2}\partial_{-}\hspace{0.01in}\partial_{3},\nonumber\\
\partial_{+}\hspace{0.01in}\partial_{-}-\partial_{-}\hspace{0.01in}%
\partial_{+}=\hspace{0.01in}\lambda\hspace{0.01in}\partial_{3}\hspace
{0.01in}\partial_{3}.
\end{gather}
The commutation relations above are invariant under conjugation if the
derivatives show the following conjugation properties:\footnote{The indices of
partial derivatives are raised and lowered in the same way as those of
coordinates [see Eq.~(\ref{HebSenInd}) in App.~\ref{KapQuaZeiEle}].}%
\begin{equation}
\overline{\partial_{A}}=-\hspace{0.01in}\partial^{A}=-g^{AB}\partial
_{B},\qquad\overline{\partial_{0}}=-\hspace{0.01in}\partial^{0}=-\hspace
{0.01in}\partial_{0}. \label{KonAbl}%
\end{equation}

There are two ways of commuting $q$-de\-formed partial derivatives with
$q$-de\-formed space-time coordinates. One is given by the following
$q$-de\-formed Leibniz rules
\cite{CarowWatamura:1990zp,Wess:1990vh,Wachter:2020A}:%
\begin{align}
\partial_{B}X^{A}  &  =\delta_{B}^{A}+q^{4}\hat{R}{^{AC}}_{BD}\,X^{D}%
\partial_{C},\nonumber\\
\partial_{A}X^{0}  &  =X^{0}\hspace{0.01in}\partial_{A},\nonumber\\
\partial_{0}\hspace{0.01in}X^{A}  &  =X^{A}\hspace{0.01in}\partial
_{0},\nonumber\\
\partial_{0}\hspace{0.01in}X^{0}  &  =1+X^{0}\hspace{0.01in}\partial_{0}.
\label{DifKalExtEukQuaDreUnk}%
\end{align}
$\hat{R}{^{AC}}_{BD}$ denotes the vector representation of the
R-ma\-trix for the three-di\-men\-sion\-al $q$-de\-formed Euclidean space \cite{Lorek:1993tq}.
Introducing
$\hat{\partial}_{A}=q^{6}\partial_{A}$ and $\hat{\partial}_{0}=\partial_{0}$,
we can write the Leibniz rules for the second differential calculus in the
following form:
\begin{align}
\hat{\partial}_{B}\hspace{0.01in}X^{A}  &  =\delta_{B}^{A}+q^{-4}(\hat{R}%
^{-1}){^{AC}}_{BD}\,X^{D}\hat{\partial}_{C},\nonumber\\
\hat{\partial}_{A}\hspace{0.01in}X^{0}  &  =X^{0}\hspace{0.01in}\hat{\partial
}_{A},\nonumber\\
\hat{\partial}_{0}\hspace{0.01in}X^{A}  &  =X^{A}\hspace{0.01in}\hat{\partial
}_{0},\nonumber\\
\hat{\partial}_{0}\hspace{0.01in}X^{0}  &  =1+X^{0}\hspace{0.01in}%
\hat{\partial}_{0}. \label{DifKalExtEukQuaDreKon}%
\end{align}

Using the Leibniz rules in Eq.$~$(\ref{DifKalExtEukQuaDreUnk}) or
Eq.$~$(\ref{DifKalExtEukQuaDreKon}), we can calculate how partial derivatives
act on nor\-mal-or\-dered monomials of noncommutative coordinates. We can
carry over these actions to commutative coordinate monomials with the help of
the Mo\-yal-Weyl mapping:%
\begin{equation}
\partial^{i}\triangleright(x^{+})^{n_{+}}(x^{3})^{n_{3}}(x^{-})^{n_{-}}%
\hspace{0.01in}t^{\hspace{0.01in}n_{0}}=\mathcal{W}^{\hspace{0.01in}%
-1}\big (\partial^{i}\triangleright(X^{+})^{n_{+}}(X^{3})^{n_{3}}%
(X^{-})^{n_{-}}(X^{0})^{n_{0}}\big ).
\end{equation}
Since the Mo\-yal-Weyl mapping is linear, we can apply the action above to
space-time functions that can be written as a power series:%
\begin{equation}
\partial^{i}\triangleright f(\mathbf{x},t)=\mathcal{W}^{\hspace{0.01in}%
-1}\big (\partial^{i}\triangleright\mathcal{W}(f(\mathbf{x},t))\big ).
\end{equation}

If we use the ordering given in Eq.~(\ref{StePro0}),
the Leibniz rules in Eq.~(\ref{DifKalExtEukQuaDreUnk})   lead to the
following operator representations \cite{Bauer:2003}:%
\begin{align}
\partial_{+}\triangleright f(\mathbf{x},t)  &  =D_{q^{4},\hspace{0.01in}x^{+}%
}f(\mathbf{x},t),\nonumber\\
\partial_{3}\triangleright f(\mathbf{x},t)  &  =D_{q^{2},\hspace{0.01in}x^{3}%
}f(q^{2}x^{+},x^{3},x^{-},t),\nonumber\\
\partial_{-}\triangleright f(\mathbf{x},t)  &  =D_{q^{4},\hspace{0.01in}x^{-}%
}f(x^{+},q^{2}x^{3},x^{-},t)+\lambda\hspace{0.01in}x^{+}D_{q^{2}%
,\hspace{0.01in}x^{3}}^{2}f(\mathbf{x},t). \label{UnkOpeDarAbl}%
\end{align}
The derivative $\partial_{0}$, however, is represented on the commutative
space-time algebra by an ordinary partial derivative:%
\begin{equation}
\partial_{0}\triangleright\hspace{-0.01in}f(\mathbf{x},t)=\partial
_{t}\triangleright\hspace{-0.01in}f(\mathbf{x},t)=\frac{\partial
f(\mathbf{x},t)}{\partial t}. \label{OpeDarZeiAblExtQuaEuk}%
\end{equation}

Using the Leibniz rules in Eq.$~$(\ref{DifKalExtEukQuaDreKon}), we get
operator representations for the partial derivatives $\hat{\partial}_{i}$. The
Leibniz rules in Eq.$~$(\ref{DifKalExtEukQuaDreUnk}) transform into those in Eq.$~$%
(\ref{DifKalExtEukQuaDreKon}) by the following
substitutions:%
\begin{gather}
q\rightarrow q^{-1},\quad X^{-}\rightarrow X^{+},\quad X^{+}\rightarrow
X^{-},\nonumber\\
\partial^{\hspace{0.01in}+}\rightarrow\hat{\partial}^{\hspace{0.01in}-}%
,\quad\partial^{\hspace{0.01in}-}\rightarrow\hat{\partial}^{\hspace{0.01in}%
+},\quad\partial^{\hspace{0.01in}3}\rightarrow\hat{\partial}^{\hspace
{0.01in}3},\quad\partial^{\hspace{0.01in}0}\rightarrow\hat{\partial}%
^{\hspace{0.01in}0}. \label{UebRegGedUngAblDreQua}%
\end{gather}
For this reason, we obtain the operator representations of the partial
derivatives $\hat{\partial}_{A}$ from those of the partial derivatives
$\partial_{A}$ [cf. Eq.~(\ref{UnkOpeDarAbl})] if we replace $q$ with $q^{-1}$
and exchange the indices $+$ and $-$:%
\begin{align}
\hat{\partial}_{-}\,\bar{\triangleright}\,f(\mathbf{x},t)  &  =D_{q^{-4}%
,\hspace{0.01in}x^{-}}f(\mathbf{x},t),\nonumber\\
\hat{\partial}_{3}\,\bar{\triangleright}\,f(\mathbf{x},t)  &  =D_{q^{-2}%
,\hspace{0.01in}x^{3}}f(q^{-2}x^{-},x^{3},x^{+},t),\nonumber\\
\hat{\partial}_{+}\,\bar{\triangleright}\,f(\mathbf{x},t)  &  =D_{q^{-4}%
,\hspace{0.01in}x^{+}}f(x^{-},q^{-2}x^{3},x^{+},t)-\lambda\hspace{0.01in}%
x^{-}D_{q^{-2},\hspace{0.01in}x^{3}}^{2}f(\mathbf{x},t). \label{KonOpeDarAbl}%
\end{align}
Once again, $\hat{\partial}_{0}$ is represented on the commutative space-time
algebra by an ordinary partial derivative:%
\begin{equation}
\hat{\partial}_{0}\,\bar{\triangleright}\,f(\mathbf{x},t)=\partial
_{t}\triangleright\hspace{-0.01in}f(\mathbf{x},t)=\frac{\partial
f(\mathbf{x},t)}{\partial t}. \label{OpeDarZeiAblExtQuaEukKon}%
\end{equation}
Due to the substitutions given in\ Eq.~(\ref{UebRegGedUngAblDreQua}), the
actions in Eqs.~(\ref{KonOpeDarAbl}) and (\ref{OpeDarZeiAblExtQuaEukKon})
refer to nor\-mal-or\-dered monomials different from those in
Eq.~(\ref{StePro0}):%
\begin{equation}
\widetilde{\mathcal{W}}\left(  t^{\hspace{0.01in}n_{0}}(x^{+})^{n_{+}}%
(x^{3})^{n_{3}}(x^{-})^{n_{-}}\right)  =(X^{0})^{n_{0}}(X^{-})^{n_{-}}%
(X^{3})^{n_{3}}(X^{+})^{n_{+}}. \label{UmNor}%
\end{equation}

We should not forget that we can also commute $q$-de\-formed partial
derivatives from the \textit{right} side of a nor\-mal-or\-dered monomial to
the left side by using the Leibniz rules. This way, we get\ so-called
\textit{right} representations of partial derivatives, for which we write
$f\,\bar{\triangleleft}\,\partial^{i}$ or $f\triangleleft\hat{\partial}^{i}$.
Note that the operation of conjugation transforms left actions of partial
derivatives into right actions and vice versa \cite{Bauer:2003}:%
\begin{align}
\overline{\partial^{i}\triangleright f}  &  =-\bar{f}\,\bar{\triangleleft
}\,\partial_{i}, & \overline{f\,\bar{\triangleleft}\,\partial^{i}}  &
=-\hspace{0.01in}\partial_{i}\triangleright\bar{f},\nonumber\\
\overline{\hat{\partial}^{i}\,\bar{\triangleright}\,f}  &  =-\bar
{f}\triangleleft\hat{\partial}_{i}, & \overline{f\triangleleft\hat{\partial
}^{i}}  &  =-\hspace{0.01in}\hat{\partial}_{i}\,\bar{\triangleright}\,\bar{f}.
\label{RegConAbl}%
\end{align}

In general, the operator representations in Eqs.~(\ref{UnkOpeDarAbl}) and
(\ref{KonOpeDarAbl}) consist of two terms, which we call $\partial
_{\operatorname*{cla}}^{A}$ and $\partial_{\operatorname*{cor}}^{A}$:%
\begin{equation}
\partial^{A}\triangleright F=\left(  \partial_{\operatorname*{cla}}%
^{A}+\partial_{\operatorname*{cor}}^{A}\right)  \triangleright F.
\end{equation}
In the undeformed limit $q\rightarrow1$, $\partial_{\operatorname*{cla}}^{A}$
becomes an ordinary partial derivative, and $\partial_{\operatorname*{cor}%
}^{A}$ disappears. We get a solution to the difference equation $\partial
^{A}\triangleright F=f$ with given $f$ by using the following formula
\cite{Wachter:2004A}:%
\begin{align}
F  &  =(\partial^{A})^{-1}\triangleright f=\left(  \partial
_{\operatorname*{cla}}^{A}+\partial_{\operatorname*{cor}}^{A}\right)
^{-1}\triangleright f\nonumber\\
&  =\sum_{k\hspace{0.01in}=\hspace{0.01in}0}^{\infty}\left[  -(\partial
_{\operatorname*{cla}}^{A})^{-1}\partial_{\operatorname*{cor}}^{A}\right]
^{k}(\partial_{\operatorname*{cla}}^{A})^{-1}\triangleright f.
\end{align}
Applying the above formula to the operator representations in
Eq.~(\ref{UnkOpeDarAbl}), we get%
\begin{align}
(\partial_{+})^{-1}\triangleright f(\mathbf{x},t)  &  =D_{q^{4},\hspace
{0.01in}x^{+}}^{-1}f(\mathbf{x},t),\nonumber\\
(\partial_{3})^{-1}\triangleright f(\mathbf{x},t)  &  =D_{q^{2},\hspace
{0.01in}x^{3}}^{-1}f(q^{-2}x^{+},x^{3},x^{-},t), \label{InvParAbl1}%
\end{align}
and%
\begin{gather}
(\partial_{-})^{-1}\triangleright f(\mathbf{x},t)=\nonumber\\
=\sum_{k\hspace{0.01in}=\hspace{0.01in}0}^{\infty}q^{2k\left(  k\hspace
{0.01in}+1\right)  }\left(  -\lambda\,x^{+}D_{q^{4},\hspace{0.01in}x^{-}}%
^{-1}D_{q^{2},\hspace{0.01in}x^{3}}^{2}\right)  ^{k}D_{q^{4},\hspace
{0.01in}x^{-}}^{-1}f(x^{+},q^{-2\left(  k\hspace{0.01in}+1\right)  }%
x^{3},x^{-},t). \label{InvParAbl2}%
\end{gather}
Note that $D_{q,\hspace{0.01in}x}^{-1}$ stands for a Jackson integral with $x$
being the variable of integration \cite{Jackson:1908}. The explicit form of
this Jackson integral depends on its limits of integration and the value for
the deformation parameter $q$. If $x>0$ and $q>1$, for example, the following
applies:%
\begin{equation}
\int_{0}^{\hspace{0.01in}x}\text{d}_{q}z\hspace{0.01in}f(z)=(q-1)\hspace
{0.01in}x\sum_{j=1}^{\infty}q^{-j}f(q^{-j}x).
\end{equation}
The integral for the time coordinate is an ordinary integral since
$\partial_{0}$ acts on the commutative space-time algebra like an ordinary
partial derivative [cf. Eq.~(\ref{OpeDarZeiAblExtQuaEuk})]:%
\begin{equation}
(\partial_{0})^{-1}\triangleright f(\mathbf{x},t)\hspace{0.01in}=\int
\text{d}t\,f(\mathbf{x},t).
\end{equation}

The above considerations also apply to the partial derivatives with a hat.
However, we can obtain the representations of $\hat{\partial}_{i}$ from those
of the derivatives $\partial_{i}$ if we replace $q$ with $q^{-1}$ and exchange
the indices $+$ and $-$. Applying these substitutions to the expressions in
Eqs.~(\ref{InvParAbl1}) and (\ref{InvParAbl2}), we immediately get the
corresponding results for the partial derivatives $\hat{\partial}_{i}$.

By successively applying the integral operators given in
Eqs.~(\ref{InvParAbl1}) and (\ref{InvParAbl2}), we can explain an integration
over all space \cite{Wachter:2004A,Wachter:2007A}:%
\begin{equation}
\int_{-\infty}^{+\infty}\text{d}_{q}^{3}\hspace{0.01in}x\,f(x^{+},x^{3}%
,x^{-})=(\partial_{-})^{-1}\big |_{-\infty}^{+\infty}\,(\partial_{3}%
)^{-1}\big |_{-\infty}^{+\infty}\,(\partial_{+})^{-1}\big |_{-\infty}%
^{+\infty}\triangleright f.
\end{equation}
On the right-hand side of the above relation, the different integral operators
can be simplified to Jackson integrals \cite{Wachter:2004A,Jambor:2004ph}:%
\begin{equation}
\int_{-\infty}^{+\infty}\text{d}_{q}^{3}\hspace{0.01in}x\,f(\mathbf{x}%
)=D_{q^{2},\hspace{0.01in}x^{-}}^{-1}\big |_{-\infty}^{+\infty}\,D_{q,x^{3}%
}^{-1}\big |_{-\infty}^{+\infty}\,D_{q^{2},\hspace{0.01in}x^{+}}%
^{-1}\big |_{-\infty}^{+\infty}\,f(\mathbf{x}).
\end{equation}
Note that the Jackson integrals in the formula above refer to a smaller
$q$-lat\-tice. Using such a smaller $q$-lat\-tice ensures that our integral
over all space is a scalar with trivial braiding properties
\cite{Kempf:1994yd}. Finally, we mention that the $q$-in\-te\-gral over all
space behaves as follows under quantum space conjugation:%
\begin{equation}
\overline{\int_{-\infty}^{+\infty}\text{d}_{q}^{3}\hspace{0.01in}x\,f}%
=\int_{-\infty}^{+\infty}\text{d}_{q}^{3}\hspace{0.01in}x\,\bar{f}.
\label{KonEigVolInt}%
\end{equation}

\section{Exponentials and translations\label{KapExp}}

A $q$-de\-formed exponential is an eigenfunction of each partial derivative of
a given $q$-de\-formed quantum space
\cite{Majid:1993ud,Schirrmacher:1995,Wachter:2004ExpA}. In the following, we
consider $q$-de\-formed exponentials that are eigenfunctions for left-actions
or right-actions of partial derivatives:%
\begin{align}
\text{i}^{-1}\partial^{A}\triangleright\exp_{q}(\mathbf{x}|\text{i}%
\mathbf{p})  &  =\exp_{q}(\mathbf{x}|\text{i}\mathbf{p})\circledast
p^{A},\nonumber\\
\exp_{q}(\text{i}^{-1}\mathbf{p}|\hspace{0.01in}\mathbf{x})\,\bar
{\triangleleft}\,\partial^{A}\text{i}^{-1}  &  =p^{A}\circledast\exp
_{q}(\text{i}^{-1}\mathbf{p}|\hspace{0.01in}\mathbf{x}). \label{EigGl1N}%
\end{align}
The $q$-ex\-po\-nen\-tials are uniquely defined by the above eigenvalue
equations and the following normalization conditions:%
\begin{align}
\exp_{q}(\mathbf{x}|\text{i}\mathbf{p})|_{x\hspace{0.01in}=\hspace{0.01in}0}
&  =\exp_{q}(\mathbf{x}|\text{i}\mathbf{p})|_{p\hspace{0.01in}=\hspace
{0.01in}0}=1,\nonumber\\
\exp_{q}(\text{i}^{-1}\mathbf{p}|\hspace{0.01in}\mathbf{x})|_{x\hspace
{0.01in}=\hspace{0.01in}0}  &  =\exp_{q}(\text{i}^{-1}\mathbf{p}%
|\hspace{0.01in}\mathbf{x})|_{p\hspace{0.01in}=\hspace{0.01in}0}=1.
\label{NorBedExp}%
\end{align}

Using the operator representation in Eq.~(\ref{UnkOpeDarAbl}), we have found the following expressions for the $q$%
-ex\-po\-nen\-tials of the three-di\-mensional Euclidean quantum space
\cite{Wachter:2004ExpA}:%
\begin{align}
\exp_{q}(\mathbf{x}|\text{i}\mathbf{p})  &  =\sum_{\underline{n}%
\,=\,0}^{\infty}\frac{(x^{+})^{n_{+}}(x^{3})^{n_{3}}(x^{-})^{n_{-}}%
(\text{i}p_{-})^{n_{-}}(\text{i}p_{3})^{n_{3}}(\text{i}p_{+})^{n_{+}}%
}{[[\hspace{0.01in}n_{+}]]_{q^{4}}!\,[[\hspace{0.01in}n_{3}]]_{q^{2}%
}!\,[[\hspace{0.01in}n_{-}]]_{q^{4}}!},\nonumber\\
\exp_{q}(\text{i}^{-1}\mathbf{p}|\mathbf{x})  &  =\sum_{\underline{n}%
\,=\,0}^{\infty}\frac{(\text{i}^{-1}p^{+})^{n_{+}}(\text{i}^{-1}p^{3})^{n_{3}%
}(\text{i}^{-1}p^{-})^{n_{-}}(x_{-})^{n_{-}}(x_{3})^{n_{3}}(x_{+})^{n_{+}}%
}{[[\hspace{0.01in}n_{+}]]_{q^{4}}!\,[[\hspace{0.01in}n_{3}]]_{q^{2}%
}!\,[[\hspace{0.01in}n_{-}]]_{q^{4}}}. \label{ExpEukExp}%
\end{align}
If we substitute $q$ with $q^{-1}$ in both expressions of Eq.~(\ref{ExpEukExp}%
), we get two more $q$-exponentials, which we designate $\overline{\exp}%
_{q}(x|$i$\mathbf{p})$ and $\overline{\exp}_{q}($i$^{-1}\mathbf{p}|x)$. We
obtain the eigenvalue equations and normalization conditions of these two
$q$-exponentials by applying the following substitutions to
Eqs.~(\ref{EigGl1N}) and (\ref{NorBedExp}):%
\begin{equation}
\exp_{q}\rightarrow\overline{\exp}_{q},\qquad\triangleright\,\rightarrow
\,\bar{\triangleright},\qquad\bar{\triangleleft}\,\rightarrow\,\triangleleft
,\qquad\partial^{A}\rightarrow\hat{\partial}^{A}. \label{ErsRegQExp}%
\end{equation}

We can use $q$-ex\-po\-nen\-tials to generate $q$-trans\-la\-tions. If we
replace the momentum coordinates in the expressions for $q$-ex\-po\-nen\-tials
with derivatives, it applies \cite{Carnovale:1999,Majid:1993ud,Wachter:2007A}%
\begin{align}
\exp_{q}(x|\partial_{y})\triangleright g(\hspace{0.01in}y)  &  =g(x\,\bar
{\oplus}\,y),\nonumber\\
\overline{\exp}_{q}(x|\hat{\partial}_{y})\,\bar{\triangleright}\,g(\hspace
{0.01in}y)  &  =g(x\oplus y), \label{q-TayN}%
\end{align}
and%
\begin{align}
g(\hspace{0.01in}y)\,\bar{\triangleleft}\,\exp_{q}(-\hspace{0.01in}%
\partial_{y}|\hspace{0.01in}x)  &  =g(\hspace{0.01in}y\,\bar{\oplus
}\,x),\nonumber\\
g(\hspace{0.01in}y)\triangleleft\hspace{0.01in}\overline{\exp}_{q}%
(-\hspace{0.01in}\hat{\partial}_{y}|\hspace{0.01in}x)  &  =g(\hspace
{0.01in}y\oplus x). \label{q-TayRecN}%
\end{align}
In the case of the three-di\-men\-sion\-al $q$-de\-formed Euclidean space, for
example, we can get the following formula for calculating $q$-trans\-la\-tions
\cite{Wachter:2004phengl}:%
\begin{align}
f(\mathbf{x}\oplus\mathbf{y})=  &  \sum_{i_{+}=\hspace{0.01in}0}^{\infty}%
\sum_{i_{3}=\hspace{0.01in}0}^{\infty}\sum_{i_{-}=\hspace{0.01in}0}^{\infty
}\sum_{k\hspace{0.01in}=\hspace{0.01in}0}^{i_{3}}\frac{(-q^{-1}\lambda
\lambda_{+})^{k}}{[[2k]]_{q^{-2}}!!}\frac{(x^{-})^{i_{-}}(x^{3})^{i_{3}%
-\hspace{0.01in}k}(x^{+})^{i_{+}+\hspace{0.01in}k}\,(y^{-})^{k}}%
{[[i_{-}]]_{q^{-4}}!\,[[i_{3}-k]]_{q^{-2}}!\,[[i_{+}]]_{q^{-4}}!}\nonumber\\
&  \qquad\times\big (D_{q^{-4},\hspace{0.01in}y^{-}}^{i_{-}}D_{q^{-2}%
,\hspace{0.01in}y^{3}}^{i_{3}+\hspace{0.01in}k}\hspace{0.01in}D_{q^{-4}%
,\hspace{0.01in}y^{+}}^{i_{+}}f\big )(q^{2(k\hspace{0.01in}-\hspace
{0.01in}i_{3})}y^{-},q^{-2i_{+}}y^{3}).
\end{align}

\begin{figure}
[ptb]
\begin{center}
\centerline{\psfig{figure=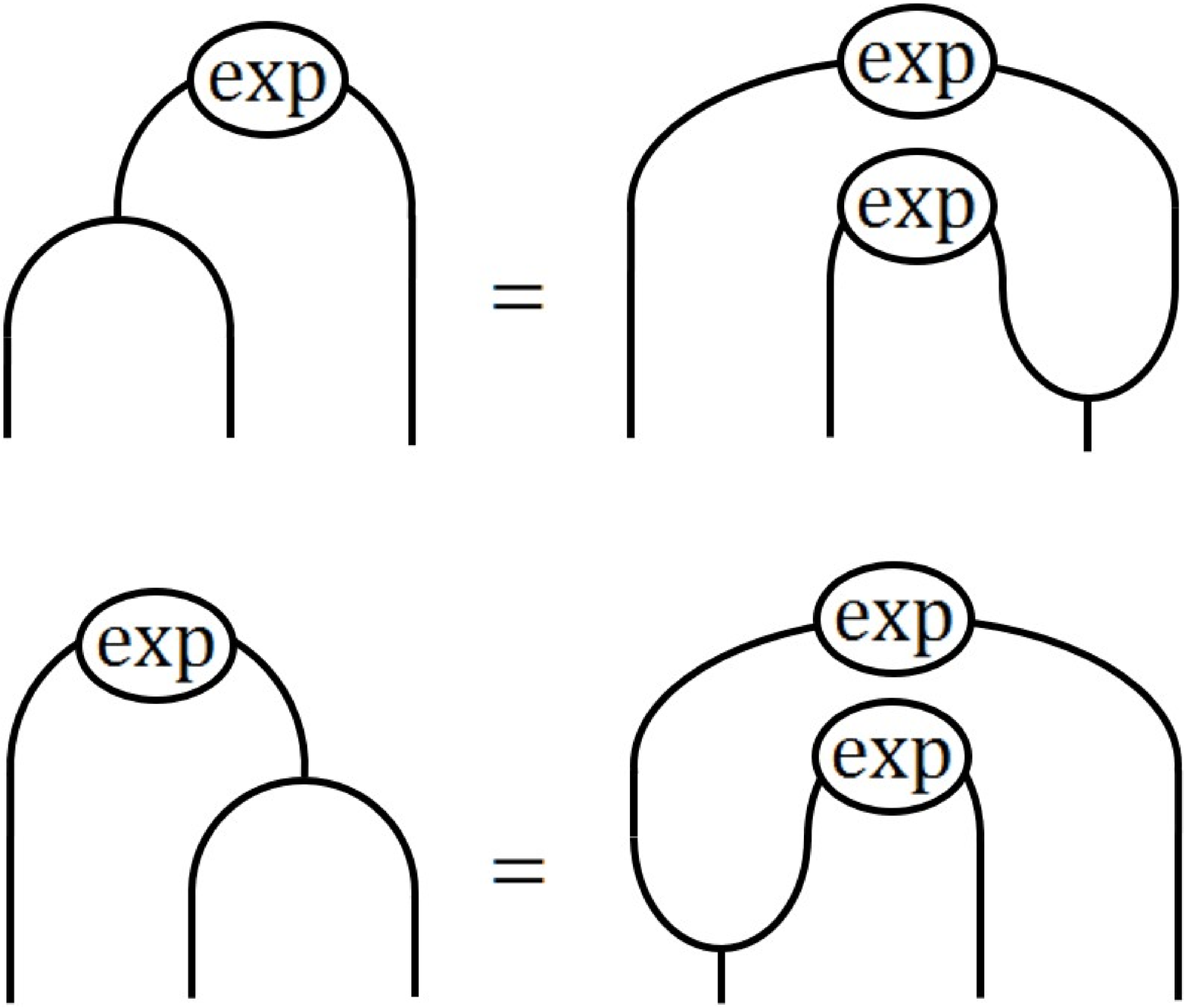,width=1.8827in}}%
\caption{Addition theorems for $q$-exponentials.}%
\label{Fig2}%
\end{center}
\end{figure}

In analogy to the undeformed case, $q$-ex\-po\-nen\-tials satisfy addition
theorems \cite{Majid:1993ud,Schirrmacher:1995,Wachter:2007A}. Concretely, we
have:%
\begin{align}
\exp_{q}(\mathbf{x}\,\bar{\oplus}\,\mathbf{y}|\text{i}\mathbf{p})  &
=\exp_{q}(\mathbf{x}|\exp_{q}(\hspace{0.01in}\mathbf{y}|\text{i}%
\mathbf{p})\circledast\text{i}\mathbf{p}),\nonumber\\
\exp_{q}(\text{i}\mathbf{x}|\mathbf{p}\,\bar{\oplus}\,\mathbf{p}^{\prime})  &
=\exp_{q}(\mathbf{x}\circledast\exp_{q}(\mathbf{x}|\hspace{0.01in}%
\text{i}\mathbf{p})|\hspace{0.01in}\text{i}\mathbf{p}^{\prime}),
\label{AddTheExp}%
\end{align}
and%
\begin{align}
\overline{\exp}_{q}(\mathbf{x}\oplus\mathbf{y}|\text{i}\mathbf{p})  &
=\overline{\exp}_{q}(\mathbf{x}|\overline{\exp}_{q}(\hspace{0.01in}%
\mathbf{y}|\text{i}\mathbf{p})\circledast\text{i}\mathbf{p}),\nonumber\\
\overline{\exp}_{q}(\text{i}\mathbf{x}|\mathbf{p}\oplus\mathbf{p}^{\prime})
&  =\overline{\exp}_{q}(\mathbf{x}\circledast\overline{\exp}_{q}%
(\mathbf{x}|\text{i}\mathbf{p})|\hspace{0.01in}\text{i}\mathbf{p}^{\prime}).
\end{align}
Note that we obtain further addition theorems from the above identities by
substituting position coordinates with momentum coordinates and vice versa.
For a better understanding of the meaning of the two addition theorems in
Eq.~(\ref{AddTheExp}), we have given their graphic representation in
Fig.~\ref{Fig2}.

The $q$-de\-formed quantum spaces we have considered so far are so-called
braided Hopf algebras \cite{Majid:1996kd}. From this point of view, the two
versions of $q$-trans\-lations are nothing else but realizations of two
braided co-pro\-ducts $\underline{\Delta}$ and $\underline{\bar{\Delta}}$ on
the corresponding commutative coordinate algebras \cite{Wachter:2007A}:%
\begin{align}
f(\mathbf{x}\oplus\mathbf{y})  &  =((\mathcal{W}^{\hspace{0.01in}-1}%
\otimes\mathcal{W}^{\hspace{0.01in}-1})\circ\underline{\Delta})(\mathcal{W}%
(f)),\nonumber\\[0.08in]
f(\mathbf{x}\,\bar{\oplus}\,\mathbf{y})  &  =((\mathcal{W}^{\hspace{0.01in}%
-1}\otimes\mathcal{W}^{-1})\circ\underline{\bar{\Delta}})(\mathcal{W}(f)).
\label{KonReaBraCop}%
\end{align}
The braided Hopf algebras also have braided antipodes $\underline{S}$ and
$\underline{\bar{S}}$, which can be realized on the corresponding commutative
coordinate algebras as well:%
\begin{align}
f(\ominus\,\mathbf{x})  &  =(\mathcal{W}^{\hspace{0.01in}-1}\circ\underline
{S}\hspace{0.01in})(\mathcal{W}(f)),\nonumber\\
f(\bar{\ominus}\,\mathbf{x})  &  =(\mathcal{W}^{\hspace{0.01in}-1}%
\circ\underline{\bar{S}}\hspace{0.01in})(\mathcal{W}(f)). \label{qInvDef}%
\end{align}
In the following, we refer to the operations in Eq.~(\ref{qInvDef})\ as
$q$\textit{-in\-ver\-sions}. In the case of the $q$-de\-formed Euclidean
space, for example, we found the following operator representation for
$q$-in\-ver\-sions \cite{Wachter:2004phengl}:%
\begin{align}
\hat{U}^{-1}f(\ominus\,\mathbf{x})=  &  \sum_{i=0}^{\infty}(-\hspace
{0.01in}q\lambda\lambda_{+})^{i}\,\frac{(x^{+}x^{-})^{i}}{[[2i]]_{q^{-2}}%
!!}\,q^{-2\hat{n}_{+}(\hat{n}_{+}+\hspace{0.01in}\hat{n}_{3})-2\hat{n}%
_{-}(\hat{n}_{-}+\hspace{0.01in}\hat{n}_{3})-\hat{n}_{3}\hat{n}_{3}%
}\nonumber\\
&  \qquad\times D_{q^{-2},\hspace{0.01in}x^{3}}^{2i}\,f(-\hspace
{0.01in}q^{2-4i}x^{-},-\hspace{0.01in}q^{1-2i}x^{3},-\hspace{0.01in}%
q^{2-4i}x^{+}).
\end{align}
Note that the operators $\hat{U}$ and $\hat{U}^{-1}$ act on a commutative
function $f(x^{+},x^{3},x^{-})$ as follows:%
\begin{align}
\hat{U}f  &  =\sum_{k\hspace{0.01in}=\hspace{0.01in}0}^{\infty}\left(
-\lambda\right)  ^{k}\frac{(x^{3})^{2k}}{[[k]]_{q^{-4}}!}\,q^{-2\hat{n}%
_{3}(\hat{n}_{+}+\hspace{0.01in}\hat{n}_{-}+\hspace{0.01in}k)}D_{q^{-4}%
,\hspace{0.01in}x^{+}}^{k}D_{q^{-4},\hspace{0.01in}x^{-}}^{k}f,\nonumber\\
\hat{U}^{-1}f  &  =\sum_{k\hspace{0.01in}=\hspace{0.01in}0}^{\infty}%
\lambda^{k}\hspace{0.01in}\frac{(x^{3})^{2k}}{[[k]]_{q^{4}}!}\,q^{2\hat{n}%
_{3}(\hat{n}_{+}+\hspace{0.01in}\hat{n}_{-}+\hspace{0.01in}k)}D_{q^{4}%
,\hspace{0.01in}x^{+}}^{k}D_{q^{4},\hspace{0.01in}x^{-}}^{k}f.
\end{align}

The braided co-prod\-ucts and braided antipodes satisfy the axioms (also see
Ref.~\cite{Majid:1996kd})%
\begin{align}
m\circ(\underline{S}\otimes\operatorname*{id})\circ\underline{\Delta}  &
=m\circ(\operatorname*{id}\otimes\,\underline{S}\hspace{0.01in})\circ
\underline{\Delta}=\underline{\varepsilon},\nonumber\\
m\circ(\underline{\bar{S}}\otimes\operatorname*{id})\circ\underline
{\bar{\Delta}}  &  =m\circ(\operatorname*{id}\otimes\,\underline{\bar{S}%
}\hspace{0.01in})\circ\underline{\bar{\Delta}}=\underline{\bar{\varepsilon}},
\label{HopfVerAnfN}%
\end{align}
and%
\begin{align}
(\operatorname*{id}\otimes\,\underline{\varepsilon})\circ\underline{\Delta}
&  =\operatorname*{id}=(\underline{\varepsilon}\otimes\operatorname*{id}%
)\circ\underline{\Delta},\nonumber\\
(\operatorname*{id}\otimes\,\underline{\bar{\varepsilon}})\circ\underline
{\bar{\Delta}}  &  =\operatorname*{id}=(\underline{\bar{\varepsilon}}%
\otimes\operatorname*{id})\circ\underline{\bar{\Delta}}. \label{HopfAxi2}%
\end{align}
In the identities above, we denote the operation of multiplication on the
braided Hopf algebra by $m$. The co-units $\underline{\varepsilon}%
,\underline{\bar{\varepsilon}}$ of the two braided Hopf structures are both
linear mappings that vanish on the coordinate generators:%
\begin{equation}
\varepsilon(X^{i})=\underline{\bar{\varepsilon}}(X^{i})=0.
\end{equation}
For this reason, we can realize the co-units $\underline{\varepsilon}$ and
$\underline{\bar{\varepsilon}}$ on a commutative coordinate algebra as
follows:%
\begin{equation}
\underline{\varepsilon}(\mathcal{W}(f))=\underline{\bar{\varepsilon}%
}(\mathcal{W}(f))=\left.  f(\mathbf{x})\right\vert _{x\hspace{0.01in}%
=\hspace{0.01in}0}=f(0). \label{ReaVerZopNeuEleKomAlg}%
\end{equation}
We can also translate the Hopf algebra axioms in Eqs.~(\ref{HopfVerAnfN}) and
(\ref{HopfAxi2}) into corresponding rules for $q$-translations and
$q$-inversions \cite{Wachter:2007A}, i.~e.%
\begin{align}
f((\ominus\,\mathbf{x})\oplus\mathbf{x})  &  =f(\mathbf{x}\oplus
(\ominus\,\mathbf{x}))=f(0),\nonumber\\
f((\bar{\ominus}\,\mathbf{x})\,\bar{\oplus}\,\mathbf{x})  &  =f(\mathbf{x}%
\,\bar{\oplus}\,(\bar{\ominus}\,\mathbf{x}))=f(0), \label{qAddN}%
\end{align}
and%
\begin{align}
f(\mathbf{x}\oplus\mathbf{y})|_{y\hspace{0.01in}=\hspace{0.01in}0}  &
=f(\mathbf{x})=f(\mathbf{y}\oplus\mathbf{x})|_{y\hspace{0.01in}=\hspace
{0.01in}0},\nonumber\\
f(\mathbf{x}\,\bar{\oplus}\,\mathbf{y})|_{y\hspace{0.01in}=\hspace{0.01in}0}
&  =f(\mathbf{x})=f(\mathbf{y}\,\bar{\oplus}\,\mathbf{x})|_{y\hspace
{0.01in}=\hspace{0.01in}0}. \label{qNeuEle}%
\end{align}

\begin{figure}
[ptb]
\centerline{\psfig{figure=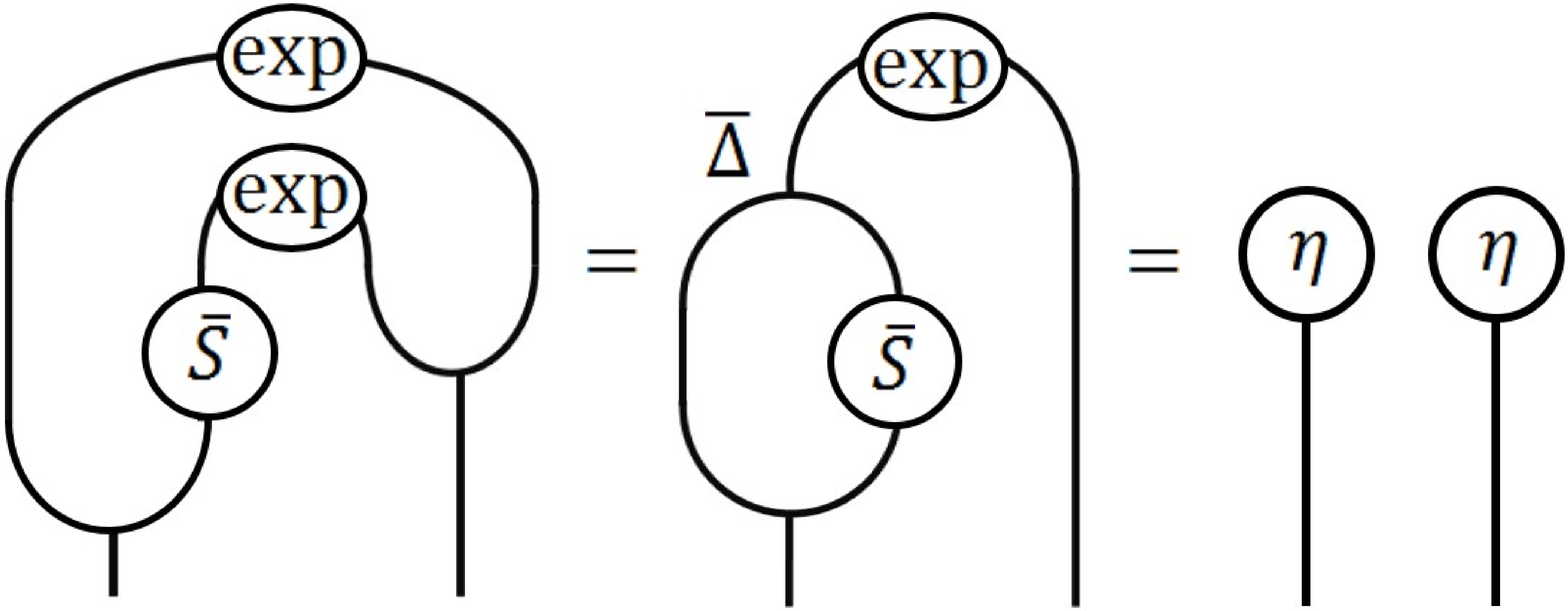,width=2.5754in}}%
\caption{Invertibility of $q$-exponentials.}%
\label{Fig3}
\end{figure}

Using $q$-in\-ver\-sions, we are able to introduce inverse $q$%
-ex\-po\-nen\-tials:%
\begin{equation}
\exp_{q}(\bar{\ominus}\,\mathbf{x}|\text{i}\mathbf{p})=\exp_{q}(\text{i}%
\mathbf{x}|\text{{}}\bar{\ominus}\,\mathbf{p}). \label{InvExpAlgDefKom}%
\end{equation}
Due to the addition theorems and the normalization conditions of our
$q$-ex\-po\-nen\-tials, the following applies:%
\begin{equation}
\exp_{q}(\text{i}\mathbf{x}\circledast\exp_{q}(\bar{\ominus}\,\mathbf{x}%
|\hspace{0.01in}\text{i}\mathbf{p})\circledast\mathbf{p})=\exp_{q}%
(\mathbf{x}\,\bar{\oplus}\,(\bar{\ominus}\,\mathbf{x})|\hspace{0.01in}%
\text{i}\mathbf{p})=\exp_{q}(\mathbf{x}|\text{i}\mathbf{p})|_{x=0}=1.
\end{equation}
For a better understanding of these identities, we have again given their
graphic representation in Fig.~\ref{Fig3}.\footnote{You find some explanations
of this sort of graphical calculations in Ref.~\cite{Majid:2002kd}.} The
conjugate $q$-ex\-po\-nen\-tials $\overline{\exp}_{q}$ are subject to similar
rules, which we obtain from the above identities by using the following
substitutions:%
\begin{equation}
\exp_{q}\rightarrow\overline{\exp}_{q},\qquad\bar{\oplus}\,\rightarrow
\,\oplus,\qquad\bar{\ominus}\,\rightarrow\,\ominus.
\end{equation}%

\begin{figure}
[ptb]
\centerline{\psfig{figure=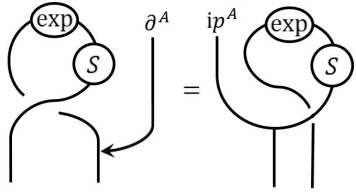,width=1.817in}}%
\caption{Eigenvalue equation of twisted $q$-exponential.}%
\label{Fig4}%
\end{figure}

Finally, we describe another way of obtaining $q$-ex\-po\-nen\-tials. For
this purpose, we exchange the two tensor factors of a $q$-ex\-po\-nen\-tial
using the inverse of the so-called universal R-ma\-trix (also see the graphic
representation in Fig.~\ref{Fig4}):%
\begin{align}
\exp_{q}^{\ast}(\text{i}\mathbf{p}|\hspace{0.01in}\mathbf{x})  &  =\tau
\circ\lbrack(\mathcal{R}_{[2]}^{-1}\otimes\mathcal{R}_{[1]}^{-1}%
)\triangleright\exp_{q}(\text{i}\mathbf{x}|\hspace{-0.03in}\ominus
\hspace{-0.01in}\mathbf{p})],\nonumber\\
\exp_{q}^{\ast}(\mathbf{x}|\text{i}\mathbf{p})  &  =\tau\circ\lbrack
(\mathcal{R}_{[2]}^{-1}\otimes\mathcal{R}_{[1]}^{-1})\triangleright\exp
_{q}(\ominus\hspace{0.02in}\mathbf{p}|\hspace{0.01in}\text{i}\mathbf{x})].
\label{DuaExp2}%
\end{align}
In the expressions above, $\tau$ denotes the ordinary twist operator. It can be
shown that the new $q$-ex\-po\-nen\-tials satisfy the following eigenvalue
equations (see Fig.~\ref{Fig4}):%
\begin{align}
\exp_{q}^{\ast}(\text{i}\mathbf{p}|\hspace{0.01in}\mathbf{x})\triangleleft
\partial^{A}  &  =\text{i}p^{A}\circledast\exp_{q}^{\ast}(\text{i}%
\mathbf{p}|\hspace{0.01in}\mathbf{x}),\nonumber\\
\partial^{A}\,\bar{\triangleright}\,\exp_{q}^{\ast}(\mathbf{x}|\text{i}%
^{-1}\mathbf{p})  &  =\exp_{q}^{\ast}(\mathbf{x}|\text{i}^{-1}\mathbf{p}%
)\circledast\text{i}p^{A}. \label{EigGleExpQueAbl}%
\end{align}

Similar considerations apply to the conjugate $q$-ex\-po\-nen\-tials. To this
end, we only need to modify Eqs.~(\ref{DuaExp2}) and (\ref{EigGleExpQueAbl})
by performing the following substitutions:%
\begin{gather}
\exp_{q}^{\ast}\rightarrow\overline{\exp}_{q}^{\ast},\qquad\mathcal{R}%
_{[2]}^{-1}\otimes\mathcal{R}_{[1]}^{-1}\rightarrow\mathcal{R}_{[1]}%
\otimes\mathcal{R}_{[2]},\qquad\ominus\,\rightarrow\,\bar{\ominus},\nonumber\\
\bar{\triangleright}\,\rightarrow\,\triangleright,\qquad\triangleleft
\,\rightarrow\,\bar{\triangleleft},\qquad\partial^{A}\rightarrow\hat{\partial
}^{A}.
\end{gather}


{\normalsize
\bibliographystyle{abbrv}
\bibliography{book,habil}
}

\end{document}